\documentclass[a4paper,11pt]{article}
\pdfoutput=1 

\usepackage{jheppub} 

\usepackage[T1]{fontenc} 
\usepackage{epsfig,multicol}
\usepackage{amsmath,amssymb,bm}

\title{\boldmath
On the
gauge invariant path-integral measure for the overlap Weyl fermions in $\underbar{16}$ of SO(10)
}

\author{Yoshio Kikukawa}
\affiliation{Institute of Physics, University of Tokyo, Tokyo 153-8902, Japan}
\emailAdd{kikukawa@hep1.c.u-tokyo.ac.jp}

\subheader{
\begin{flushright}
\rm 
UT-Komaba/17-11
\end{flushright}
}

\abstract{
We consider the lattice formulation of SO(10) chiral gauge theory with left-handed Weyl fermions in the sixteen dimensional spinor representation ($\underbar{16}$) within the framework of the Overlap fermion/the Ginsparg-Wilson relation. 
We 
define
a manifestly gauge-invariant path-integral measure for the left-handed Weyl field using all the components of the Dirac field, but the right-handed part of which is just saturated completely by inserting a suitable product of the SO(10)-invariant  't Hooft vertices in terms of the right-handed field.
The definition of the measure applies to all possible topological sectors.
The measure possesses all required transformation properties under lattice symmetries
and 
the induced effective action is CP invariant.
The global U(1) symmetry of the left-handed 
field is anomalous due to the non-trivial transformation of the measure, while that of the right-handed field is explicitly  
broken by the 't Hooft vertices.
There remains the issue of 
locality in the gauge-field dependence of the Weyl fermion measure,  but the question can be addressed   
in the weak gauge-coupling expansion at least
by Monte Carlo methods 
without encountering the sign problem.
We also discuss the relations of our formulation to 
other approaches/proposals
to decouple the species-doubling/mirror degrees of freedom.
Those include
Eichten-Preskill model, 
Ginsparg-Wilson Mirror-fermion model, 
Domain wall fermion model with the boundary Eichten-Preskill term,
4D Topological Insulator/Superconductor with gapped boundary phase, 
and
the recent studies on 
the PMS phase/``Mass without symmetry breaking''.
We clarify the similarity and
the difference in technical detail and 
show 
that our proposal is a well-defined testing ground 
for that basic question.

}

\begin{document} 
\maketitle
\flushbottom


%
\section{Introduction}
\label{sec:intro}

The experimental verification of the elementary particle spectrum of the standard model is now completed with the observation of a new particle with a mass of 
125.09 GeV by the ATLAS and CMS Collaborations through
LHC Run 1\cite{Aad:2012tfa, Chatrchyan:2012xdj,Aad:2015zhl}, 
for which all measurements of the properties, including its spin, CP properties, and coupling strengths to SM particles, are consistent within the uncertainties with those expected for the SM Higgs boson.
On the other hand,  the discoveries of neutrino mixings by 
the Super-Kamiokande Collaboration and 
the Sudbury Neutrino Observatory (SNO) Collaboration
\cite{Fukuda:1998mi,Ahmad:2001an,Ahmad:2002jz}
imply the masses of neutrinos and 
suggest 
the existence of the right-handed neutrinos.  
The consistency of the standard model predictions with experiments 
in high precision implies that
the standard model particles and 
the possible right-handed neutrinos are well-decoupled
from the fundamental scale of the Plank mass
and even from the possible scale of new physics beyond
the standard model.
And the low-lying fermionic elementary particles fit into the sixteen-dimensional irreducible representation of SO(10), $\underbar{16}$, which is complex, but free from gauge anomaly. 

Chiral gauge theories such like the SO(10) gauge theory with $\underbar{16}$s have several interesting possibilities in their own dynamics: fermion number non-conservation due to chiral anomaly\cite{'tHooft:1976up, 'tHooft:1976fv}, 
various realizations of the gauge symmetry and global flavor 
symmetry\cite{Raby:1979my, Dimopoulos:1980hn},  
the existence of massless composite fermions suggested by 't Hooft's anomaly matching condition\cite{'tHooft:1979bh}, 
the classical scale invariance and the vanishing vacuum energy\cite{Holdom:2007gg,Holdom:2009ma} 
and so on. Unfortunately, little is known so far about the actual behavior of (non-supersymmetric) chiral gauge theories beyond perturbation theory.  It is then desirable to develop a formulation to study  the non-perturbative aspect of chiral gauge theories. 

Lattice gauge theory can now
provide a framework for non-perturbative formulation of chiral gauge theories, despite the well-known problem of the species 
doubling \cite{Karsten:1980wd,Nielsen:1980rz,Nielsen:1981xu,Friedan:1982nk}.\footnote{See \cite{Luscher:1999mt,Golterman:2000hr,Neuberger:2000kg,Luscher:2000hn,Kaplan:2009yg} for the recent  reviews on this subject.}
The clue to this development is 
the construction of {\em local and gauge-covariant} lattice Dirac operators 
satisfying the Ginsparg-Wilson relation\cite{Ginsparg:1981bj,
Neuberger:1997fp,Hasenfratz:1998ri,Neuberger:1998wv,
Hasenfratz:1998jp,Hernandez:1998et}.
\begin{equation}
\gamma_5 D + D \hat \gamma_5 = 0, \qquad  \hat \gamma_5 = \gamma_5(1-2aD) . 
\end{equation}
%
An explicit example of such lattice Dirac operator is given by the overlap Dirac operator \cite{Neuberger:1997fp,Neuberger:1998wv}, 
which was derived by Neuberger from the overlap formalism 
\cite{
Narayanan:wx,Narayanan:sk,Narayanan:ss,Narayanan:1994gw,Narayanan:1993gq, Narayanan:1993gt, 
Neuberger:1997ze,
Narayanan:1998uu, Neuberger:1998qp, Neuberger:1998wf, Neuberger:1998ms,
Neuberger:1999ry,Neuberger:1999zk,Neuberger:1999jw, Neuberger:2000ab, Neuberger:2001nb, Neuberger:2003nu,
Narayanan:1995sv, Narayanan:1996cu,Huet:1996pw,
Narayanan:1997by,Kikukawa:1997qh,Neuberger:1998xn,
Narayanan:1996kz,Kikukawa:1997md,Kikukawa:1997dv, Neuberger:1998my, Neuberger:1998jk,
Neuberger:1999re, Neuberger:2000ac, Fosco:2007ry}.
The overlap formula was derived from the five-dimensional approach of domain wall fermion proposed by Kaplan\cite{Kaplan:1992bt, Golterman:1992ub}.
In the vector-like formalism of domain wall fermion by Shamir and Furman\cite{Shamir:1993zy, Furman:ky,
Blum:1996jf, Blum:1997mz}, 
the {\em local} low energy effective action of the chiral mode 
is precisely given by the overlap Dirac 
operator \cite{Vranas:1997da,Neuberger:1997bg, Kikukawa:1999sy}.
%
By the Ginsparg-Wilson relation, it is possible to realize an exact chiral symmetry 
on the lattice\cite{Luscher:1998pq,Kikukawa:1998pd,Luscher:1998kn,Fujikawa:1998if,Adams:1998eg,Suzuki:1998yz,Chiu:1998xf}
in the manner consistent 
with the no-go theorem.
It is also possible to introduce Weyl fermions on the lattice and 
this opens the possibility
to formulate anomaly-free chiral gauge theories on the lattice\cite{Luscher:1998kn,Luscher:1998du,
Luscher:1999un,Luscher:1999mt,Luscher:2000hn,Suzuki:1999qw,Neuberger:2000wq,Adams:2000yi,
Suzuki:2000ii,Igarashi:2000zi,Luscher:2000zd,Aoyama:1999hg,Kikukawa:2000kd,Kikukawa:2001jm,Kikukawa:2001mw,Kadoh:2003ii,Kadoh:2004uu,Kikukawa:2005ILFTN,Kadoh:2007wz,Kadoh:2007xb}.
In the case of U(1) chiral gauge theories,  
L\"uscher\cite{Luscher:1998du}  proved rigorously that
it is possible to construct the fermion path-integral measure 
which depends smoothly on the gauge field  and
fulfills the fundamental requirements such as 
locality, gauge-invariance and lattice symmetries.\footnote{In the above formulation 
of U(1) chiral lattice gauge theories\cite{Luscher:1998du}, 
although the proof of the existence of the fermion measure is constructive,  the resulted formula 
of the fermion measure turns out to be rather complicated for the case of the finite-volume lattice. 
It also relies on the results obtained in the infinite lattice.  Therefore it does not provide a formulation 
which is immediately usable for numerical applications. See 
\cite{Kadoh:2003ii, Kadoh:2004uu,Kikukawa:2005ILFTN,Kadoh:2007wz} for a simplified formulation toward a practical implementation.} 
This construction was extended to 
the SU(2)$\times$U(1) chiral gauge theory of the Glashow-Weinberg-Salam model\cite{Glashow:1961tr,Weinberg:1967tq,Salam:1968rm} based on the pseudo reality and anomaly-free conditions
of SU(2)$\times$U(1) by Kadoh and the author\cite{Kadoh:2007xb}.
For generic non-abelian chiral gauge theories, 
the construction in all orders of the weak gauge-coupling expansion was given by Suzuki\cite{Suzuki:2000ii,Igarashi:2000zi} and by L\"uscher\cite{Luscher:2000zd}. However, a non-perturbative construction is not obtained yet so far. 
%


In this article, 
we consider the lattice formulation of SO(10) chiral gauge theory with Weyl fermions in the sixteen dimensional spinor representation $\underbar{16}$ within the framework of the Overlap fermion/the Ginsparg-Wilson relation.\footnote{
In the continuum theory, 
Slavnov and Frolov have proposed a gauge-invariant
reguralization method
for the standard model based on the SO(10) chiral gauge theory,
employing the infinite number of the Pauli-Villars fields\cite{Frolov:1992ck}.
The close relation of this method with Kaplan's domain wall fermion
was pointed out by Narayanan and Neuberger.\cite{Narayanan:wx}
By this method, 
the parity-even part of 
the one-loop effective action induced by the chiral fermions
is indeed reguralized gauge-invariantly.
But the parity-odd part is not actually reguralized and, 
in particular, the part of gauge anomaly is indefinite.
Therefore
it is difficult to see
the cancellation of gauge anomaly unambiguously.
%
The procedure to compute the chiral anomaly with this method
has been given by Aoki and the present author\cite{Aoki:1993fz}.
%
}  We propose a manifestly gauge-invariant path-integral measure of the left-handed Weyl fermions, which is defined with all the components of the Dirac field, but the right-handed part of which is just saturated completely by inserting a suitable product of the SO(10)-invariant  't Hooft vertices in terms of the right-handed field.
The definition of the measure applies to all possible topological sectors.
The measure possesses all required transformation properties under lattice symmetries
and 
the induced effective action is CP invariant.
The global U(1) symmetry of the left-handed 
field is anomalous due to the non-trivial transformation of the measure\cite{Kikukawa:1998pd, Luscher:1998kn,Fujikawa:1998if, Adams:1998eg, Suzuki:1998yz, Chiu:1998xf}, while that of the right-handed field is explicitly  broken by the 't Hooft vertices.
%
There remains the issue of smoothness/locality in the gauge-field dependence of the Weyl fermion measure.
But 
the question is well-defined and can be addressed 
in the weak gauge-coupling limit at least
through Monte Carlo simulations without encountering the sign problem.

We also discuss the relation of our formulation with
other approaches/proposals
to decouple the species-doublers and mirror fermions. 
Those include 
the Eichten-Preskill model \cite{Eichten:1985ft, Golterman:1992yha}, 
the Mirror-fermion model
using Ginsparg-Wilson fermions \cite{Bhattacharya:2006dc,Giedt:2007qg,Poppitz:2007tu,Poppitz:2008au,Poppitz:2009gt,Poppitz:2010at,Chen:2012di,Giedt:2014pha},
Domain wall fermions with the boundary Eichten-Preskill term\cite{Creutz:1996xc, Neuberger:1997cz}, 
the recent studies on ``Mass without symmetry breaking'' \cite{Ayyar:2014eua, BenTov:2014eea, BenTov:2015gra, Ayyar:2015lrd, Ayyar:2016lxq, Catterall:2015zua, Catterall:2016dzf, Catterall:2017ogi,Schaich:2017czc}/the PMS phase \cite{Giedt:2007qg,Gerhold:2007yb,Gerhold:2007gx}
and 
4D TI/TSCs with gapped boundary phases \cite{Creutz:1994ny,Qi:2008ew,Wen:2013ppa,Wang:2013yta,You:2014oaa,You:2014vea,DeMarco:2017gcb}.
We clarify the similarity and
the difference in technical detail, trying to show 
that our proposal is a well-defined testing ground for that basic question.\footnote{See \cite{Eichten:1985ft,Swift:1984dg,Smit:1985nu,Aoki:1987cb,Aoki:1988cu,Funakubo:1987zt,Funakubo:1988tg,
Golterman:1992yha,
Golterman:1990zu,Golterman:1991re,
Bock:1991bu,Bock:1992gp,Aoki:1991es} for the former attempts to decouple the species-doublers/mirror-fermions
by strong Yukawa, Wilson-Yukawa and multi-fermion interactions. 
See also \cite{Montvay:1987ys, Montvay:1988av, Farakos:1990ex, Lin:1990ue, Lin:1990vi, Munster:1991xs, Lin:1991csa, Montvay:1992mv, Montvay:1992eg, Lin:1992qb, Lin:1993hp} for the original 
Mirror-fermion approach using Wilson fermions.
}
\footnote{
The recent proposal by Grabowska and Kaplan\cite{Grabowska:2015qpk,Grabowska:2016bis,Kaplan:2016dgz,Fukaya:2016ofi,Okumura:2016dsr,Makino:2016auf,Makino:2017pbq,Hamada:2017tny}
is ``orthogonal'' to 
the approaches 
discussed in this paper.
It is based on the original domain wall fermion by Kaplan\cite{Kaplan:1992bt}, but
coupled to the
``five-dimensional'' link field
which is obtained from the dynamical four-dimensional link field
at the target wall by the gradient flow toward the mirror wall.
This choice of the ``five-dimensional'' link field 
makes possible a chiral gauge coupling for the target and mirror walls, while keeping the system four-dimensional and gauge-invariant.
It is ``orthogonal'' in the sense that
the authors do not try to decouple the massless-modes at the mirror wall, but interpret them as physical degrees of freedom with very soft form factor caused by the gradient flow, 
and that the authors do not try (do not need) to break explicitly the continuous global symmetries with ``would-be gauge anomalies'' in the mirror-wall sector,
which would be required if one would try to decouple
the mirror-modes as claimed by Eichten and Preskill
and by the other authors\cite{Eichten:1985ft,Bhattacharya:2006dc,Wang:2013yta}.
}

It is known that a chiral gauge theory is a difficult case for numerical simulations
because the effective action induced by 
Weyl fermions 
has a non-zero imaginary part.
But 
in view of the recent studies of the 
simulation methods based on the complex Langevin dynamics\cite{Parisi:1984cs, 
Klauder:1983zm, 
Klauder:1983sp, 
Ambjorn:1985iw, 
Ambjorn:1986fz, 
Berges:2006xc,
Berges:2007nr, 
Aarts:2008rr, 
Aarts:2008wh, 
Aarts:2009dg,
Aarts:2009uq, 
Aarts:2010aq, 
Aarts:2011ax, 
Aarts:2011zn, 
Seiler:2012wz,
Pawlowski:2013pje, 
Pawlowski:2013gag, 
Aarts:2013uxa, 
Sexty:2013ica, 
Aarts:2013fpa,
Giudice:2013eva,
Mollgaard:2013qra,
Sexty:2014zya,
Hayata:2014kra, 
Splittorff:2014zca, 
Aarts:2015oka,
Fodor:2015doa,
Salcedo:2015jxd,
Hayata:2015lzj, 
Li:2016srv,
Aarts:2016qrv,
Abe:2016hpd,
Ito:2016efb,
Salcedo:2016kyy, 
Aarts:2017vrv, 
Fujii:2017oti} 
and the complexified path-integration on Lefschetz thimbles\cite{
Pham:1983, 
Kaminski:1994, 
Howls:1997, 
Witten:2010cx,
Cristoforetti:2012su, 
Cristoforetti:2012uv, 
Cristoforetti:2013wha, 
Mukherjee:2013aga, 
Fujii:2013sra, 
Cherman:2014ofa,
Cristoforetti:2014gsa, 
Mukherjee:2014hsa, 
Aarts:2014nxa, 
Tanizaki:2014xba, 
Nishimura:2014kla, 
Tanizaki:2014tua, 
Kanazawa:2014qma, 
Behtash:2015kna, 
Tanizaki:2015pua, 
DiRenzo:2015foa, 
Behtash:2015kva,
Fukushima:2015qza, 
Tanizaki:2015rda, 
Fujii:2015bua, 
Fujii:2015vha,
Behtash:2015zha, 
Alexandru:2015xva, 
Behtash:2015loa,
Scorzato:2015qts, 
Alexandru:2015sua, 
Alexandru:2016lsn, 
Alexandru:2016gsd, 
Alexandru:2016san, 
Fujimori:2016ljw,
Alexandru:2016ejd, 
Tanizaki:2016xcu, 
Fujimori:2017oab,
Fukuma:2017fjq, 
Alexandru:2017oyw,  
Nishimura:2017vav,
Mori:2017zyl,
Fujimori:2017osz,
Tanizaki:2017yow,
Bedaque:2017epw}, 
it would be still interesting and even useful to 
develop a formulation of chiral lattice gauge theories by which one can work out 
fermionic observables numerically 
as the functions of link field with exact gauge invariance.

This article is organized as follows. 
In section~\ref{sec:SO10-on-lattice},
we introduce our lattice formulation of SO(10) gauge
theory with left-handed Weyl field in 
$\underbar{16}$ at the classical level.
In section~\ref{sec:path-integration-measure},
we define the  path-integral 
measures of 
the left-handed Weyl field 
and discuss its properties.
In section~\ref{sec:saturation-right-handed-measure},
we examine in detail the the saturation of 
the right-handed part of 
the fermion measure by 't Hooft vertices.
In section~\ref{sec:other-theories},
we discuss the cases of other anomalous
and anomaly-free chiral gauge theories.
Section~\ref{sec:rel-other-approaches} is devoted to the discussions
of the relations to other approaches/proposals.
In section~\ref{sec:conclusion}, we conclude with 
a summary and discussions.


\section{The SO(10) chiral lattice gauge theory with 
overlap Weyl fermions
}
\label{sec:SO10-on-lattice}

In this section, we describe a construction of the SO(10) chiral gauge theory  on the lattice
within the framework of chiral lattice gauge theories 
based on the lattice Dirac operator satisfying the Ginsparg-Wilson 
relation \cite{Luscher:1998du,Luscher:1999un}.
We assume a local, gauge-covariant lattice Dirac operator $D$ which satisfies the Ginsparg-Wilson relation. An explicit example of such lattice Dirac operator is given by the overlap Dirac operator \cite{Neuberger:1997fp,Neuberger:1998wv}, 
which was derived from the overlap formalism 
\cite{
Narayanan:wx,Narayanan:sk,Narayanan:ss,Narayanan:1994gw,Narayanan:1993gq,
Neuberger:1999ry,Narayanan:1996cu,Huet:1996pw,
Narayanan:1997by,Kikukawa:1997qh,Neuberger:1998xn}.
In this case, our formulation is equivalent to the overlap formalism  
for chiral lattice gauge theories\footnote{
The overlap formalism gives a well-defined partition function of Weyl fermions on the lattice,
which nicely reproduces the fermion zero mode and the fermion-number
violating observables ('t Hooft vertices) \cite{Narayanan:1996kz,Kikukawa:1997md,Kikukawa:1997dv}. 
The gauge-invariant construction by L\"uscher \cite{Luscher:1998du} 
provides a procedure to fix the ambiguity of  the complex phase of the overlap formula
 in a gauge-invariant manner for anomaly-free U(1) chiral gauge theories.
}
or  the domain wall fermion approach \cite{Kikukawa:2001mw, Aoyama:1999hg}.

In the followings, we consider the four-dimensional lattice $\Lambda$ of the finite size $L$ and choose lattice units $a=1$:
\begin{equation}
\Lambda =
 \left\{x=(x_1, x_2, x_3, x_4)  \in \mathbb{Z}^4 \, \vert \, \, 0 \le x_\mu < L \,  (\mu = 0,1,2,3) \right\} . 
\end{equation}
The unit vector in the directions $\mu (=0, 1, 2, 3)$ are denoted as $\hat \mu$.

\subsection{Gauge field of SO(10)}

The gauge field of SO(10) is defined as the link field on the lattice $\Lambda$. The SO(10) link variables are at first introduced
in the (reducible) spinor
representaion as the thirty-two dimensional special unitary matrices, $ U(x,\mu) \, \in \text{Spin(10)} $. The generators of Spin(10) are given by $\Sigma_{ab}= -  \frac{i}{4}\big[ \Gamma^a, \Gamma^b \big]$, where $\{ \Gamma^a \, \vert \, a=1,2,\cdots,10 \}$ form the Clifford algebra, 
$\Gamma^a \Gamma^b + \Gamma^b \Gamma^a = 2 \delta^{ab} 
\, (a,b =1,2,\cdots, 10)$. An explicit representation for $\{ \Gamma^a \, \vert \, a=1,2,\cdots,10 \}$ is given in the appendix~\ref{sec:appendix-so10-representation}.
The link variables are then parametrized as
\begin{equation}
U(x,\mu) = {\rm e}^{ i \theta^{ab}(x,\mu) \Sigma_{ab} /2} \quad \in \, \text{Spin(10)}.
\end{equation}
We require the admissibility condition on the gauge field, 
\begin{eqnarray}
\|1- P(x,\mu,\nu)  \|  < \epsilon , 
\end{eqnarray}
for all $x$, $\mu, \nu$, where the plaquette variables are defined by 
\begin{eqnarray}
P(x,\mu,\nu)&=&
U(x,\mu)
U(x+\hat\mu,\nu)
U(x+\hat\nu,\mu)^{-1}
U(x,\nu)^{-1} .  
\end{eqnarray}
This condition ensures that 
the overlap Dirac operator\cite{Neuberger:1997fp,Neuberger:1998wv}, which is assumed to act on the fermion fields in the spinor representations of SO(10), is a
smooth and local function of  the gauge field 
if $ \epsilon <1/30$\cite{Hernandez:1998et}.

To impose the admissibility condition dynamically, we adopt the following action for the gauge field:
\begin{eqnarray}
S_{\rm G} &=& \frac{1}{g^2} \sum_{x\in \Gamma} \sum_{\mu,\nu}
 {\rm tr}\{ 1-\tilde P(x,\mu,\nu) \}  
  \left[ 1 -   {\rm tr}\{ 1-\tilde P(x,\mu,\nu) \}/ 10\epsilon^2 \right]^{-1} , 
\end{eqnarray}
where the SO(10) link variables are represented 
in the defining representation as the ten-dimensional special orthogonal matrices, $\tilde  U(x,\mu) \, \in \text{SO(10)}$. 
The generators of SO(10) in the defining representation are given by $\{\tilde \Sigma_{ab}\}_{cd} = i(\delta_{ac} \delta_{bd} - \delta_{ad} \delta_{bc} )$ and the link variables are  represented with the same parameters as 
\begin{equation}
\tilde U(x,\mu) = {\rm e}^{ i \theta^{ab}(x,\mu) \tilde \Sigma_{ab} /2} \quad \in \, \text{SO(10)}.
\end{equation}





\subsection{Weyl field in 16-dimensional spinor representation of SO(10) }

%
The left-handed Weyl field in the 16-dimensional (irreducible) spinor representation of SO(10) is defined on the lattice $\Lambda$ based on the Ginsparg-Wilson relation.
First we introduce a Dirac field on the lattice in the 16-dimensional spinor representation of SO(10),
\begin{equation}
\psi(x) = {\rm P}_+ \psi (x), \qquad  \bar \psi(x) = \bar \psi (x)  {\rm P}_+ , 
\end{equation}
where 
\begin{equation}
 {\rm P}_+= \frac{1+\Gamma^{11}}{2} ,
\qquad \Gamma^{11} = - i \Gamma^1 \Gamma^2 \cdots \Gamma^{10}.
\end{equation}
We also introduce 
the overlap Dirac operator $D$ acting on $\psi(x)$ as
\begin{equation}
\label{eq:def-overlap-D-X}
D = \frac{1}{2} \left( 1 + X \frac{1}{\sqrt{X^\dagger X}} \right),
\qquad
X = \gamma_\mu  \frac{1}{2}\big( \nabla_\mu - \nabla_\mu^\dagger \big) +  \frac{1}{2} \nabla_\mu \nabla_\mu^\dagger -m_0 ,
\end{equation}
where $\nabla_\mu$ is the covariant difference operator which  acts on $\psi(x)$ as 
$\nabla_\mu \psi(x) = U(x,\mu) \psi(x+\hat\mu) - \psi(x)$ and $0 < m_0 < 2$. 
Under the admissibility condition, $D$ is a local, gauge-covariant lattice Dirac operator.
It also satisfies the Ginsparg-Wilson relation, 
\begin{equation}
\label{eq:GW-rel}
\gamma_5 D + D \hat \gamma_5  = 
0, 
\end{equation}
where 
\begin{equation}
\label{eq:hat-gamma5} 
 \hat \gamma_5 \equiv  \gamma_5(1- 2 D), \qquad (\hat \gamma_5 )^2 = \mathbb{I}. 
\end{equation}
Then we define the left-handed Weyl fermions 
in the 16-dimensional spinor representation of SO(10) 
by the eigenstates of the chiral operators, $\hat \gamma_5$ for the field and $\gamma_5$ for the anti-fields:
\begin{equation}
\psi_-(x) = \hat P_- \psi(x) ,  \qquad
\bar \psi_-(x) = \bar \psi(x) P_+,    
\end{equation}
where $\hat P_\pm$ and $P_\pm$ are the chiral projection operators given by 
\begin{equation}
\hat P_{\pm} = \left( \frac{1\pm \hat \gamma_5}{2} \right) , \quad
 P_{\pm} = \left( \frac{1\pm  \gamma_5}{2} \right) .
 \end{equation}
We note that $ \big[ \hat P_\pm , {\rm P}_\pm \big] = 0$ and $ \big[ P_\pm , {\rm P}_\pm \big] = 0$.
 
The action of the left-handed Weyl field in the 16-dimensional spinor representation of SO(10) is given by
\begin{equation}
S_{\rm W}[\psi_-, \bar \psi_-] 
= \sum_{x\in\Lambda} \bar \psi_- (x) D \psi_-(x)  
= \sum_{x\in\Lambda} \bar \psi (x) P_+ D 
\psi(x),
\end{equation}
where we note the relation $P_+ D \hat P_- = P_+ D$.
This action is manifestly invariant under the SO(10) lattice gauge transformations. It is also invariant under 
the global U(1) transformation of the left-handed fields, 
\begin{eqnarray}
 \delta_\alpha \psi_-(x)  &=& \, \, \, \,\,  i \alpha\,  \psi_-(x)  \quad \big[\,  \text{or} \quad  \delta \psi(x) = \, \, \, \,\, i \alpha\,  \hat P_-  \psi(x) \big] , \\
 \delta_\alpha  \bar \psi_-(x) &=& -  i \alpha\,  \bar \psi_-(x) \quad \big[\,  \text{or}  \quad \delta \bar \psi(x) = - i \alpha\,  \bar \psi(x)  P_+  \big].  
\end{eqnarray}
This global U(1) symmetry is, 
as we will see below,  
broken due to the non-trivial transformation property of the Weyl field path-integral measure 
and the non-vanishing vacuum expectation values of 't Hooft vertices, 
\begin{eqnarray}
\label{eq:'tHooft-vertex-field}
T_-(x) &=& \frac{1}{2}\, V_-^a(x) V_-^a(x) , \quad 
V_-^a(x) = \frac{1}{2}\,\psi_-(x)^{\rm T} i \gamma_5 C_D {\rm T}^a \psi_- (x), 
\\
\label{eq:'tHooft-vertex-antifield}
\bar T_-(x) &=& \frac{1}{2}\, \bar V_-^a(x) \bar V_-^a(x) , \quad 
\bar V_-^a(x) = \frac{1}{2}\,\bar \psi_-(x) i \gamma_5 C_D {{\rm T}^a }^\dagger\bar\psi_- (x)^{\rm T}, 
\end{eqnarray}
in the topologically nontrivial 
sectors of the gauge field.
Here ${\rm T}^a$ $(a=1,2,\cdots, 10)$ are the operators  acting on the SO(10) spinor space, 
${\rm T}^a =  {\rm C} \Gamma^a$. The explicit representations of ${\rm C}$ and $\{ {\rm T}^a \vert a=1,\cdots, 10\}$ are given in the 
appendix~\ref{sec:appendix-so10-representation}. 
The action also possesses all required transformation properties under 
lattice symmetries: translations, rotations, reflections and charge conjugation.
In particular, under $\rm P$ (space reflections) and $\rm C$ (charge conjugation)  the action is not invariant, while 
under $\rm CP$ the action is transformed into the same form, but 
the definitions of the chiral projection for the fields and anti-fields are interchanged:
\begin{eqnarray}
&& \psi_-(x) = \hat P_- \psi(x)  \quad \Rightarrow \quad  \psi_-(x) = P_- \psi(x) ,  \\
&& \bar \psi_-(x) = \bar \psi P_+ (x) \quad \Rightarrow \quad  \bar \psi_-(x) = \bar \psi \{ \gamma_5 \hat P_+ \gamma_5\} (x) .
\end{eqnarray}
But the effective action of the gauge field turns out to be CP invariant. 
This {\rm CP} transformation property of the model will be discussed below.

%

\subsection{Topology of the SO(10) lattice gauge fields}
The admissibility condition ensures that 
the overlap Dirac operator\cite{Neuberger:1997fp,Neuberger:1998wv}
is a smooth and local function of  the gauge field \cite{Hernandez:1998et}. 
Moreover, the Ginsparg-Wilson relation implies the index theorem 
\begin{equation}
{\rm Index} \, D =  {\rm Tr}\gamma_5 ( 1- D ).
\end{equation}
Then,  through the lattice Dirac operator $D$, 
it is possible to define a topological 
charge of the gauge fields \cite{Narayanan:sk,Narayanan:ss,Narayanan:1993gq,
Hasenfratz:1998ri,Luscher:1998pq}: 
for the admissible SO(10) gauge fields, one has
\begin{equation}
Q     = -\frac{1}{8}{\rm Tr} \gamma_5(1- D)
        = - \frac{1}{8} \sum_{x  \in \Gamma}  
        {\rm tr}\left\{ \gamma_5(1- D) \right\}(x,x) , 
\end{equation}
where $D(x,y)$ is the kernel of the lattice Dirac operator $D$. 
(Our convention for the gamma matrices is such that 
$\gamma_0 \gamma_1 \gamma_2 \gamma_3 \gamma_5 =1$.)
Then the admissible SO(10) gauge fields can be classified by the topological numbers $Q$.\footnote{
Strictly speaking, the complete topological classification of the space of admissible SO(10) gauge fields is not known yet.  
We assume that it is classified with $Q$ as in the continuum theory.
}
We denote the space of  the admissible SO(10) gauge fields with a given topological charge 
$Q$ by $\mathfrak{U}[Q]$.



The instanton solutions of SU(2) gauge field can be embedded into 
the mutually commuting SU(2) subgroups of the Spin(10)
gauge field. In such a case, 
the index 
counts as
\begin{equation}
{\rm Index} \, D = \sum_{\rm SU(2)} m \, q , 
\end{equation}
where $q$ is the topological charge of the embedded instanton solution, and $m$ is
the multiplicity of the doublets ($\underbar{2}$s) of the embeding SU(2) subgroup, which is an integer multiple of $4$
for the sixteen-dimensional irreducible representation
of Spin(10).


\section{Path Integration -- a proposal for the gauge-invariant measure}
\label{sec:path-integration-measure}


\subsection{Definition of the path integration measures}

The path-integral measures for the link field and the Weyl field  are formulated as follows. For the link field $U(x,\mu)$, it is defined with the group-invariant Haar measure
as usual:
\begin{eqnarray}
{\cal D}[U] 
&\equiv& \prod_{x \in \Lambda} \prod_{\mu=0}^3 d U(x,\mu) .
\end{eqnarray}
For the Weyl field $\psi_-(x)$, $\bar \psi_-(x)$, it is defined by using all the components of the original Dirac field
$\psi_{\alpha s}(x) (\alpha=1,\cdots, 4; s=1,\cdots, 16)$ not as usual,
but the right-handed part of the measure is just saturated completely by inserting
a suitable product of the 't Hooft vertexes in terms of the right-handed fields, 
\begin{eqnarray}
T_+(x) &=& \frac{1}{2}\, V_+^a(x) V_+^a(x) , \quad 
V_+^a(x) = \frac{1}{2}\,\psi_+(x)^{\rm T} i \gamma_5 C_D {\rm T}^a \psi_+(x), 
\\
\bar T_+(x) &=& \frac{1}{2}\, \bar V_+^a(x) \bar V_+^a(x) , \quad 
\bar V_+^a(x) = \frac{1}{2}\,\bar \psi_+(x) i \gamma_5 C_D {\rm T}^{a \dagger}\bar\psi_+ (x)^{\rm T}.
\end{eqnarray}
Namely, the Weyl field measure is defined as
\begin{eqnarray}
\label{eq:def-weyl-measure}
{\cal D}[\psi_-] {\cal D}[\bar \psi_-]  
&\equiv&
{\cal D}[\psi] {\cal D}[\bar \psi]  
\, \prod_{x \in \Lambda} F( T_+(x) )
\, \prod_{x \in \Lambda}  F( \bar T_+(x) ), 
\end{eqnarray}
where
\begin{eqnarray}
\label{eq:def-dirac-measure}
{\cal D}[\psi] {\cal D}[\bar \psi]  
&\equiv& 
\prod_{x \in \Lambda} \prod_{\alpha=1}^4\prod_{s=1}^{16} 
\, d \psi_{\alpha s}(x) 
\,
\prod_{x \in \Lambda} \prod_{\alpha=1}^4\prod_{s=1}^{16} 
\, d \bar \psi_{\alpha s }(x) ,
\end{eqnarray}
and $F(w)$ is the certain function to represent the product of the 't Hooft vertexes, $T_+(x)$ and $\bar T_+(x)$.
The Weyl field measure so defined depends on the link field 
$U(x,\mu)$ 
through the chiral projection $\hat P_+$ to define $T_+(x)$
in terms of the right-handed field 
$\psi_+(x) = \hat P_+ \psi(x)$.
Note that we use the four-spinor notation in the definition of the 't Hooft vertexes and 
the factor 
${\hat P_+}^{T} i \gamma_5 C_D  {\rm T}^a E^a(x) \hat P_+$, not 
${\hat P_+}^{T}  \{ i \gamma_5 C_D  P_+ {\rm T}^a E^a(x) \} \hat P_+$, appears for the field $\psi_+(x)$, while 
${P_-} i \gamma_5 C_D  {\rm T}^{a\dagger} \bar E^a(x) {P_-}^T =
{P_-} \{ i \gamma_5 C_D  {P_-}^T {\rm T}^{a\dagger} \bar E^a(x) \} {P_-}^T$ 
for the anti-field $\bar \psi_+(x)$.\footnote{This point is crucial for our proposal.
If one includes the factor $P_+$ in the definition 
of the 't Hooft operator for the field $\psi_+(x)$, one has
$
{\hat P_+}^{T} i \gamma_5 C_D P_+ {\rm T}^a E^a(x) \hat P_+
=
{(1-D)}^{T} i \gamma_5 C_D P_+ {\rm T}^a E^a(x) (1-D)
$. 
The factor $(1-D)$ projects out
the modes with the momenta $\pi_\mu^{(A)}$ ($A=1,\cdots, 15$),
where
$\pi^{(1)} \equiv (\pi, 0, 0, 0), 
\pi^{(2)} \equiv (0,\pi,0,0), \cdots, \pi^{(15)} \equiv(\pi,\pi,\pi,\pi)$.
This type of the operator cannot saturate the right-handed
part of the measure completely. 
Therefore it is not acceptable for our purpose.
This point will be discussed later in relation to other formulations.
See section~\ref{sec:rel-other-approaches}.
} 
Our choice for $F(w)$ is 
\begin{equation}
\label{eq:def-of-F}
F(w) \equiv 4!\,(z/2)^{-4} I_4(z) \, 
\Big\vert_{(z/2)^2=w}
= 4!\, \sum_{k=0}^\infty \frac{w^k}{k! (k+4)!} , 
\end{equation}
where $I_\nu(w)$ is the modified Bessel function of the first kind.
It has the integral representation as
\begin{equation}
\label{eq:integral-rep-F-modBessel}
F(w) \Big\vert_{w=(1/2) u^a u^a} 
= (\pi^5/12)^{-1} \, \int \prod_{a=1}^{10} d e^a \delta( \sqrt{e^b e^b} -1)\, {\rm e}^{e^c u^c}
\end{equation}
and allows us to prove the CP invariance of the effective action of the lattice model, as discussed bellow.\footnote{One possible  choice for $F(w)$ is simply
 $F(w) = {\rm e}^w =\sum_{k=0}^\infty \frac{w^k}{k!}$. It also has the integral representation,
\begin{equation}
F(w) \Big\vert_{w=(1/2) u^a u^a} 
= (2\pi)^{-5}\int \prod_{a=1}^{10} d x^a \, {\rm e}^{-(1/2) x^c x^c +x^c u^c}
\end{equation}
In this case, 
however, we do not succeed yet in proving the CP invariance of the effective action of the lattice model.}
%

The partition function of our lattice model for the SO(10) chiral
Gauge theory is then given as follows,
\begin{eqnarray}
\label{eq:def-Z}
Z 
&\equiv& 
\int {\cal D}[U]  \, 
 {\rm e}^{- S_G[U] + \Gamma_W[U] }  , 
\end{eqnarray}
where $\Gamma_W [U]$ is the effective action induced by the path-integration of the Weyl field, 
\begin{eqnarray}
\label{eq:def-effective-action}
{\rm e}^{\Gamma_W [U]}  
& = & \int {\cal D}[\psi_-] {\cal D}[\bar \psi_-]  \,  {\rm e}^{-S_W[\psi_-, \bar\psi_-]}    \nonumber\\
&\equiv&\int {\cal D}[\psi] {\cal D}[\bar \psi] \,
\, \prod_{x \in \Lambda} F( T_+(x) )
\, \prod_{x \in \Lambda}  F( \bar T_+(x) ) \, 
 {\rm e}^{- S_W[\psi_-, \bar \psi_-] } \nonumber\\
&=&
\int 
{\cal D}[\psi] {\cal D}[\bar \psi] \,
{\cal D}[E] {\cal D}[\bar E] \,
\, {\rm e}^{- S_W[\psi_-, \bar \psi_-] + \sum_{x \in \Lambda} \{ 
E^a(x) 
V_+^a(x) 
+ \bar E^a(x)
\bar V_+^a(x)
\}[\psi_+, \bar\psi_+]
} . \nonumber\\
\end{eqnarray}
In the last equation, the integral representation of $F(w)$ is used and 
the path-integrations over the SO(10)-vector real spin fields with unit length, $E^a(x)$ and $\bar E^a(x)$, are introduced:
\begin{eqnarray}
\label{eq:path-integral-measure-E-bar-E}
{\cal D}[E] 
&=&
\prod_{x \in \Lambda} 
(\pi^5/12)^{-1}\,\prod_{a=1}^{10} d E^a(x)\delta(\sqrt{E^b(x) E^b(x)} -1) \\
{\cal D}[\bar E] 
&=&
\prod_{x \in \Lambda} 
(\pi^5/12)^{-1} \,\prod_{a=1}^{10} d \bar E^a(x)\delta(\sqrt{\bar E^b(x) \bar E^b(x)} -1).
\end{eqnarray}

Defined with 
all the components of the Dirac field $\psi(x)$, $\bar \psi(x)$, 
the Weyl field measure is manifestly invariant under the SO(10) gauge transformation.
It also possesses all required transformation properties under 
lattice symmetries: translations, rotations, reflections and charge conjugation.
As to the global U(1) fermion symmetry of the left-handed 
field $\psi_-(x)$, $\bar \psi_-(x)$,
the fermionic measure transforms as  
\begin{equation}
\delta_\alpha {\cal D}[\psi_-] {\cal D}[\bar \psi_-]  =- i \sum_{x \in \Gamma}  \alpha (x) {\rm tr}\{ \hat P_- - P_+ \}(x,x) \times {\cal D}[\psi_-] {\cal D}[\bar \psi_-] 
\end{equation}
with a local parameter $\alpha(x)$, and it gives rise to the non-trivial chiral anomaly in the U(1) Ward-Takahashi relation.
One may consider the similar global U(1) fermion symmetry of the right-handed field $\psi_+(x)$, $\bar \psi_+(x)$, but it is broken explicitly by the 't Hooft vertexes, 
$T_+(x)$ and $\bar T_+(x)$, down to Z$_4$ $\times$ Z$_4$, one $Z_4$ for the field $\psi_+(x)$ and the other $Z_4$ for the anti-field $\bar\psi_+(x)$. The reason for the two independent Z$_4$ is that
the bilinear kinetic term of the right-handed field, $ \sum_{x \in \Lambda} \bar \psi_+ (x) D \psi_+(x)$, is not introduced here. Conversely, this  Z$_4$ $\times$ Z$_4$ symmetry prohibits such bilinear terms of  the right-handed field to appear, as long as it is not broken spontaneously.

\subsection{Chiral determinant and 't Hooft-vertex pfaffians}

The path-integral weight for the effective action
defined by eq.~(\ref{eq:def-effective-action})
consists of 
$S_W[\psi_-,\bar\psi_-]$,
the action of the left-handed fields, 
and 
$\sum_{x \in \Lambda} \{ 
E^a(x) 
V_+^a(x) 
+ \bar E^a(x)
\bar V_+^a(x)
\}[\psi_+, \bar\psi_+]$,
the 't Hooft vertex terms of the right-handed fields.
These two terms
can be written
in terms of the Dirac fields $\psi(x)$, $\bar\psi(x)$,
as follows:
\begin{eqnarray}
&&
S_W[\psi_-,\bar\psi_-]
-
\sum_{x \in \Lambda} \{ 
E^a(x) 
V_+^a(x) 
+ \bar E^a(x)
\bar V_+^a(x)
\}[\psi_+, \bar\psi_+]
\nonumber\\
&&=
\frac{1}{2}
\sum_{x\in\Lambda}
\left(
\begin{array}{cc} \psi^T & \bar \psi \end{array} 
\right)(x)
\left(
\begin{array}{cc}
-  \hat P_+^T i \gamma_5 C_D {\rm T}^a E^a \hat P_+ &
- \hat P_-^T D^T P_+^T \\
P_+ D \hat P_- &
-   P_- i \gamma_5 C_D {{\rm T}^a}^\dagger \bar E^a P_-^T 
\end{array}
\right)
\left(
\begin{array}{c} \psi \\ \bar \psi^T \end{array} 
\right) (x).
\nonumber\\
\end{eqnarray}
Then the path-integration of the fermion fields
with the Dirac field measure ${\cal D}[\psi] {\cal D}[\bar \psi]$ gives rise 
to the pfaffian of the above gauge-covariant anti-symmetric operator,  
\begin{eqnarray}
\label{eq:covariant-pfaffian}
{\rm pf} 
\left(
\begin{array}{cc}
-  \hat P_+^T i \gamma_5 C_D {\rm T}^a E^a \hat P_+ &
- \hat P_-^T D^T P_+^T \\
P_+ D \hat P_- &
-   P_- i \gamma_5 C_D {{\rm T}^a}^\dagger \bar E^a P_-^T 
\end{array}
\right)
. 
\end{eqnarray}

This pfaffian factorises into the chiral determinant
of the left-handed fields and the 't Hooft-vertex pfaffians
of the right-handed fields
in the chiral bases for the field $\psi(x)$ and the anti-field $\bar\psi(x)$ where $\hat \gamma_5$ and $\gamma_5$ are
diagonalized, respectively.
One can introduce the four-spinor
vectors of the chiral bases as 
\begin{eqnarray}
\label{eq:chiral-basis-vector-hat-gamma5}
&&  {\rm P}_+ \otimes \hat P_{+} u_i(x) = u_i(x) \quad \, (i=1, \cdots, n/2-8Q) ;    \quad ( u_i,  u_j) = \delta_{ij} ,  \\
&& {\rm P}_+ \otimes \hat P_{-} v_i(x) = v_i(x) \quad \, (i=1, \cdots, n/2+8Q) ;    \quad ( v_i,  v_j) = \delta_{ij} , 
\end{eqnarray}
\begin{eqnarray}
\label{eq:chiral-basis-vector-gamma5}
&& \bar u_k (x) P_{-} \otimes {\rm P}_+  = \bar u_k (x) \quad (k=1, \cdots, n/2) ; \quad \quad \quad \, (\bar u_k, \bar u_l) = \delta_{kl} , \\
&& \bar v_k (x) P_{+} \otimes {\rm P}_+   = \bar v_k (x) \quad (k=1, \cdots, n/2) ; \quad \quad \quad \, (\bar v_k, \bar v_l) = \delta_{kl} 
\end{eqnarray}
in the given topological sector $\mathfrak{U}[Q]$, 
where $n= \text{dim}\, \Lambda \times 4 \times 16$.
The basis vectors $u_i(x)$ and  $v_i(x)$ depend on the gauge field through the chiral projectors $\hat P_{\pm} $, while 
the basis vectors $\bar u_k(x)$ and  $\bar v_k(x)$ can be chosen 
so that they are independent of the gauge field. For example, 
${\bar u_k}(x)_{\alpha s} = \delta_{x x'} \delta_{\alpha, \sigma+2} \delta_{s t}$ for 
$k=\{ x' \in \Lambda; \sigma = 1,2 ; t=1,\cdots ,16 \}$ and 
${\bar v_k}(x)_{\alpha s} = \delta_{x x'} \delta_{\alpha \sigma} \delta_{s t}$ for 
$k=\{ x' \in \Lambda; \sigma= 1,2 ; t=1,\cdots ,16 \}$, assuming $\gamma_5 =\text{diag}(1,1,-1,-1)$. 
%
One can always choose the bases of the Dirac fields, $ \{ u_j(x), v_j(x) \}$,
and $\{\bar u_k(x), \bar v_k(x) \}$, 
so that the jacobian factors,  $\det (u_j(x), v_j(x) )$, 
and $\det (\bar u_j(x), \bar v_j(x) )$, are 
unity independent of the gauge field. 
In this choice of the chiral bases, the pfaffian can be evaluated as
\begin{eqnarray}
&&
{\rm pf} 
\left(
\begin{array}{cc}
-  \hat P_+^T i \gamma_5 C_D {\rm T}^a E^a \hat P_+ &
- \hat P_-^T D^T P_+^T \\
P_+ D \hat P_- &
-   P_- i \gamma_5 C_D {{\rm T}^a}^\dagger \bar E^a P_-^T 
\end{array}
\right)
\nonumber\\
&&\nonumber\\
&=&
{\rm pf} 
\left(
\begin{array}{cccc}
-  (u^T i \gamma_5 C_D {\rm T}^a E^a u)  & 0 &0 & 0 \\
0 & 0 & 0 & - (v^T D^T \bar v^T) \\
0 & 0& 
-  (\bar u \, i\gamma_5 C_D {{\rm T}^a}^\dagger \bar E^a \bar u^T )& 0 \\
0 & (\bar v D v) & 0 & 0
\end{array}
\right)
\\
&&\nonumber\\
&=&
\det(\bar v D v) \times \,
{\rm pf} (u^T i \gamma_5 C_D {\rm T}^a E^a u) \times \,
{\rm pf}  (\bar u \, i \gamma_5 C_D {{\rm T}^a}^\dagger \bar E^a \bar u^T ) ,
\end{eqnarray}
where the matrices
$(\bar v D v )$,
$( u^{\rm T}\, i \gamma_5 C_D {\rm T}^a E^a u )$ 
and 
$(\bar u \, i \gamma_5 C_D {{\rm T}^a}^\dagger \bar E^a\bar u^{\rm T} )$
are given by
\begin{eqnarray}
\label{eq:D-in-left-haded-bases}
(\bar v D v ) _{k i} 
&\equiv& \sum_{x \in \Lambda} \bar v_k(x) D v_i(x)  \qquad ( k=1,\cdots, n/2 ;  i=1,\cdots, n/2+8Q), 
\nonumber\\
&&\\
\label{eq:the-first-matrix}
\big( u^{\rm T} i \gamma_5 C_D {\rm T}^a E^a u \big)_{ij} 
&\equiv& \sum_{x \in \Lambda} u_i(x)^{\rm T} i \gamma_5 C_D {\rm T}^a E^a(x) u_j(x) \nonumber\\
&& \qquad\qquad\qquad\qquad\qquad (i,j=1,\cdots, n/2-8Q) , \\
&& \nonumber\\
\label{eq:the-second-matrix}
\big( \bar u i \gamma_5 C_D {{\rm T}^a}^\dagger \bar E^a\bar u^{\rm T}  \big)_{kl} 
&\equiv&  \sum_{x \in \Lambda} \bar u_k(x)  i \gamma_5 C_D {{\rm T}^a}^\dagger \bar E^a(x) \bar u_l(x)^{\rm T} \nonumber\\
&& \qquad\qquad\qquad\qquad\qquad (k,l=1,\cdots,n/2) ,
\end{eqnarray}
and 
$( u^{\rm T}\, i \gamma_5 C_D {\rm T}^a E^a u )$ and $(\bar u \, i \gamma_5 C_D {{\rm T}^a}^\dagger \bar E^a\bar u^{\rm T} )$ are  
anti-symmetric complex matrices.

Therefore, the effective action eq.~(\ref{eq:def-effective-action}) is now given by
\begin{eqnarray}
\label{eq:def-effective-action-with-covariant-pfaffian}
{\rm e}^{\Gamma_W [U]}  
&=& 
\det (\bar v D v)  \, \times \, 
\int {\cal D}[ E] \, 
{\rm pf}( u^T  i \gamma_5 C_D  T^a E^a u) \,
\int {\cal D}[\bar E] \,
{\rm pf}  (\bar u \, i \gamma_5 C_D {{\rm T}^a}^\dagger \bar E^a \bar u^T ) . 
\nonumber\\
\end{eqnarray}
Thus the effective action $\Gamma_W[U]$, with our definition of the Weyl field measure eq.~(\ref{eq:def-weyl-measure}),
has 
the extra factor of logarithm of the
pfaffians, ${\rm pf}( u^T  i \gamma_5 C_D  {\rm T}^a E^a u)$ and
${\rm pf} \big(  \bar u  \, i \gamma_5 C_D {{\rm T}^a}^\dagger \bar E^a\bar u^{\rm T} \big)$, 
integrated over the auxiliary spin fields
in addition to the usual effective action given by
the logarithm of the chiral determinant,
$\ln \det (\bar v D v)$ \cite{Narayanan:sk,Narayanan:ss, Narayanan:1994gw}
\cite{Luscher:1998du,Luscher:1999un}.

The first factor in the r.h.s. of eq.~(\ref{eq:def-effective-action-with-covariant-pfaffian}) is nothing but
the chiral determinant in the overlap formalism.\cite{Narayanan:sk,Narayanan:ss, Narayanan:1994gw}
In the weak gauge-coupling limit,  
the matrix $(\bar v D v)$ shows the massless singularity associated with the free left-handed Weyl field.
With the periodic boundary condition, in particular, 
$(\bar v D v)$ is not invertible because
there appear the zero modes in the eigenvalues of $D$, which  have zero index $n_+ - n_- =0$. 
In the  topologically non-trivial sectors, the matrix $(\bar v_k D v_i)$  is not a square matrix and $\det (\bar v D v )$ vanishes identically.
This is due to the appearance of the chiral zero modes with a
non-trivial index 
$n_+ -n_- =-8Q \not = 0$. 
These zeromodes, saturated by the insetion of the 't Hooft vertices in terms of the left-handed fields
$T_-(x)$, $\bar T_-(x)$ given by eqs. (\ref{eq:'tHooft-vertex-field}) and (\ref{eq:'tHooft-vertex-antifield}), 
give rise to the non-vanishing vacuum expectation values of
the 't Hooft vertices and break the global U(1) fermion symmetry.
These are the robust results of the overlap formalism/the index theorem followed from the Ginsparg-Wilson relation\cite{Narayanan:sk,Narayanan:ss,Narayanan:1994gw}\cite{Hasenfratz:1998ri}.\footnote{
It has been known that there is a problem about
the fermion number violation\cite{Banks:1991sh,Banks:1992af}
in the gauge-fixing approach to construct chiral lattice gauge
theories. (See \cite{Golterman:2000hr} and references there in.)
In the gauge-fixing approach, 
to remove the species doublers,
one uses 
non-gauge-covariant Wilson terms
for the pairs of 
the target (left-handed) Weyl fermions and 
the spectator (right-handed) Weyl fermions.
But the action and the path-integral measure of 
the Wilson fermions are invariant exactly under the vector-like
fermion number symmetry, and this causes a problem in
describing the process of fermion number violation\cite{Banks:1991sh,Banks:1992af}.
One needs, as shown by Golterman and Shamir 
in \cite{Golterman:2002ns},
to introduce
the explicit symmetry-breaking term such as 
the non-gauge-invariant Majorana mass term for the target
(left-handed) 
fermions 
to reproduce the non-vanishing vacuum expectation values of 't Hooft vertex operators.
%
%
In \cite{Golterman:2002ns}, 
the case of the SO(10) chiral gauge theory was considered
and 
the semi-classical continuum limit 
of the lattice model was examined
for the background gauge fields with
a SU(2) instanton solution embedded in all possible ways
consistent with the SO(10) global symmetry.
%
}

As to the second and third factors, on the other hand, 
we note that the first matrix 
$( u^{\rm T}\, i \gamma_5 C_D {\rm T}^a E^a u )$
changes its size as $n/2 - 8Q$ depending on the topological charge $Q$, but remains to be a square matrix, while 
the second one 
$(\bar u \, i \gamma_5 C_D {{\rm T}^a}^\dagger \bar E^a\bar u^{\rm T} )$
is the square matrix of the fixed size $n/2$.
Therefore the pfaffians of these matrices do not vanish identically
in general 
in any topological
sector $\mathfrak{U}[Q]$.
And
the path-integration of the pfaffians over the spin fields $E^a(x)$ and $\bar E^a(x)$ gives a certain non-zero functional of the admissible link field $U(x,\mu)$,
as we will discuss more in detail in the following sections~\ref{subsec:saturation-right-handed-measure}
and \ref{sec:saturation-right-handed-measure}.
We refer this fact as ``saturation of the right-handed part of  the fermion measure by the 't Hooft vertices''.

\begin{figure}[thb]
\begin{center}
\includegraphics[width =125mm]{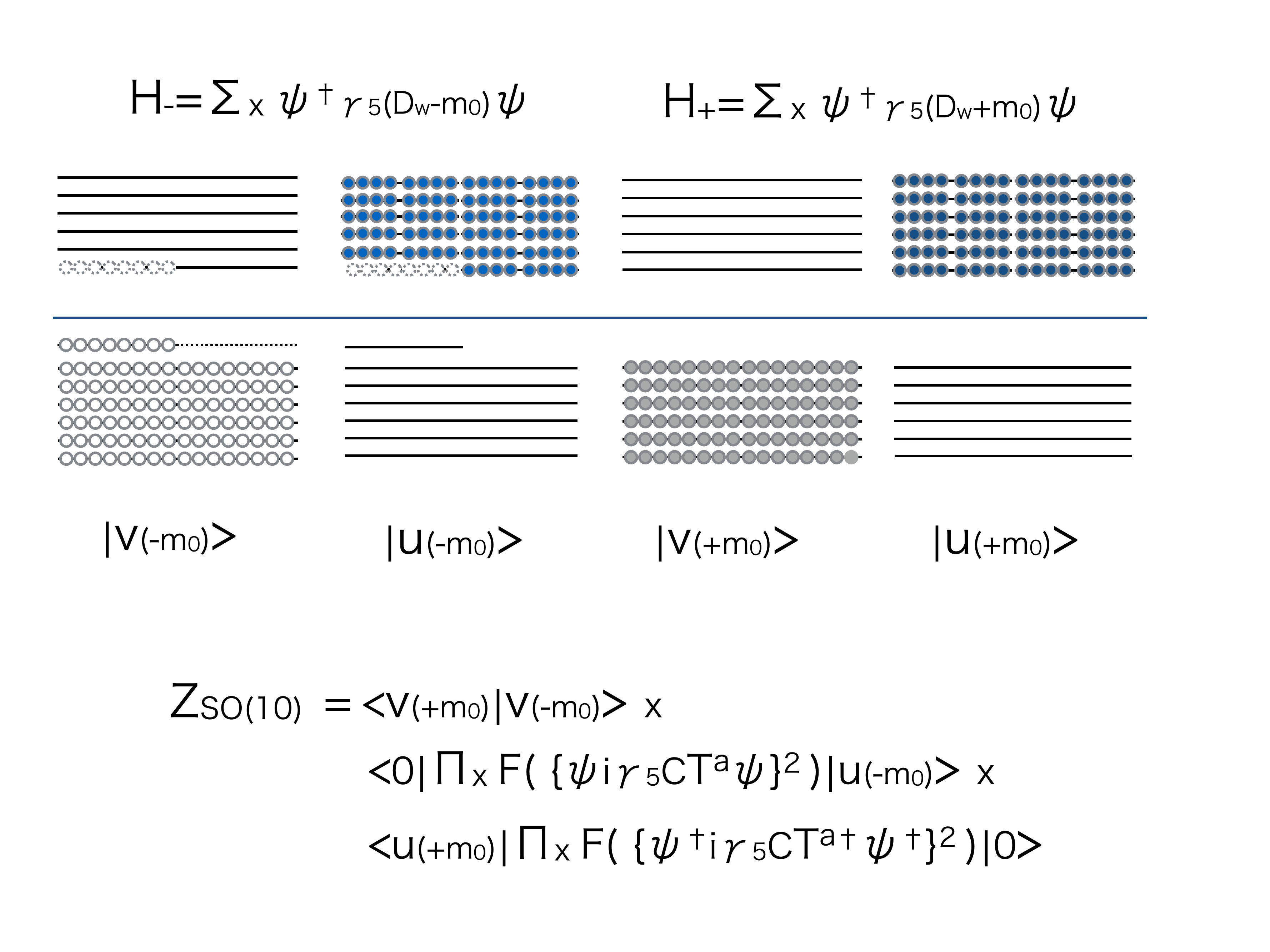} 
\caption{The schematic picture of the SO(10) model in terms of 
the overlap formalism}
\label{fig:overlap-picture-SO(10)}
\end{center}
\end{figure}

This result on the saturation of the right-handed part of the  measure
is in sharp contrast to that on the left-handed part of the measure, 
which gives rize to the chiral determinant.
Thanks to this result, 
the total Weyl field measure defined by eqs.~(\ref{eq:def-weyl-measure}) and (\ref{eq:def-dirac-measure})
reproduces the correct
zero-modes associated with only the left-handed Weyl fields
in any topologically non-trivial sectors
$\mathfrak{U}[Q]$,
while it is manifestly gauge-invariant and 
there is no phase-ambiguity which depends on the gauge field due to the choice of the chiral basis vectors.
We hope that the above result on the saturation of the right-handed part of the measure for the anomaly-free multiplet of 
$\underbar{16}$ of SO(10) is another robust property of the overlap fermion/the Ginsparg-Wilson relation.
%
%
The schematic picture of our SO(10) model is shown in fig.~\ref{fig:overlap-picture-SO(10)} in terms of 
the overlap formalism\cite{Narayanan:sk,Narayanan:ss,Narayanan:1994gw}.

\subsection{The Weyl field measure in terms of chiral basis}

In the definition of the Weyl field measure, eqs.~(\ref{eq:def-weyl-measure}) and (\ref{eq:def-dirac-measure}), 
the part of the Dirac field measure, 
${\cal D}[\psi] {\cal D}[\bar \psi]$, 
may be formulated 
in chiral components
by using the chiral bases defined with the chiral projectors 
$\hat P_\pm$ and $P_\pm$.
In the given topological sector $\mathfrak{U}[Q]$, it reads 
\begin{eqnarray}
\label{eq:fermion-measure-left-right-in-chiral-basis}
{\cal D}_\star[\psi_-] {\cal D}_\star[\bar \psi_-]  
{\cal D}_\star[\psi_+] {\cal D}_\star[\bar \psi_+]
&=& \prod_{j=1}^{n/2+8Q} d c_j  \prod_{k=1}^{n/2} d \bar c_k   \prod_{j=1}^{n/2-8Q} d b_j  \prod_{k=1}^{n/2} d \bar b_k ,   
\end{eqnarray}
where $n= \text{dim}\, \Lambda \times 4 \times 16$ and 
$\{c_j ,  \bar c_k \}$  and $\{ b_j , \bar b_k\}$ are the Grassmann coefficients
in the expansion of the chiral component fields, 
\begin{equation}
\psi_-(x) = \sum_j  v_j(x) c_j , \quad  \bar \psi_-(x) = \sum_k \bar c_k \bar v_k(x) , 
\end{equation}
\begin{equation}
\psi_+(x) = \sum_j  u_j(x) b_j , \quad  \bar \psi_+(x) = \sum_k \bar b_k \bar u_k(x) , 
\end{equation}
in terms of the chiral orthonormal bases defined by 
eqs.~(\ref{eq:chiral-basis-vector-hat-gamma5})
and (\ref{eq:chiral-basis-vector-gamma5}).\cite{Luscher:1998du,Luscher:1999un}
Since the original measure 
${\cal D}[\psi] {\cal D}[\bar \psi]$ does not depend on the gauge field,  it follows that one can always choose the basis of the Dirac field, $ \{ u_j(x), v_j(x) \}$,
so that the jacobian factor,  $\det (u_j(x), v_j(x) )$, 
is independent of the gauge field. For the infinitesimal variation of the link field $\delta_\eta U(x,\mu) = i \eta_\mu(x) U(x,\mu)$, this condition is given by
\begin{eqnarray}
\label{eq:basis-independent-on-link}
&&  \sum_j ( u_j, \delta_\eta u_j) + \sum_j ( v_j, \delta_\eta v_j)  = 0 . 
\end{eqnarray}
Adjusting the overall constant phase factors of the Jacobian
as $\det (u_j(x), v_j(x) ) = 1$, 
one obtains
\begin{equation}
{\cal D}_\star[\psi_-] {\cal D}_\star[\bar \psi_-]
{\cal D}_\star[\psi_+] {\cal D}_\star[\bar \psi_+]
=
{\cal D}[\psi] {\cal D}[\bar \psi] .
\end{equation}

Using this chiral decomposition of ${\cal D}[\psi] {\cal D}[\bar \psi]$ and the integral representation of $F(w)$, 
the Weyl field measure eq.~(\ref{eq:def-weyl-measure}) now reads 
in the factorized form,
\begin{eqnarray}
\label{eq:def-weyl-measure-in-chiral-basis}
{\cal D}[\psi_-] {\cal D}[\bar \psi_-]  
&=&
{\cal D}_\star[\psi_-] {\cal D}_\star[\bar \psi_-] \,\times 
{\cal D}_\star[\psi_+] {\cal D}_\star[\bar \psi_+]
\, \prod_{x \in \Lambda} F( T_+(x) )
\, \prod_{x \in \Lambda}  F( \bar T_+(x) )
\nonumber\\
&=& 
{\cal D}_\star[\psi_-] {\cal D}_\star[\bar \psi_-] \,\times 
\nonumber\\
&& 
{\cal D}_\star[\psi_+] {\cal D}_\star[\bar \psi_+]
\,
\int {\cal D}[E] {\cal D}[\bar E] \,
\, {\rm e}^{\, \sum_{x \in \Lambda} \{ 
E^a(x) 
V_+^a(x) 
+ \bar E^a(x)
\bar V_+^a(x)
\}[\psi_+, \bar\psi_+]
} .
\nonumber\\
\end{eqnarray}
Moreover, using the chiral bases, 
't Hooft vertex terms of the right-handed fields $\psi_+(x), \bar\psi_+(x)$ are written as follows,
\begin{eqnarray}
\sum_{x \in \Lambda} \{ 
E^a(x) 
V_+^a(x) 
\}[\psi_+]
&=&
\sum_{i,j} 
\frac{1}{2}\,
b_i ( u^{\rm T}\, i \gamma_5 C_D {\rm T}^a E^a u )_{ij} b_j ,
\\
\sum_{x \in \Lambda} \{ 
\bar E^a(x)
\bar V_+^a(x)
\}[\bar\psi_+]
&=&
\sum_{k,l} 
\frac{1}{2}\,
\bar b_k (\bar u \, i \gamma_5 C_D {{\rm T}^a}^\dagger \bar E^a\bar u^{\rm T} )_{kl}\bar b_l .
\end{eqnarray}
%
Then the path-integration over the right-handed fields $\psi_+$, $\bar\psi_+$ can be performed explicitly as
\begin{eqnarray}
\label{eq:def-weyl-measure-in-chiral-basis-pfaffian}
{\cal D}[\psi_-] {\cal D}[\bar \psi_-]  
&=& 
{\cal D}_\star[\psi_-] {\cal D}_\star[\bar \psi_-] \,\times 
\nonumber\\
&& 
\int {\cal D}[E] \, 
{\rm pf}  \big(u^{\rm T} \,  i \gamma_5 C_D {\rm T}^a E^a u \big)
\,
\int {\cal D}[\bar E] \,
{\rm pf} \big(  \bar u  \, i \gamma_5 C_D {{\rm T}^a}^\dagger \bar E^a\bar u^{\rm T} \big) , \nonumber\\
\end{eqnarray}
where 
the symbol ${\rm pf}$ stands for the pfaffians of these anti-symmetric matrices. 
Thus, in our definition of the Weyl field measure eq.~(\ref{eq:def-weyl-measure}),
the insertion of 
the product of the 't Hooft vertexes in terms of the right-handed fields, $T_+(x)$ and $\bar T_+(x)$, adds the extra factors of the
pfaffians integrated over the auxiliary spin fields to the usual definition of the (left-handed) Weyl field measure, ${\cal D}_\star[\psi_-] {\cal D}_\star[\bar \psi_-]= \prod_{j=1}^{n/2+8Q} d c_j  \prod_{k=1}^{n/2} d \bar c_k $ \cite{Luscher:1998du,Luscher:1999un}.

As we mentioned before, the first matrix 
$( u^{\rm T}\, i \gamma_5 C_D {\rm T}^a E^a u )$ 
given by eq.~(\ref{eq:the-first-matrix}) changes its size as $n/2 - 8Q$ depending on the topological charge $Q$, but remains to be a square matrix, while 
the second one $(\bar u \, i \gamma_5 C_D {{\rm T}^a}^\dagger \bar E^a\bar u^{\rm T} )$ given by eq.~(\ref{eq:the-second-matrix}) is the square matrix of the fixed size $n/2$.
Therefore these pfaffians do not vanish identically
in general 
in any given topological sector $\mathfrak{U}[Q]$, 
and
the path-integration of the pfaffians over the spin fields $E^a(x)$ and $\bar E^a(x)$ gives a certain non-zero functional of the admissible link field $U(x,\mu)$,
as we will argue in the following 
sections~\ref{subsec:saturation-right-handed-measure}
and \ref{sec:saturation-right-handed-measure}.

\subsection{Saturation of the right-handed part of the fermion measure by 't Hooft vertices}
\label{subsec:saturation-right-handed-measure}

The pfaffian of the second matrix eq.~(\ref{eq:the-second-matrix}) turns out to be unity.
This is because the matrix is represented as
\begin{eqnarray}
(\bar u i \gamma_5 C_D {{\rm T}^a}^\dagger \bar E^a \bar u^{\rm T})_{kl}
&=& i \, \epsilon_{\sigma \sigma'}  \delta_{x x'} 
\big( {{\rm T}^a}^\dagger {\rm P}_+\big)_{t t'}\bar E^a(x')
\end{eqnarray}
for $k=\{x, \sigma, t \}$ and $l=\{x', \sigma', t' \}$,
in the bases $\gamma_5 =\text{diag}(1,1,-1,-1)$, 
${\bar u_k}(x)_{\alpha s} = \delta_{x x'} \delta_{\alpha \sigma+2} \delta_{s t}$ for 
$k=\{ x' \in \Lambda; \sigma = 1,2 ; t=1,\cdots ,16 \}$.
Then the pfaffian of the matrix is evaluated as 
\begin{eqnarray}
{\rm pf} \big(\bar u i \gamma_5 C_D {\rm T}^a \bar E^a \bar u^{\rm T}\big)
&=& \prod_{x} {\rm det} \big({\rm P}_- + {\rm P}_+  i \, {{\rm T}^a}^\dagger \bar E^a(x) \big) \nonumber \\
&=& \prod_{x} {\rm det} \big(  i \, \check{\rm T}^a{}^\dagger \bar E^a(x) \big) \nonumber \\
&=& \prod_{x} {\rm det}   \big(  i {\check C}^\dagger [ E^{10}(x) + i \check{\Gamma}^{a'} \bar E^{a'}(x) ] \big)  \nonumber \\
&=& 1.
\end{eqnarray}
Note that $\det(i {\check C}{}^\dagger)$ and ${\rm det}\big([ E^{10}(x) + i \check{\Gamma}^{a'} \bar E^{a'}(x) ] \big)$ are both equal to $+1$ and the latter, in particular, is independent of $\bar E^a(x)$.
Then the path-integration over $\bar E^a(x)$ simply gives
\begin{equation}
\label{eq:path-integral-pfaffian-nonzero-for-antifield}
\int {\cal D}[\bar E] \,
{\rm pf} \big(  \bar u \,  i \gamma_5 C_D {{\rm T}^a}^\dagger \bar E^a\bar u^{\rm T} \big) = 1 .
\end{equation}
Thus the measure of the right-handed anti-field, 
${\cal D}_\star[\bar \psi_+]$, is indeed saturated 
completely by inserting the product of the 't Hooft vertex 
$\bar T_+(x) [ \bar \psi_+]$.
This is actually the known result which was first shown 
by Eichten and Preskill in \cite{Eichten:1985ft}, where the effects of the generalized Wilson-terms were studied in the strong coupling limit.
In fact, our result reads 
\begin{eqnarray}
\label{eq:path-integral-pfaffian-nonzero-for-antifield-2}
&&
\int {\cal D}_\star[\bar \psi_+] F\big( \bar T_+(x)[\bar \psi_+] \big) 
\nonumber\\
&=&
\int \,
\prod_{x \in \Lambda} \prod_{\alpha=3}^4\prod_{s=1}^{16} 
\, d \bar \psi_{\alpha s }(x) \,
\prod_{x \in \Lambda} 
\frac{4!}{8! 12!}
\left\{
\frac{1}{2^3}\,
\bar \psi(x) P_-  i \gamma_5 C_D {\rm T}^{a \dagger} \bar\psi(x)^{\rm T} \, 
\bar \psi(x) P_- i \gamma_5 C_D {\rm T}^{a \dagger} \bar\psi (x)^{\rm T}
\right\}^{8}   
\nonumber\\
&=& 1 ,
\end{eqnarray}
and it provides the explicit normalization for the constant in the result there\cite{Eichten:1985ft}. 

The pfaffian of the first matrix eq.~(\ref{eq:the-first-matrix}),  on the other hand, is a complex number in general, which depends on the spin field $E^a(x)$ as well as the link field $U(x,\mu)$. 
We do not have a rigorous proof so far that the path-integration of the pfaffian over $E^a(x)$ is non-zero for any admissible link fields. 
But there are typical examples of link field configurations
where one can argue that it is indeed the case.
%
Those include the case in the weak gauge-coupling limit where the link variables are set to unity, $U(x,\mu)=1$, and the cases
of the SU(2)(=Spin(3)) link fields with non-zero topological charges $Q (\not = 0)$, which represent the non-trivial topological sectors $\mathfrak{U}[Q]$.
These cases will be discussed further in detail
in the following section~\ref{sec:saturation-right-handed-measure}.

%
%
%
%
%
%
%
%
%

In summary, the effective action 
$\Gamma_W [U]$ is obtained in the chiral basis
as follows.
\begin{eqnarray}
\label{eq:def-effective-action-in-chiral-basis}
{\rm e}^{\Gamma_W [U]}  
& = & 
\int {\cal D}[\psi_-] {\cal D}[\bar \psi_-]  \,  {\rm e}^{-S_W[\psi_-, \bar\psi_-]}    \nonumber\\
&\equiv &\int {\cal D}[\psi] {\cal D}[\bar \psi] \,
\, \prod_{x \in \Lambda} F( T_+(x) )
\, \prod_{x \in \Lambda}  F( \bar T_+(x) ) \, 
 {\rm e}^{- S_W[\psi_-, \bar \psi_-] } \nonumber\\
&=&
\int 
{\cal D}_\star[\psi_-] {\cal D}_\star[\bar \psi_-]
{\cal D}_\star[\psi_+] {\cal D}_\star[\bar \psi_+] \,
{\cal D}[E] {\cal D}[\bar E] \, \times
\nonumber\\
&& \qquad \qquad \qquad \qquad \qquad
\, {\rm e}^{- S_W[\psi_-, \bar \psi_-] + \sum_{x \in \Lambda} \{ 
E^a(x) 
V_+^a(x) 
+ \bar E^a(x)
\bar V_+^a(x)
\}[\psi_+, \bar\psi_+]
} \nonumber\\
&=& \det (\bar v D v)  \,
 \int {\cal D}[ E] \, {\rm pf}( u^T  i \gamma_5 C_D  {\rm T}^a E^a u) . 
\end{eqnarray}
%
%
%
%
%
%
For later convenience, 
we introduce the abbreviation 
$\big\langle \cdots \big\rangle_{F}$ for
the path-integration of only the fermion fields and the spin fields with the link field fixed as a background field:
\begin{eqnarray}
\big\langle {\cal O} \big\rangle_{F}
& \equiv & \int {\cal D}[\psi_-] {\cal D}[\bar \psi_-]  \,  {\rm e}^{-S_W[\psi_-, \bar\psi_-]}  \,  {\cal O}  \nonumber\\
&=&\int {\cal D}[\psi] {\cal D}[\bar \psi] \,
\, \prod_{x \in \Lambda} F( T_+(x) )
\, \prod_{x \in \Lambda}  F( \bar T_+(x) ) \, 
 {\rm e}^{- S_W[\psi_-, \bar \psi_-] } \, {\cal O} \nonumber\\
&=&
\int 
{\cal D}[\psi] {\cal D}[\bar \psi] \,
{\cal D}[E] {\cal D}[\bar E] \,
\, {\rm e}^{- S_W[\psi_-, \bar \psi_-] + \sum_{x \in \Lambda} \{ 
E^a(x) 
V_+^a(x) 
+ \bar E^a(x)
\bar V_+^a(x)
\}[\psi_+, \bar\psi_+]
} \, {\cal O} . \nonumber\\
\end{eqnarray}
We also use the abbreviation 
$\big\langle \cdots \big\rangle_{E}$
and
$\big\langle \cdots \big\rangle'_{E}$
for
the path-integration of the spin field $E^a(x)$:
\begin{eqnarray}
\big\langle {\cal O} \big\rangle_{E}
& \equiv & 
\int {\cal D}[ E] \, 
{\rm pf}( u^T  i \gamma_5 C_D  T^a E^a u) \,
{\cal O} , \\
\big\langle {\cal O} \big\rangle'_{E}
& \equiv & 
\int {\cal D}[ E] \, 
{\cal O} .
\end{eqnarray}
With these abbreviations, 
the effective action eq.~(\ref{eq:def-effective-action-in-chiral-basis}) reads
\begin{eqnarray}
{\rm e}^{\Gamma_W [U]} 
&=& 
\big\langle 1 \big\rangle_{F} [U]
\nonumber\\
&=&
\det (\bar v D v) \,\,
\big\langle
1
\big\rangle_{E} [U] 
\nonumber\\
&=&
\det (\bar v D v) \,\,
\big\langle
{\rm pf}( u^T  i \gamma_5 C_D  {\rm T}^a E^a u)
\big\rangle'_{E} [U]  \, .
\end{eqnarray}

\subsection{CP invariance}

We define
CP transformation as
\begin{eqnarray}
\label{eq:cp-transf-of-gauge-field}
U(x, \mu) &\longrightarrow& 
U(x, \mu)^{CP} = 
\left( U(x^P, 0)^\ast, U(x^P-\hat k, k)^{\ast \, -1} \right) ,
\\
\label{eq:cp-transf-of-dirac-field}
\psi(x) &\longrightarrow&
\psi(x)^{CP} \quad \, = \quad
+ ({\cal P} \gamma_0)^{-1}  C_D^{-1} \,  \bar \psi(x)^T ,
\\
\label{eq:cp-transf-of-dirac-field}
\bar \psi(x) &\longrightarrow&
\bar \psi(x)^{CP} \quad \, = \quad
- \psi(x)^T  \, C_D  {\cal P} \gamma_0  ,
\\
\label{eq:cp-transf-of-spin-field}
E^a(x) &\longrightarrow& { E^a(x)}^{CP} \,\,\,\, = \quad \,
(-1)^a \, \bar E^a(x^P) \qquad (a=0,1,\cdots, 9) ,
\end{eqnarray} 
wherer $ {\cal P} : x \longrightarrow x^P \equiv (x_0,-x_k)$.
Then, 
the overlop Dirac operator obeys the CP-conjugation relation,
\begin{eqnarray}
D[U^{CP}] &=&
({\cal P} \gamma_0)^{-1}  C_D^{-1} D[U]^T C_D  \,{\cal P} \gamma_0  \\
&=&
({\cal P} \gamma_0)^{-1}  ( \gamma_5 C_D )^{-1} D[U]^\ast \gamma_5 C_D  \, {\cal P} \gamma_0  , 
\end{eqnarray}
and accordingly, the chiral projection operators satisfy the CP-conjugation relations given by
\begin{eqnarray}
\hat P_\pm [U^{CP}] 
&=&
({\cal P} \gamma_0)^{-1}  ( \gamma_5 C_D )^{-1} \hat P_\mp[U]^T \gamma_5 C_D  {\cal P} \gamma_0  , 
\\
P_\pm 
&=&
({\cal P} \gamma_0)^{-1}  ( \gamma_5 C_D )^{-1}  P_\mp ^T \gamma_5 C_D  {\cal P} \gamma_0  .
\end{eqnarray}
Under the CP transformation, the action of the left-handed fields, $\psi_-(x)$ and $\bar\psi_-(x)$, is transformed as
\begin{equation}
S_{\rm W}=\sum_{x\in\Lambda} \bar \psi (x) P_+ D \psi(x) \quad
\longrightarrow \quad 
S'_{\rm W} = \sum_{x\in\Lambda} \bar \psi (x) D P_-\psi(x) ,
\end{equation}
%
%
while the 't Hooft  vertices of the right-handed fields, 
$\psi_+(x)$ and $\bar\psi_+(x)$, 
are transformed as
\begin{eqnarray}
&&
T_+(x) = \frac{1}{2}\, V_+^a(x) V_+^a(x) , \quad\,\,\,
V_+^a(x) = 
\frac{1}{2}\,
\psi^{\rm T}(x)  
{\hat P_+}^T i \gamma_5 C_D {\rm T}^a \hat P_+ 
\psi(x)
\nonumber\\
& \longrightarrow &
T'_+(x) = \frac{1}{2}\, V_+^{' a}(x) V_+^{' a}(x) , \quad 
V_+^{' a}(x) = (-1)^a \,
\frac{1}{2}\,
\bar \psi(x)  
\{\gamma_5\hat P_- \gamma_5\} i \gamma_5 C_D {{\rm T}^a}^\dagger
\{ \gamma_5 \hat P_- \gamma_5 \}^T
\bar \psi(x)^T,
\nonumber\\
&&\\
&&
\bar T_+(x) = \frac{1}{2}\, \bar V_+^a(x) \bar V_+^a(x) , \quad\,\,\, \,\,\,
\bar V_+^a(x) = 
\frac{1}{2}\,
\bar \psi(x)
{P_-} i \gamma_5 C_D {{\rm T}^a}^\dagger \bar E^a  P_-^T 
\bar \psi(x)^T 
\nonumber\\
&\longrightarrow&
\bar T'_+(x) = \frac{1}{2}\, \bar V_+^{' a}(x) \bar V_+^{' }a(x) , \quad 
\bar V_+^{' a}(x) = (-1)^a \,
\frac{1}{2}\,
\psi^{\rm T}(x)  
{ P_+}^T i \gamma_5 C_D {\rm T}^a P_+ 
\psi(x) .
\end{eqnarray}
%
Therefore, in our model, CP invariance is not manifest\cite{Narayanan:1994gw, Suzuki:2000ku, Fujikawa:2002vj, Fujikawa:2002up}. Instead,
the definition of the chiral projection for the fields and anti-fields are interchanged as
\begin{eqnarray}
&& \psi_-(x) = \hat P_- \psi(x)  \quad \Rightarrow \quad  \psi_-(x) = P_- \psi(x) ,  \\
&& \bar \psi_-(x) = \bar \psi P_+ (x) \quad \Rightarrow \quad  \bar \psi_-(x) = \bar \psi \{ \gamma_5 \hat P_+ \gamma_5\} (x) .
\end{eqnarray}
\begin{eqnarray}
&& \psi_+(x) = \hat P_+ \psi(x)  \quad \Rightarrow \quad  \psi_+(x) = P_+ \psi(x) ,  \\
&& \bar \psi_+(x) = \bar \psi P_- (x) \quad \Rightarrow \quad  \bar \psi_+(x) = \bar \psi \{ \gamma_5 \hat P_- \gamma_5\} (x) .
\end{eqnarray}

This transformation property implies immediately that
the fermion expectation value 
$\big\langle 1 \big\rangle_F [U] ( = {\rm e}^{\Gamma_W [U]} )$
is trasformed in the following manner.
\begin{eqnarray}
&& 
\big\langle 1 \big\rangle_F \,\big[U\big]
\quad\,\, =
\det (\bar v D v)  \,
 \int {\cal D}[ E] \, {\rm pf}( u^T  i \gamma_5 C_D  T^a E^a u) 
\nonumber\\
&\longrightarrow&
\big\langle 1 \big\rangle_F \,\big[U^{CP}\big]
=
\det (u^\dagger \gamma_5 D \gamma_5 \bar u^\dagger )  \,
 \int {\cal D}[ \bar E] \, {\rm pf}( v^\dagger \gamma_5
 i \gamma_5 C_D  {T^a}^\dagger \bar E^a \gamma_5 v^\ast ) 
\nonumber\\
&& 
\phantom{
\big\langle 1 \big\rangle_F \,[U^{CP}]
}
=
\Big\{
\det (\bar u D u )  \,
 \int {\cal D}[E] \, {\rm pf}( v^T
 i \gamma_5 C_D   T^a E^a  v ) 
\Big\}^\ast .
\end{eqnarray}
Therefore a necessary and sufficient condition for the CP invariance of the effective action, 
$\Gamma_W [U^{CP}] =\Gamma_W [U]$, is formulated by the following identity,
\begin{equation}
\label{eq:nece-suf-condition-CP}
\det (\bar v D v)  \,
 \int {\cal D}[ E] \, {\rm pf}( u^T  i \gamma_5 C_D  T^a E^a u) 
= \Big\{
\det (\bar u D u )  \,
 \int {\cal D}[E] \, {\rm pf}( v^T
 i \gamma_5 C_D   T^a E^a  v ) 
\Big\}^\ast .
\end{equation}
Here we assume that
the (background) link field  is in the topologically trivial sector, where $\big\langle 1 \big\rangle_F \,[U]$ is not vanishing and the effective action is well-defined.

To prove the identity eq.~(\ref{eq:nece-suf-condition-CP}),
we consider the two unitary matrices of the size $n (=
{\rm dim} \Lambda  \times 4 \times 16)$ defined by
\begin{eqnarray}
&& \left( \begin{array}{cc} 
            \left( \bar u u\right)  &  \left( \bar u v\right) \\
            \left( \bar v u\right)  &  \left( \bar v v\right)
            \end{array} 
     \right) , \\
&&\left( \begin{array}{cc} 
            \left(  u^T i\gamma_5 C_D T^a E^a u\right)  &  
            \left(  u^T i\gamma_5 C_D T^a E^a v\right)  \\
            \left(  v^T i\gamma_5 C_D T^a E^a u\right)   &                                     \left(  v^T i\gamma_5 C_D T^a E^a v\right) 
            \end{array} 
     \right) ,
\end{eqnarray}
where $\{ u_j, v_j \}$ and $\{ \bar u_j, \bar v_j \}$
consist the complete orthonormal bases of the Dirac fields
$\psi(x)$ and $\bar \psi(x)$, respectively (cf. \cite{Narayanan:1994gw}).
One can choose the bases so that the determinant of the first matrix is unity, while the determinant of the second one is unity independent of the choice.
Note that
$\left( \bar u u\right) = 
\left( \bar u  D u\right)$ and $\left( \bar v v\right) = \left( \bar v D v\right)$.
Note also that the second matrix is unitary because of the constraint
on the spin field, 
$E^a(x) E^a(x) = 1$.
If a unitary matrix $U$ has the block structure as
\begin{equation}
U
=\left( \begin{array}{cc} 
             N &  O \\
             P &  M
            \end{array} 
     \right), 
\end{equation}
where $N$ and $M$ are non-singular square matrices,
it follows that 
\begin{eqnarray}
\det U &=&\det N \times \det \left( M-P N^{-1} O \right)   \nonumber\\
       &=&\det N / \det M^\dagger.
\end{eqnarray}
In the second equality, the relations
$N^{-1}=-P^\dagger(M^\dagger)^{-1}O^{-1}$ and 
$MM^\dagger+PP^\dagger=1$ are used, which follow from the unitarity of $U$. This result implies that
\begin{eqnarray}
\det (\bar v D v)  &=& \big\{ \det (\bar u D u ) \big\}^\ast ,\\
{\rm pf}( u^T  i \gamma_5 C_D  T^a E^a u) 
&=& \pm \big\{
{\rm pf}( v^T
 i \gamma_5 C_D   T^a E^a  v ) 
\big\}^\ast .
\end{eqnarray}
The signature in the second result is the constant independent
of 
the link field 
and the spin field, 
and it can be fixed at 
$U(x,\mu)=1$ and $E^a(x) = \delta^{a 10}$
to be $+1$. Since the path-integration
over the real spin field commute with the complex conjugation,
the identity eq.~(\ref{eq:nece-suf-condition-CP}) now follows.

Therefore, we have
\begin{equation}
\big\langle 1 \big\rangle_F \,\big[U^{CP}\big] =
\big\langle 1 \big\rangle_F \,\big[U\big] ,
\end{equation}
and the effective action is indeed CP invariant\cite{Narayanan:1994gw}, 
\begin{equation}
\Gamma_W [U^{CP}] =\Gamma_W [U] .
\end{equation}
Thus the Weyl field measure defined by eqs.~(\ref{eq:def-weyl-measure}) and (\ref{eq:def-dirac-measure})
respects the CP symmetry.

\subsection{Schwinger-Dyson equations and Correlation functions}

The Schwinger-Dyson equations for the link field and the Weyl field can be derived from the path-integral definition of the partition function,
eqs.~(\ref{eq:def-Z}) and (\ref{eq:def-effective-action}).
With respect to the local variation of the link field, 
$\delta_\eta U(x,\mu) = i \eta_\mu(x) U(x,\mu)$,
the simplest non-trivial example is given by 
\begin{eqnarray}
\label{eq:SD-eq-U}
&& \left\langle \Big[  
  - \delta_\eta S_G[U] 
  - \sum_{x\in\Lambda} \bar \psi (x) P_+ \delta_\eta D \psi(x) 
  + \sum_{x\in\Lambda} \psi^{\rm T}  {\hat P_+}^T i \gamma_5 C_D {\rm T}^a E^a \delta_\eta \hat P_+ \psi(x)
   \Big]  \right\rangle =0 , \nonumber\\ 
\end{eqnarray}
The operators in the bracket $[ \cdots ]$ in the l.h.s. are all the local operators with respect to the variation point $x$
and therefore  the equation of motion is local.
We note that the third term comes from the link field dependence
of the Weyl field measure.
With respect to the local variations of the fermion fields
$\delta \psi(x)$, $\delta \bar \psi(x)$ and of the spin field 
$\delta E^a(x)$, one can derive the following non-trivial examples.
\begin{eqnarray}
\label{eq:SD-eq-psi}
\left\langle 
   \psi(y) \, 
   \Big[  
   \bar \psi P_+ D (x)
  -  \psi^{\rm T}  {\hat P_+}^T i \gamma_5 C_D {\rm T}^a E^a \hat P_+ (x)
   \Big] \,    \right\rangle_F  &=&  \delta_{xy} 
\big\langle 1 \big\rangle_{F} , \\
\label{eq:SD-eq-barpsi}
\left\langle \Big[  
   P_+ D \psi(x)
  -   { P_-} i \gamma_5 C_D {{\rm T}^a}^\dagger \bar E^a {P_-}^T \bar \psi^{\rm T} (x)
   \Big] \, \bar \psi(y) \right\rangle_F  &=& \delta_{xy}
\big\langle 1 \big\rangle_{F} , \\
\label{eq:SD-eq-E}
\frac{1}{2}\,
\left\langle 
   \psi^{\rm T}  {\hat P_+}^T i \gamma_5 C_D 
{\rm C}[\Sigma_{bc} , \Gamma^a] E^a(x) \hat P_+ \psi
\,    \right\rangle_F &=&  0.
\end{eqnarray}
The first two equations can be decomposed into the chiral components by
noting $ P_+ D = D {\hat P}_-$ and 
$\delta_{xy} = (P_+ + P_- ) \delta_{xy} = \hat P_+(x,y) + \hat P_-(x,y)$. We finally obtain
\begin{eqnarray}
\label{eq:two-point-function-lefthanded}
\left\langle  \psi_- (x) \, \bar \psi_-(y)              \right\rangle_F  &=&   {\hat P}_- D^{-1} 
P_+ (x,y) \big\langle 1 \big\rangle_{F} , \\
\label{eq:SD-eq-psi+}
\left\langle 
   \psi_+ (y) \, 
   \Big[  
  \psi_+^{\rm T}   i \gamma_5 C_D {\rm T}^a E^a \hat P_+ (x)
   \Big] \,    \right\rangle_F  
&=& - \hat P_+(y,x)\big\langle 1 \big\rangle_{F} , \\
\label{eq:SD-eq-barpsi+}
\left\langle \Big[  
   { P_-} i \gamma_5 C_D {{\rm T}^a}^\dagger \bar E^a  \bar \psi_+^{\rm T} (x)
   \Big] \, \bar \psi_+(y) \right\rangle_F  
&=& - P_- \delta_{xy} \big\langle 1 \big\rangle_{F},
\end{eqnarray}
assuming that $D$ is invertible. 

As long as $\big\langle 1 \big\rangle_{F}$
is finite and well-defined, these results imply the 
following facts about the particle spectrum
in the channel of the $\underline{16}$ representation of SO(10) symmetry:
the left-handed fields $\psi_-(x)$, $\bar \psi_-(x)$ 
support the massless Weyl fermions
and have long-range correlations,
while
the right-handed fields $\psi_+(x)$, $\bar \psi_+(x)$
are decoupled each other and 
have short-range correlations of order the several lattice spacings
with the composite operators 
$\big[\psi_+^{\rm T}   i \gamma_5 C_D {\rm T}^a E^a \hat P_+ (x)
 \big]$ and 
$\big[{ P_-} i \gamma_5 C_D {{\rm T}^a}^\dagger \bar E^a  \bar \psi_+^{\rm T} (x)\big]$, respectively.\cite{Hernandez:1998et}
As to the right-handed field $\psi_+(x)$, however,
the information of yet another correlation function 
$\big\langle 
   \psi_+ (y) \, 
   \big[  
  \psi_+^{\rm T}   i \gamma_5 C_D {\rm T}^a E^a \hat P_- (x)
   \big] \,    \big\rangle_F$
is also required before deducing a definite conclusion.

%

\subsection{Gauge field dependence of the Weyl field measure -- Locality issue remaining}

The variation of the effective action $\Gamma_W [U]$ 
w.r.t. the link field can be derived
from the path-integral definition eq.~(\ref{eq:def-effective-action}) as follows.
\begin{eqnarray}
\delta_\eta \Gamma_W [U] 
&=&
 \Big\langle 
  - \sum_{x\in\Lambda} \bar \psi(x) P_+ \delta_\eta D \psi(x) 
  +  \sum_{x\in\Lambda} \psi^{\rm T}(x)  {\hat P_+}^T i \gamma_5 C_D {\rm T}^a E^a \delta_\eta \hat P_+ \psi(x)
\Big\rangle_F  \, 
\big\slash 
\big\langle 1 \big\rangle_F 
\nonumber\\ 
&=&
{\rm Tr}
\big\{
P_+ \delta_\eta D \big\langle \psi_- \bar \psi_- \big\rangle_F
\big\}\, 
\big\slash 
\big\langle 1 \big\rangle_F 
- \,{\rm Tr} \big\{ 
\delta_\eta \hat P_+  
 \big\langle 
\psi_+ \big[ \psi_+^{\rm T} i \gamma_5 C_D {\rm T}^a E^a 
\big] 
\big\rangle_F  \big\}\, 
\big\slash 
\big\langle 1 \big\rangle_F .
\nonumber\\ 
%
\end{eqnarray}
The first term can be rewritten further
using the result of 
the two-point correlation function of the left-handed
fields eq.~(\ref{eq:two-point-function-lefthanded}) as 
\begin{equation}
%
{\rm Tr}
\big\{
P_+ \delta_\eta D \big\langle \psi_- \bar \psi_- \big\rangle_F
\big\}\, 
\big\slash 
\big\langle 1 \big\rangle_F 
= {\rm Tr}\{ P_+ \delta_\eta D D^{-1} \} .
\end{equation}
It is identified as the physical contribution of the left-handed
Weyl fermions.
The second term, on the other hand, represents 
the gauge field dependence of the Weyl field measure eq.~(\ref{eq:def-weyl-measure}) through the right-handed 't Hooft vertices.
It replaces
the measure term 
$-i \mathfrak{L}_\eta = \sum_j( v_j, \delta_\eta v_j)$\cite{Luscher:1998du, Luscher:1999un}. 
So we denote this term with $-i \mathfrak{T}_\eta$,
\begin{equation}
 -i \mathfrak{T}_\eta 
\equiv
- \,{\rm Tr} \big\{ 
\delta_\eta \hat P_+  
 \big\langle 
\psi_+ \big[ \psi_+^{\rm T} i \gamma_5 C_D {\rm T}^a E^a 
\big] 
\big\rangle_F  \big\}\, 
\big\slash 
\big\langle 1 \big\rangle_F .
\end{equation}
Then the variation of the effective action 
is written as
\begin{eqnarray}
\delta_\eta \Gamma_W [U] 
&=&
{\rm Tr}\{ P_+ \delta_\eta D D^{-1} \} 
 -i \mathfrak{T}_\eta .
\end{eqnarray}

For the gauge transformation, $\delta_\eta U(x,\mu) =
i \{ \omega(x) U(x, \mu) - U(x,\mu) \omega(x + \hat \mu) \}$ and
$ \eta_\mu(x) = \omega(x)- U(x,\mu) \omega(x + \hat \mu) U(x,\mu)^{-1} = - D_\mu \omega (x)$,
the first term gives the gauge anomaly term,
\begin{equation}
{\rm Tr}\{ P_+ \delta_\eta D D^{-1} \} 
\big\vert_{\eta_\mu = - D_\mu \omega} 
= - i  {\rm Tr}\{ \omega \gamma_5 D \},
\end{equation}
where,
in the weak gauge-coupling expansion, 
the leading non-trivial term is vanishing because of the
anomaly cancellation condition for 
the 16-dimensional (irreducible) spinor representation of SO(10),
${\rm Tr} 
\big\{ 
{\rm P_+} \Sigma_{a_1 b_1} 
[   \Sigma_{a_2 b_2}  \Sigma_{a_3 b_3}  
+   \Sigma_{a_3 b_3}  \Sigma_{a_2 b_2} 
] 
\big\} = 0  .
$
The second term gives
\begin{eqnarray}
-i \mathfrak{T}_\eta \big\vert_{\eta_\mu = - D_\mu \omega} 
&=& 
-i  \,{\rm Tr} \big\{ 
 [\omega, \hat P_+] 
 \big\langle 
\psi_+ \big[ \psi_+^{\rm T} i \gamma_5 C_D {\rm T}^a E^a 
\big] 
\big\rangle_F  \big\}\, 
\big\slash 
\big\langle 1 \big\rangle_F 
\nonumber\\
&=&
\Big(
-i \, \frac{1}{2}\, \,{\rm Tr} \big\{ 
\big\langle 
\psi_+ \big[ \psi_+^{\rm T} i \gamma_5 C_D {\rm C}[\Gamma^a , \omega] E^a \big]
\big\rangle_F 
 \big\}\, 
\nonumber\\
&&\quad
+ i \,{\rm Tr} \big\{ 
 \big\langle 
\psi_+ \big[ \psi_+^{\rm T} i \gamma_5 C_D {\rm T}^a E^a
\hat P_+ \big]
\big\rangle_F \omega 
 \big\}\, 
\Big)
\big\slash 
\big\langle 1 \big\rangle_F 
\nonumber\\
&=&
\label{eq:gauge-variation-pfaffian}
+ i  {\rm Tr}\{ \omega \gamma_5 D \},
\end{eqnarray}
where  the Schwinger-Dyson equations eqs.~(\ref{eq:SD-eq-E}) and (\ref{eq:SD-eq-psi+}) are used
at the last equality. Thus we can check that the effective action is gauge-invariant.

The measure term $-i \mathfrak{T}_\eta$ 
is required to be a smooth and local function of the link field variables, since it appears
as an operator of the link field in the Schwinger-Dyson equation
w.r.t. the link field, eq.~(\ref{eq:SD-eq-U}).
In the weak gauge-coupling expansion, in particular, the vertex functions are derived from this term as
\begin{eqnarray}
-i \mathfrak{T}_\eta
&=&
\sum_{m=0}^\infty
 \frac{(ig)^{1+m}}{V^{1+m} \, m!}\, 
\sum_{k,p_1,\cdots,p_m} \tilde \eta^{ab}_\mu(-k)\,
{\mathfrak C}_{\mu \nu_1 \cdots \nu_m}^{a b a_1 b_1 \cdots a_m b_m} (k,p_1,\cdots,p_m ) \, \times
\nonumber\\
&& \qquad\qquad\qquad\qquad\qquad\qquad
\qquad\qquad
 \tilde A^{a_1 b_1}_{\nu_1}(p_1)\cdots  \tilde A^{a_m b_m}_{\nu_m}(p_m) 
\end{eqnarray}
and they are required to be analytic w.r.t. the external momenta.
Since the gauge field dependence of the Weyl field measure
is induced by the path-integrations of the right-handed Weyl field $\psi_+(x)$ and the spin field $E^a(x)$, 
it is required 
at least that
these fields have short range correlations with
the correlation lengths of order the lattice spacing.
A necessary and sufficient condition for this requirement is that
the corrlation function 
$\big\langle 
\psi_+(x) 
\big[ \psi_+^{\rm T} i \gamma_5 C_D {\rm T}^a E^a (y)
\big] 
\big\rangle_F $ is the smooth function of the (background) link field variables 
and it 
satisfies the locality bound,
\begin{eqnarray}
\big\| 
\big\langle \psi_+(x) 
\big[ \psi_+^{\rm T} i \gamma_5 C_D {\rm T}^a E^a (y)
\big] 
\big\rangle_F 
\big\slash 
\big\langle 1 \big\rangle_F 
\big\| 
& \, \, \le \, \, & C \, |x-y|^\sigma \, {\rm e}^{-|x-y|/ \xi } 
\end{eqnarray}
for certain constants $C > 0$, $\sigma > 0$ and $\xi > 0$
and the similar bounds for its variations 
w.r.t.~the link field. (cf. \cite{Hernandez:1998et}) 
We note that the above condition is satisfied by the part
$\big\langle \psi_+(x) 
\big[ \psi_+^{\rm T} i \gamma_5 C_D {\rm T}^a E^a \hat P_+ (y) \big] \big\rangle$ because of eq.~(\ref{eq:SD-eq-psi+}) 
and is therefore about the part of
$\big\langle \psi_+(x) 
\big[ \psi_+^{\rm T} i \gamma_5 C_D {\rm T}^a E^a \hat P_- (y) \big] \big\rangle$. 
The locality range $\xi$ then determines the effective cutoff scale $\Lambda$  of the lattice model as $\Lambda =\pi (\xi a )^{-1}$.

This question is of not quite dynamical but non-perturbative nature
in our model, 
involving
the path-integration of the spin field $E^a(x)$ with the 
weight ${\rm pf}( u^T  i \gamma_5 C_D  T^a E^a u) $
which is complex in general. 
And we do not have yet a rigorous proof on the smoothness and locality of the measure term for any admissible link fields. 
But 
this question is well-defined. 
It can be addressed 
in the weak gauge-coupling limit at least because
the pfaffian 
${\rm pf}( u^T  i \gamma_5 C_D  T^a E^a u)$ 
is positive semi-definite in this case, as we will argue in the following section, and 
Monte Carlo methods are applicable 
to evaluate the correlation functions and the vertex functions. 
We leave this important and interesting  question
for our future study.

For later use, we express 
the measure term $-i \mathfrak{T}_\eta$ in terms of the chiral basis, although it does not actually depend on the choice of the basis. 
For this, we first note that the correlation function 
$\big\langle 
\psi_+(x) 
\big[ \psi_+^{\rm T} i \gamma_5 C_D {\rm T}^a E^a (y)
\big] 
\big\rangle_F $ 
is written explicitly in the chiral basis as
\begin{eqnarray}
&& \big\langle \psi_+(x) 
\big[ \psi_+^{\rm T} i \gamma_5 C_D {\rm T}^a E^a (y)
\big] 
\big\rangle_F 
\nonumber\\
&& =
- 
\big\langle 
u_i(x) 
( u^T  i \gamma_5 C_D  T^a E^a u)^{-1}_{ij}
( u_j^T(y) i \gamma_5 C_D  T^a E^a (y)
\big\rangle_E / \big\langle 1 \big\rangle_E .
\end{eqnarray}
Then
the measure term $-i \mathfrak{T}_\eta$
can be also expressed in terms of the chiral basis as
\begin{eqnarray}
-i \mathfrak{T}_\eta 
&\equiv&
- \,{\rm Tr} \big\{ 
\delta_\eta \hat P_+  
\big\langle 
\psi_+ \big[ \psi_+^{\rm T} i \gamma_5 C_D {\rm T}^a E^a 
\big] 
\big\rangle_F  
\big\}\, 
\big\slash 
\big\langle 1 \big\rangle_F
\nonumber\\
&=& 
\big \langle
{\rm Tr} \{ 
( u^T  i \gamma_5 C_D  T^a E^a  \delta_\eta \hat P_+ u) 
( u^T  i \gamma_5 C_D  T^a E^a u)^{-1}
 \}
\big\rangle_E
\big\slash
\big\langle 1 \big\rangle_E .
\end{eqnarray}
In this respect, we find it interesting to see that 
such quantity like the integral of pfaffian
$\int {\cal D}[ E] \,
 {\rm pf}( u^T  i \gamma_5 C_D  T^a E^a u)$ 
can reproduce the gauge anomaly term, 
$ i {\rm Tr}\{ \omega \gamma_5 D \}$, 
besides the chiral determinant $\det (\bar v D v) $
when the anomaly cancellation condition is fulfilled.
The variation of the integral of pfaffian is evaluated as
\begin{eqnarray}
&&\delta_\eta \ln
\big\{
\int {\cal D}[ E] \,
 {\rm pf}( u^T  i \gamma_5 C_D  T^a E^a u)
\big\} 
\nonumber\\
&=&
\sum_j (u_j , \delta_\eta u_j)
+
\big \langle
{\rm Tr} \{ 
( u^T  i \gamma_5 C_D  T^a E^a  \delta_\eta \hat P_+ u) 
( u^T  i \gamma_5 C_D  T^a E^a u)^{-1}
 \}
\big\rangle_E
\big\slash
\big\langle 1 \big\rangle_E 
\nonumber\\
&=&
\sum_j (u_j , \delta_\eta u_j)
-i \mathfrak{T}_\eta .
\end{eqnarray}
The first term in the r.h.s. is followed from the property of the pfaffian as
${\rm pf} \{ Q^T A Q \} = {\rm pf} A \times \det Q$ where $Q$ is unitary.
It sums up with
the term 
$\sum_j( v_j, \delta_\eta v_j) (= -i \mathfrak{L}_\eta)$
from the variation of chiral determinant $\det (\bar v D v) $
to zero, because of the condition eq.~(\ref{eq:basis-independent-on-link}). Thus the gauge-variation of the integral of pfaffian  leads to the result 
eq.~(\ref{eq:gauge-variation-pfaffian}).\footnote{If one makes the other choice for $F(w)$ as
 $F(w) = {\rm e}^w =\sum_{k=0}^\infty \frac{w^k}{k!}$,
the integral of pfaffian  $\int {\cal D}[ E] \,
 {\rm pf}( u^T  i \gamma_5 C_D  T^a E^a u)$ 
is replaced by the hyper-pfaffian, ${\rm hpf} A$, of
the rank-four complete anti-symmetric tensor
$A_{ijkl} \equiv T_{ijkl} + T_{iklj} + T_{iljk}$ where
$T_{ijkl} =\sum_x (1/2) u^T_i(x) i \gamma_5 C_D T^a u_j(x) \,u^T_k(x) i \gamma_5 C_D T^a u_l(x) $. It can also reproduce the gauge anomaly term.
We wonder if it is possible to interpret these quantities
from the point of view of topological field theory.
}

\section{More on the saturation of 
the fermion measure by 't Hooft vertices}
\label{sec:saturation-right-handed-measure}

As discussed in the previous section, 
the pfaffian of the matrix eq.~(\ref{eq:the-first-matrix}) is in general a complex number which depends on the spin field $E^a(x)$ as well as the link field $U(x,\mu)$. 
And we do not have yet  a rigorous proof that the path-integration of the pfaffian over $E^a(x)$,
\begin{eqnarray}
\big\langle 1 \big\rangle_{E}
& = & 
\int {\cal D}[ E] \, 
{\rm pf}( u^T  i \gamma_5 C_D  T^a E^a u) \, ,
\end{eqnarray}
is non-zero for any admissible link fields. 
But there are typical examples where one can argue that it is indeed the case.
Those include the case in the weak gauge-coupling limit where the link variables are set to unity, $U(x,\mu)=1$, and the cases
of the ${\rm SU(2)}$ link fields with non-zero topological charges $Q (\not = 0)$, which represent the non-trivial topological sectors $\mathfrak{U}[Q]$.
We will examine these cases in detail.

\subsection{Property of the functional pfaffian for 
the link fields in Spin(9) subgroup}

For this purpose, let us assume that
the link field
$U(x,\mu)$ is in the ${\rm SO(9)}$ subgroup
and 
commutes with 
$\Gamma^{10}$,
%
\begin{equation}
\left[\, \Gamma^{10} , U(x,\mu) \, \right]  = 0 
\end{equation}
and, accordingly, 
\begin{eqnarray}
 \Gamma^{10} \, \hat P_{+} [U] \, \Gamma^{10} =  \hat P_{+} [U] .
\end{eqnarray}
Then the charge-conjugation relation,
\begin{eqnarray}
{\rm  C }^{-1}( \gamma_5 C_D)^{-1} \, \hat P_{+}[U]{}^T \,
(\gamma_5 C_D ) {\rm C}
 =  \hat P_{+} [U],
\end{eqnarray}
imlpies that one can choose the basis vectors $\{ u_j(x) \}$ so that they satisfy the relation 
\begin{eqnarray}
 &&  { u_j(x)}^T i \gamma_5 C_D  \, {\rm C} \Gamma^{10} 
= {\cal C}_{jk} \,  { u_k(x) }^\dagger,  \\
 &&
 {\cal C}^{-1} = {\cal C}^\dagger =
 {\cal C}^T = - \, {\cal C} .
\end{eqnarray}
${\cal C}$ is then given by the expression
${\cal C}_{jk}
= ( u^T  i \gamma_5 C_D  \, {\rm C} \Gamma^{10} u)_{jk}$
and the matrix eq.~(\ref{eq:the-first-matrix}) can be represented as 
\begin{eqnarray}
( u^{\rm T}\, i \gamma_5 C_D {\rm T}^a E^a u )
&=& 
 {\cal C} \times
( u^\dagger \Gamma^{10} \Gamma^{a} E^{a} u ) 
\nonumber\\
&=& 
( u^\dagger \Gamma^{10} \Gamma^{a} E^{a} u )^T
 \times
{\cal C} .
\end{eqnarray}
This implies that
while the eigenvalues of 
$( u^{\rm T}\, i \gamma_5 C_D {\rm T}^a E^a u )$ appear in pair with the opposite signatures as
$ \{ (\tilde \lambda_i, - \tilde \lambda_i) \, \vert \,  i=1,\cdots, n/4-4Q\}$, 
the eigenvalues of 
$( u^\dagger \Gamma^{10} \Gamma^{a} E^{a} u )$ 
degenerate with the multiplicity two at least as
$\{ (\lambda_i, \lambda_i) \, \vert \, i=1,\cdots, n/4-4Q \}$.

Then the pfaffian of the matrix $( u^{\rm T}\, i \gamma_5 C_D {\rm T}^a E^a u )$ can be written as
\begin{eqnarray}
\label{eq:pfaffian-by-half-product}
{\rm pf}( u^{\rm T}\, i \gamma_5 C_D {\rm T}^a E^a u )
&=& 
{\rm pf} ( u^T  i \gamma_5 C_D  \, {\rm C} \Gamma^{10} u)
\times
{\prod_{i=1}^{n/4-4Q}} \lambda_i .
\end{eqnarray}
Since
the space of the spin field configurations, which we
denote with ${\cal V}_E$, is the direct product of 
multiple $S^9$,   ${\cal V}_E =S^9 \times \cdots \times S^9 $ 
(by ${\rm dim} \, \Lambda$ times), it is pathwise connected 
and
any configuration of the spin field $E^a(x)$ can be reached
from the constant configuration 
$E_0^a(x) = \delta^{a, 10} = {\rm const.}$ 
through a continuous deformation.
For the constant configuration we have $\lambda_i = 1$ ($i=1,\cdots,n/4-4Q$).
And this fixes the signature of the above formula
(as long as the pfaffian is not vanishing).

The half product 
of the eigenvalues of 
$( u^\dagger \Gamma^{10} \Gamma^{a} E^{a} u )$, 
${\prod_{i=1}^{n/4-4Q}} \lambda_i $
in eq.~(\ref{eq:pfaffian-by-half-product}), 
is independent of the choice of the basis vectors. 
Its square or the full product can be expressed 
in the basis-independent manner as folows,
\begin{eqnarray}
{\prod_{i=1}^{n/4-4Q}} \lambda_i^2
&=&
\det ( u^\dagger \Gamma^{10} \Gamma^{a} E^{a} u ) 
\nonumber\\
&=& 
\det \Big(
{\rm P}_- 
+
{\rm P}_+ 
\big[
\hat P_- 
+ \hat P_+ \Gamma^{10} \Gamma^{b} E^b 
\big]
\Big) .
\end{eqnarray}

\subsection{The case of trivial link field in the weak gauge-coupling limit}

In the weak gauge-coupling limit where the link variables are set to unity, $U(x,\mu)=1$, 
one can choose the basis vectors $\{ u_j(x) \}$ as
\begin{eqnarray}
\label{eq:chiral-basis-u-free}
u_j(x) = \frac{1}{\sqrt{L^4}} { \rm e}^{ i p x } \, 
u_\alpha (p, \sigma) \, \delta_{s,t}    \qquad   ( j=\{ p, \sigma, t \} ) ,
\end{eqnarray}
where $\{ u_\alpha (p, \sigma) \}$ are the four-spinor eigenvectors
of the free hermitian Wilson-Dirac operator 
$H_{\rm w} = \gamma_5 (D_{\rm w} - m_0) \,(0 < m_0 < 2)$
with the negative eigenvalues
in the plane-wave basis given by
\begin{eqnarray}
\label{eq:chiral-basis-u-planewave}
u_\alpha (p, \sigma) &=&
\left\{ 
\begin{array}{lcl}
\left( \begin{array}{c} 
-c(p)\chi_\sigma \\ 
( \omega(p)+ b(p) )\chi_\sigma  \end{array}\right) 
               / \sqrt{2 \omega(p) ( \omega(p)+ b(p) )}
 && ( p \not = 0 )
\\
&&\\
\left( \begin{array}{c} \chi_\sigma \\ 
                          0     \end{array} \right) 
%
&& ( p = 0 )
\end{array}
\right.
\end{eqnarray}
and
\begin{eqnarray}
b(p) &=&  \sum_\mu (1- \cos p_\mu) - m_0 , \\
c(p) &=& I  \{ i \sin p_0 \} - \sum_k \sigma_k \sin p_k , \\
\omega(p) &=& \sqrt{{\scriptstyle \sum_\mu} \sin^2 ( p_\mu )   
+ \big\{ {\scriptstyle \sum_\mu } (1- \cos (p_\mu) ) - m_0 \big\}^2} .
\end{eqnarray}
The four-momentum $p_\mu$ is given by 
$p_\mu = 2\pi n_\mu / L \,\, (n_\mu \in \mathbb{Z})$ 
for the periodic boundary condition and 
$p_\mu = 2\pi (n_\mu + 1/2) / L \,\, (n_\mu \in \mathbb{Z})$
for the anti-periodic boundary condition.
The zero modes with $p_\mu=0$ in eq.~(\ref{eq:chiral-basis-u-planewave}) exist only for the periodic boundary condition.
(See the appendix~\ref{app:chiral-basis-free} for detail.)

For the constant configuration of the spin field, 
$E^a_0(x) = \delta^{a,10}$, 
$( u^{\rm T}\, i \gamma_5 C_D {\rm T}^a E^a_0 u )_{jk} (= 
{\cal C}_{jk} ) =
 \delta_{p+p',0} \epsilon_{\sigma, \sigma'} \,  i {\check C}_{t t'}$ ($j = \{p, \sigma, t\}, k=\{ p', \sigma', t'\}$) and 
$\tilde \lambda_i = \pm i$, 
while
$ \big( u^\dagger \Gamma^{10} \Gamma^{a} E^{a}_0 u \big)_{jk} =
\delta_{jk}$ and $\lambda_i = + 1$.
Then ${\rm pf} ( u^{\rm T}\, i \gamma_5 C_D {\rm T}^a E^a_0 u )$ is unity.

\begin{figure}[th]
\includegraphics[width =75mm]{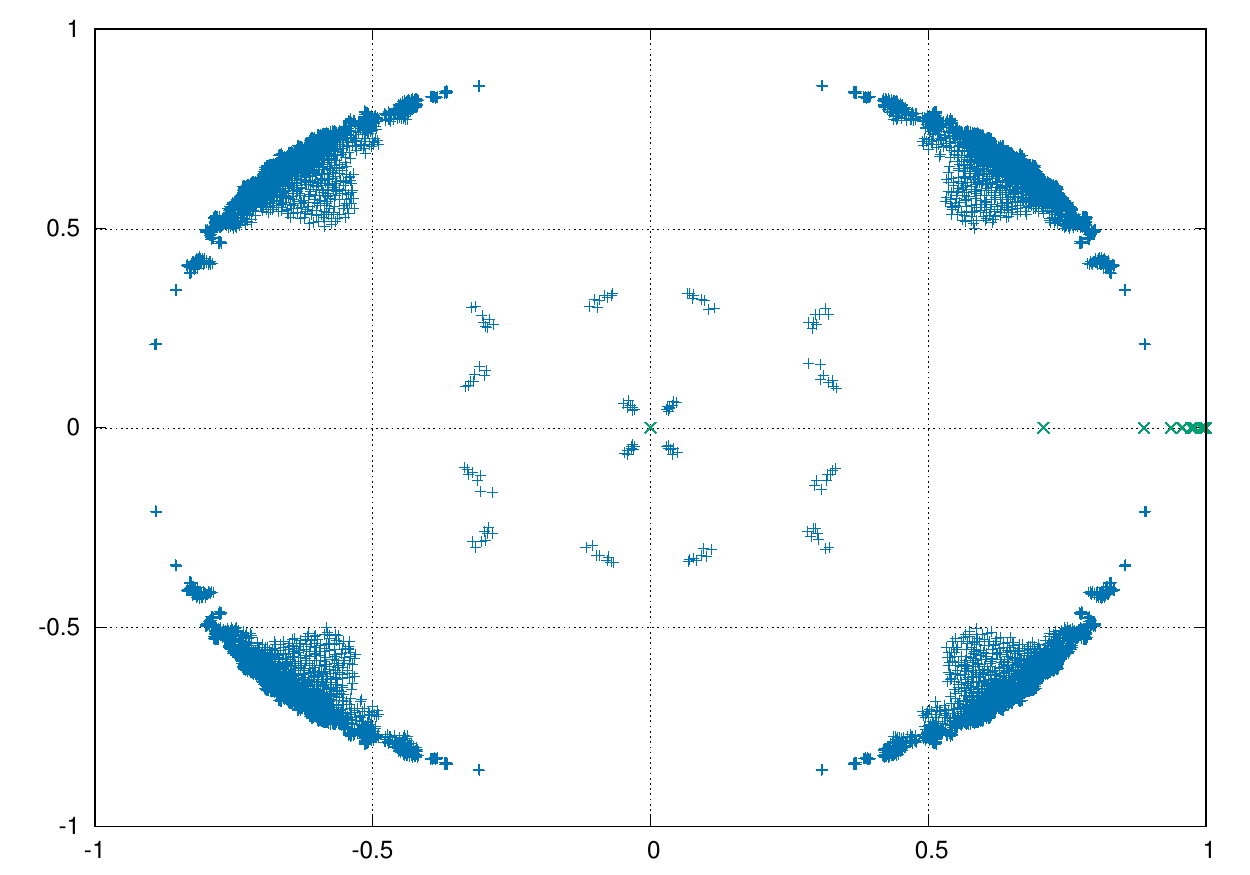} 
\includegraphics[width =75mm]{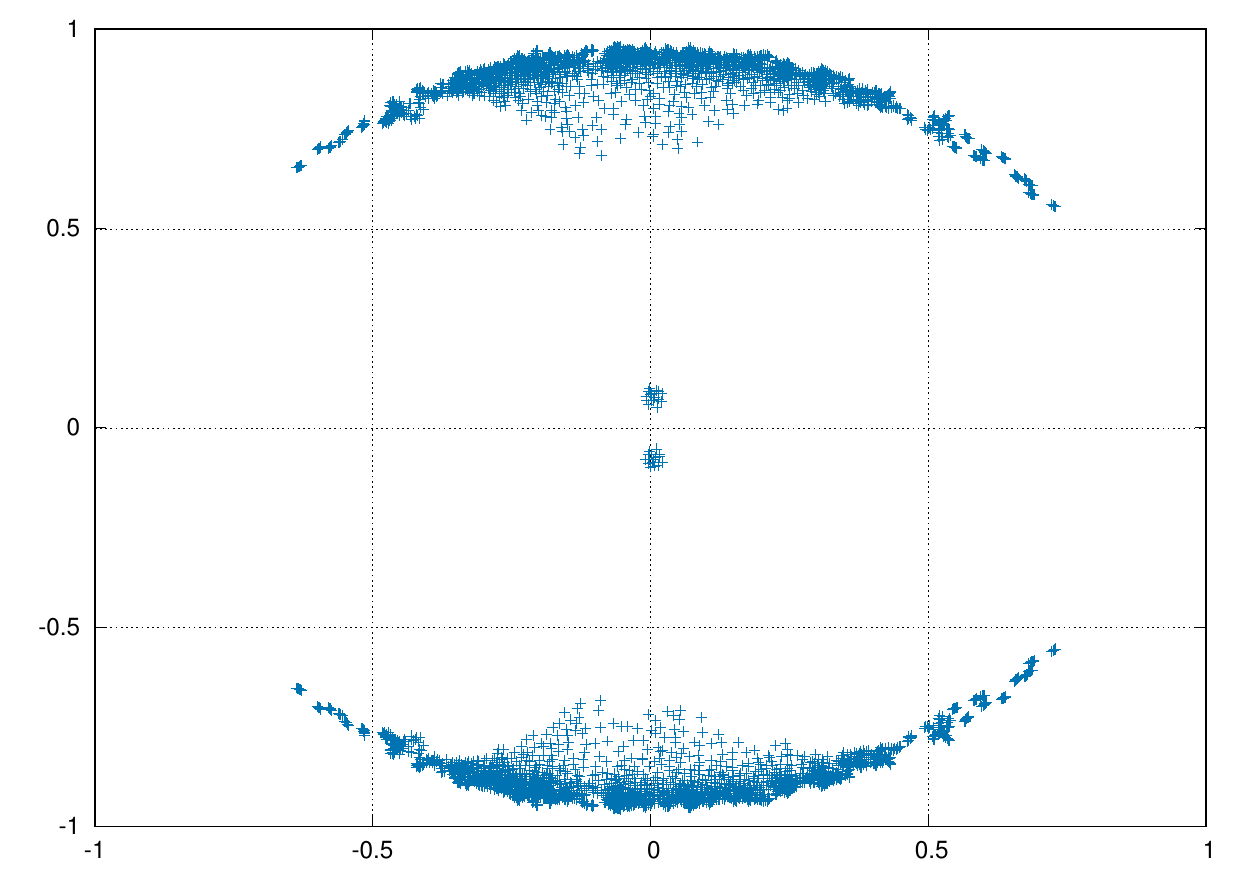} 
\caption{The eigenvalue spectra 
of the matrices
$ \big(u^{\rm T} \,  i \gamma_5 C_D {\rm T}^a E^a u \big) $ and
$ \big( u^\dagger \Gamma^{10} \Gamma^{a} E^{a} u \big) $ 
with a randomly generated spin-field configuration
for the case of the trivial link field. 
The lattice size is $L=4$ and
the boundary condition for the fermion field is periodic.
For reference, the eigenvalue spectrum of the matrix $(\bar v_k D v_i)$  
is also shown with green x symbol for the same boundary condition. 
}
\label{fig:eigenvalues-L=4-Q=0E=R}
\end{figure}

For randomly generated spin-field configurations
with the lattice sizes up to $L=4$, 
we computed numerically
the eigenvalues of the matrices
$ \big(u^{\rm T} \,  i \gamma_5 C_D {\rm T}^a E^a u \big) $ and
$ \big( u^\dagger \Gamma^{10} \Gamma^{a} E^{a} u \big) $,
the pfaffian 
${\rm pf} \big(u^{\rm T} \, i\gamma_5 C_D {\rm T}^a E^a u \big)$ and the half product
${\prod_{i=1}^{n/4-4Q}} \lambda_i$.
We checked that the eigenvalue spectra of the these matrices 
have the structures of 
$ \{ (\tilde \lambda_i, - \tilde \lambda_i) \, \vert \,  i=1,\cdots, n/4-4Q\}$
and
$\{ (\lambda_i, \lambda_i) \, \vert \, i=1,\cdots, n/4-4Q \}$,
respectively.
All eigenvalues turn out to be non-zero.
But 
there appear relatively small eigenvalues for the periodic boundary condition. 
We found that both the pfaffian 
and 
the half product
stay real-positive.
The typical examples of the eigenvalue spectra 
are shown in fig.~\ref{fig:eigenvalues-L=4-Q=0E=R} for $L=4$ with the periodic boundary condition.
One can see how the half product remains real-positive:
the eigenvalues of the matrix appear closely but not exactly in degenerate complex-conjugate pairs $\{ \lambda, \lambda, \lambda^\ast, \lambda^\ast \}$ and the complex phases of the ``half set'' of the eigenvalues sum up still exactly to zero mod $2\pi$. Accordingly, the eigenvalues of $ \big(u^{\rm T} \,  i \gamma_5 C_D {\rm T}^a E^a u \big) $ appear closely but not exactly in the quartet structure $\{ \tilde \lambda, - \tilde \lambda , \tilde \lambda^\ast, -\tilde \lambda^\ast \}$ and the pfaffian is still exactly real-positive.

\begin{figure}[thb]
\includegraphics[width =75mm]{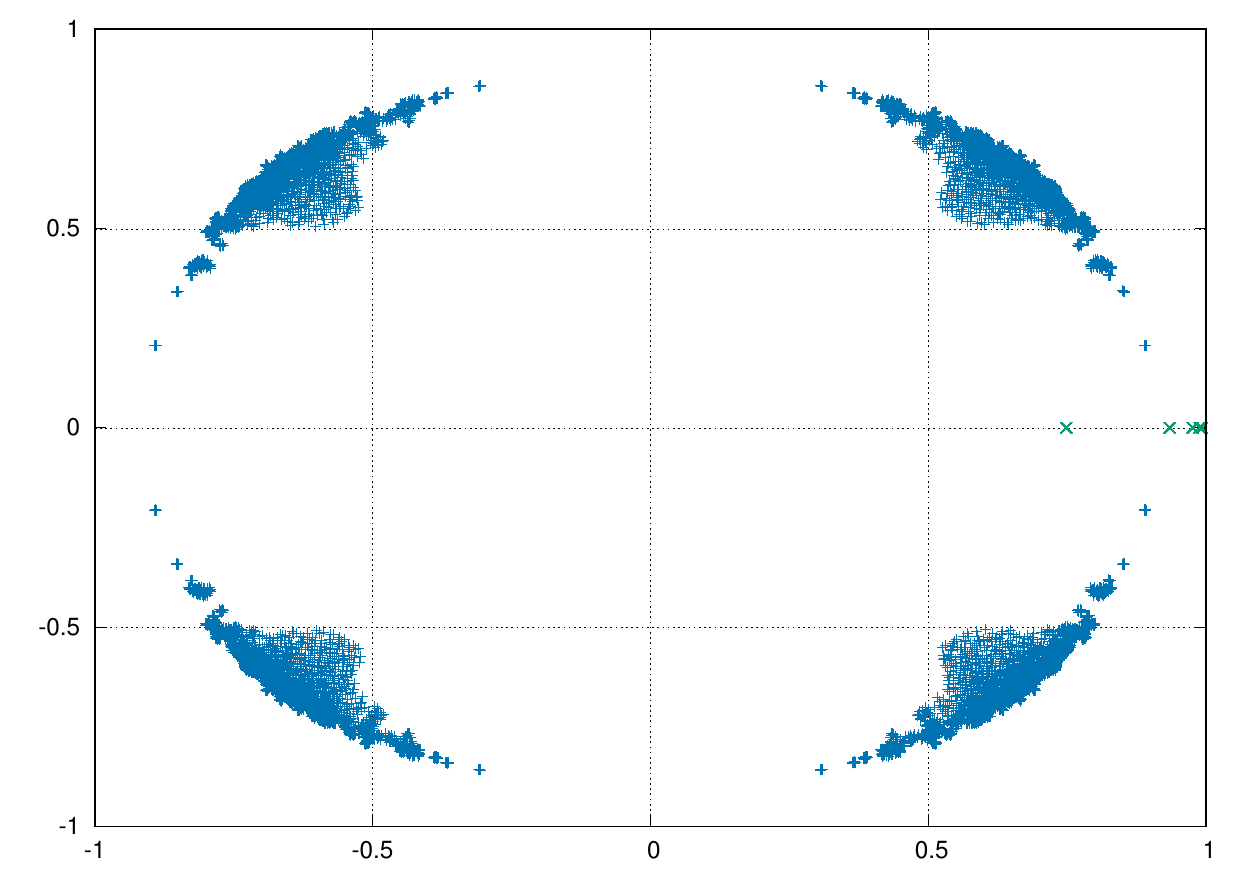} 
\includegraphics[width =75mm]{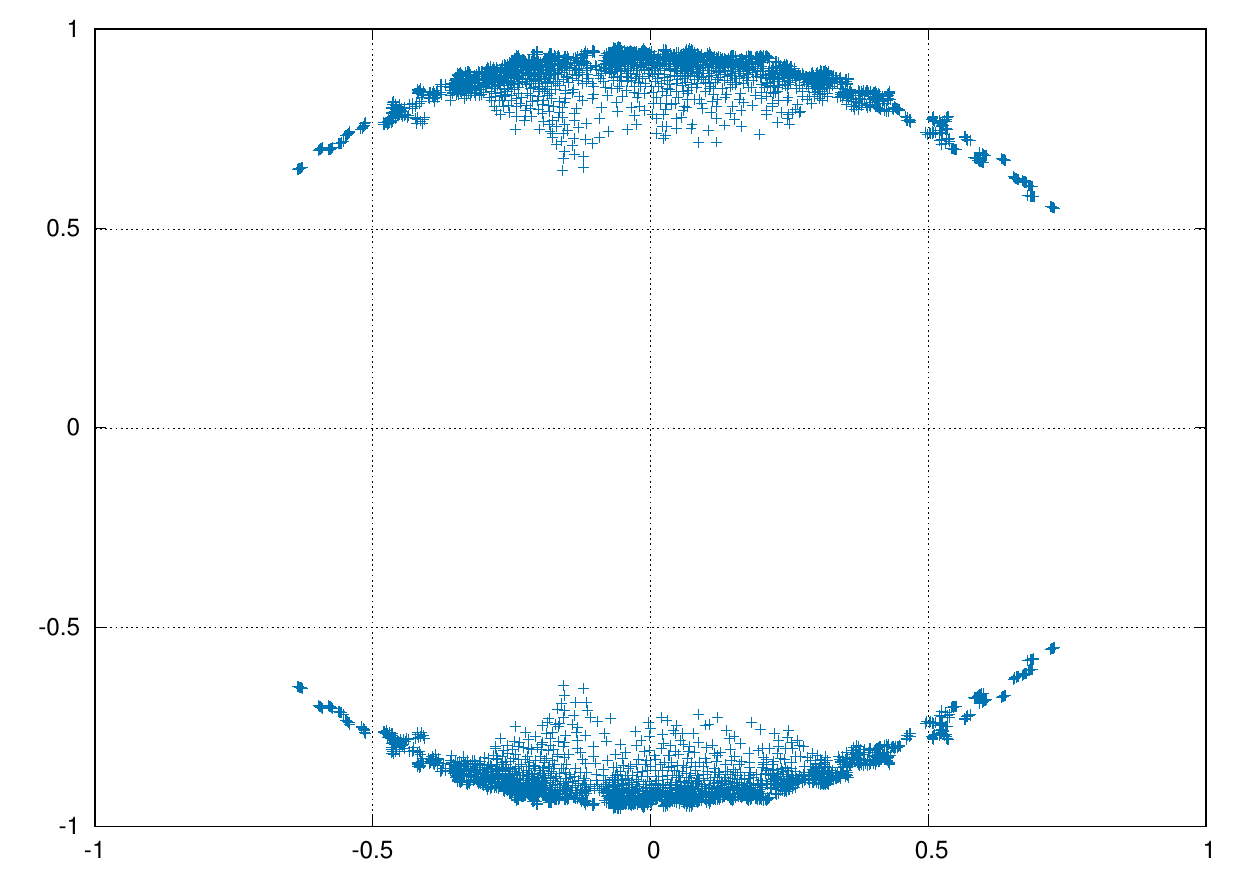} 
\caption{The eigenvalue spectra 
of the matrices
$ \big(u^{\rm T} \,  i \gamma_5 C_D {\rm T}^a E^a u \big) $ and
$ \big( u^\dagger \Gamma^{10} \Gamma^{a} E^{a} u \big) $ 
with a randomly generated spin-field configuration
for the case of the trivial link field. 
The lattice size is $L=4$ and
the boundary condition for the fermion field is anti-periodic.
For reference, the eigenvalue spectrum of the matrix $(\bar v_k D v_i)$  
is also shown with green x symbol for the same boundary condition. 
}
\label{fig:eigenvalues-L=4-Q=0E=R-apbc}
\end{figure}

\begin{figure}[htb]
\begin{center}
\includegraphics[width =75mm]{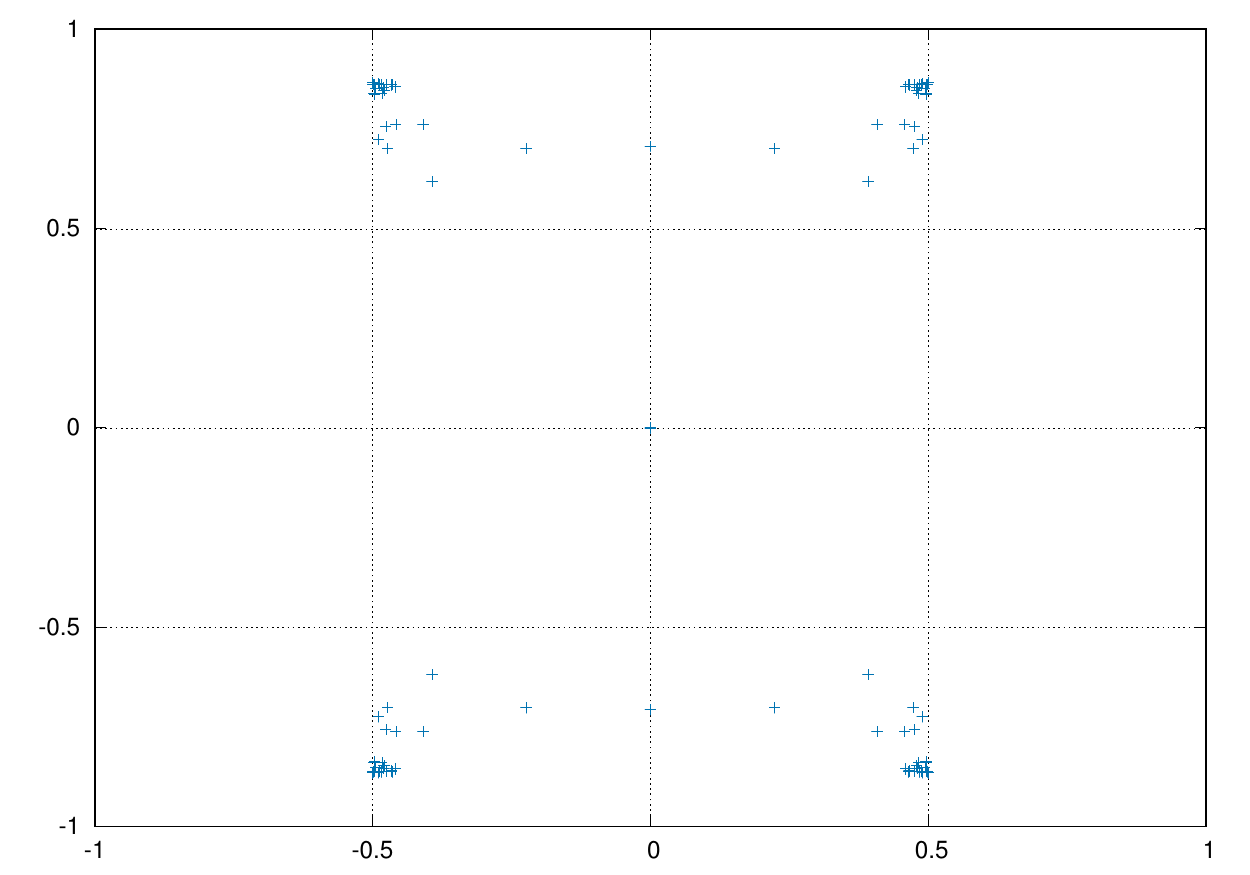} 
\end{center}
\caption{The eigenvalue spectrum
of the matrix
$ \big( u^\dagger \Gamma^{10} \Gamma^{a} E^{a} u \big) $ 
with a spin-field configuration in the class $E_\ast^a(x)$
for the case of the trivial link field. 
The lattice size is $L=4$ and
the boundary condition for the fermion field is periodic.}
\label{fig:eigenvalues-L=4-Q=0E=E*}
\end{figure}

As to the relatively small eigenvalues observed 
for the trivial link field and randomly generated spin-field configurations
with the periodic boundary condition,
they are attributed
to the zero modes with $p_\mu =0 $ 
in eqs.~(\ref{eq:chiral-basis-u-free}) and (\ref{eq:chiral-basis-u-planewave}) and their mixing-partners.
This is because the 
number of these small eigenvalues always counts to 64 ($= 32 \times 2$) and such small eigenvalues do not appear with the anti-periodic boundary condition (for up to $L=4$) as shown in fig.~\ref{fig:eigenvalues-L=4-Q=0E=R-apbc}.
The non-zero components of the zero modes' vectors are right-handed as $(\chi_\sigma, 0)^T$ $(\sigma=1,2)$,
while the never-vanishing components of the other modes' vectors  are left-handed.
The relevant matrix elements of 
$( u^\dagger \Gamma^{10} \Gamma^{a} E^{a} u )$ for the mixing
then read
\begin{equation}
- \chi_\sigma^T c(p') \chi_{\sigma'}
\delta_{0,p'+ k} \Gamma^{10}\Gamma^a \tilde E^a(k) /V\sqrt{2 \omega(p') ( \omega(p')+ b(p') )},
\end{equation}
where $ \tilde E^a(k)$ is the fourier-components of $E^a(x)$  defined by
$\tilde E^a(k) \equiv \sum_{x} \, {\rm e}^{-i k x} \,E^a(x)  $ with the constraints, $\sum_{k_\mu} \tilde E^a(k)^\ast \tilde E^a(k) = V^2$ and
$\sum_{k_\mu} \tilde E^a(k)^\ast \tilde E^a(k+ p) = 0$ $(p \not = 0)$.
%
Therefore the zero modes mix with a linear-combination of the modes $\{ p'_\mu ,\sigma' \}$ for which
$\chi_\sigma^T c(p') \chi_{\sigma'} \delta_{p'+ k, 0} \tilde E^a(k) \not = 0$ ,
but they decouple completely
from the modes with the
momenta $p'_\mu = \pi_\mu^{(A)}$ $(A=1,\cdots,15)$ 
where
$\pi^{(1)} \equiv (\pi, 0, 0, 0), 
\pi^{(2)} \equiv (0,\pi,0,0), \cdots, \pi^{(15)} \equiv(\pi,\pi,\pi,\pi)$.
This implies that 
the mixing of the zero modes is completely
suppressed 
for the following class of the spin configurations,
\begin{eqnarray}
E_\ast^a(x) &=& \frac{1}{V} \sum_{A} \, 
\cos(\pi^{(A)} x )\, 
\tilde E^a(\pi^{(A)}), \\
&& \sum_{A} \tilde E^a(\pi^{(A)})\tilde E^a(\pi^{(A)}) = V^2 , \\
&& \sum_{A \not = B} \tilde E^a(\pi^{(A)}) \tilde E^a(\pi^{(A)}+ \pi^{(B)}) = 0 \qquad (B =1, \cdots, 15 ).
\end{eqnarray}
In this case, zero eigenvalues appear and the multiplicity of the zero eigenvalues is at least 64. 
This explains why the relatively small eigenvalues appear
for randomly generated spin-field configurations
with the periodic boundary condition.
One can verify numerically the appearance of zero eigenvalues for $E_\ast^a(x)$.
The example of the eigenvalue spectra of
$ \big( u^\dagger \Gamma^{10} \Gamma^{a} E_\ast^{a} u \big) $
are shown in fig.~\ref{fig:eigenvalues-L=4-Q=0E=E*} for 
$L=4$ with the periodic boundary condition.

Based on 
the analytical results in the previous subsection
and the above numerical observations, 
we assume that the half product
${\prod_{i=1}^{n/4-4Q}} \lambda_i$ 
stays real 
independently of the spin field $E^a(x)$.
Then
we can argue that the pfaffian 
is positive semi-definite for any spin-field configuration $E^a(x)$
in the weak gauge-coupling limit: 
\begin{eqnarray}
{\rm pf} (u^T \, i\gamma_5 C_D  {\rm T}^a E^a u) 
=
{\prod_{i=1}^{n/4-4Q}} \lambda_i 
\, \ge 0  \qquad ( Q=0 \, ; \, g \rightarrow 0 ) .
\end{eqnarray}
We first note, as we discussed before,  that
the space of the SO(10)-vector spin field configurations, 
which we denote with ${\cal V}_E$, 
is the direct product of 
multiple $S^9$
and is pathwise connected.
Then any configuration of the spin field $E^a(x)$ can be reached
from the constant configuration 
$E_0^a(x) = \delta^{a, 10}$ 
through a continuous deformation.
Since 
it is unity for the constant configuration, 
the half product
${\prod_{i=1}^{n/4-4Q}} \lambda_i$
should be positive for a given configuration $E^a(x)$
as long as
there exists a path 
to $E^a(x)$ 
from $E_0^a(x) (= \delta^{a, 10})$ 
such that
the half product never vanish along the path.
%
On the other hand, 
for the spin configurations with which 
the half product is zero, 
a certain subset in the eigenvalue spectrum
of $(u^\dagger \, i \Gamma^{10} \Gamma^a E^a u)$
are zero.
Along the path which goes though 
such a spin configuration, 
the eigenvalue spectrum 
flow 
and the subset of would-be zeros pass the origin in the complex plane.
Then 
the half product
can change discontinuously
in its signature(phase).
Since the signature(phase) 
stays constant as far as the half product is nonzero,
this could happen if and only if
the subspace of the configurations with the vanishing determinant, 
which we denote with ${\cal V}_{E}^0$,
can divide
the entire space of the spin configurations ${\cal V}_E$ 
into the subspaces 
which are disconnected each other.
And the divided disconnected subspaces of ${\cal V}_E \, \backslash \, {\cal V}_{E}^0$ should be classified by the 
values of the signature(phase) of the half product.
%
In this respect, however, one notes that 
$\pi_k(S^9) = 0 \, (k < 9)$ and 
any topological obstructions and
the associated topological terms 
are not known in the continuum limit
for the SO(10)-vector spin field $E^a(x)$ 
on the four-dimensional spacetime $S^4$ or $T^4$.
In particular, 
any topologically non-trivial configurations/defects of the SO(10)-vector spin field and the associated
fermionic massless excitations are not known
in the continuum limit.
Then it seems reasonable to assume that 
${\cal V}_{E}^0$ consists of lattice artifacts and 
in particular it is given solely by the subspace of the configurations $E^a_\ast(x)$, which we denote with ${\cal V}_{E}^\ast$. If one assumes that ${\cal V}_{E}^0 = {\cal V}_{E}^\ast$, 
the multiplicity of the zero eigenvalues are 64
and the would-be zero eigenvalues have the approximate
structure 
$\{ (\lambda_i, \lambda_i, \lambda_i^\ast, \lambda_i^\ast) \, \vert \,  i=1,\cdots, 16 \}$.
Then
the signature(phase) of the half product 
${\prod_{i=1}^{n/4-4Q}} \lambda_i$
does not change
in passing ${\cal V}_{E}^0  ( = {\cal V}_{E}^\ast)$.
Therefore 
the pfaffian 
${\rm pf}( u^{\rm T}\, i \gamma_5 C_D {\rm T}^a E^a u )$ 
is positive semi-definite.

It then follows that the
path-integration of the pfaffian 
is real and positive in the weak gauge-coupling limit:
\begin{eqnarray}
\big\langle
1
\big\rangle_{E}
&=&
\int {\cal D}[E^a] \,
\det (u^T \, i\gamma_5 C_D \check{\rm T}^a E^a u) 
 \, \, > \,\,  0 \qquad (Q=0 \, ; \,g \rightarrow 0) .
\end{eqnarray}

\subsection{The case of representative SU(2) link fields 
of topologically non-trivial sectors}

As for the case of the $SU(2)$ link fields with non-zero 
topological charges $Q \not = 0$, 
we take
the following link field which gives the topological charge $Q = 2 \, m_{01} m_{23}$ ($m_{01}, m_{23} \in \mathbb{Z}$) \cite{Smit:1986fn,GonzalezArroyo:1997uj,Hamanaka:2002ha}:
\begin{eqnarray}
U(x,\mu) = {\rm e}^{i \theta_{12}(x,\mu) \Sigma^{12}},
\end{eqnarray}
where
\begin{eqnarray}
\theta_{12}(x,0) &=& 
 \left\{
\begin{array}{cc}
0 & (x_0 < L-1)  \\ -   F_{01}  L x_1 & (x_0 = L-1)
\end{array}
\right. , 
\qquad\quad\,\,\,
\theta_{12}(x,1) =   F_{01} \, x_0 , \\
\theta_{12}(x,3) &=& 
 \left\{
\begin{array}{cc}
0 & (x_2 < L-1)\\ -  F_{23} L x_3 & (x_2 = L-1)
\end{array}
\right. , 
\qquad\quad\,\,\,
\theta_{12}(x,4) =  F_{23} \, x_2 , 
\end{eqnarray}
and
\begin{eqnarray}
&& F_{01} = \frac{4 \pi m_{01}}{L^2}, \qquad F_{23} = \frac{ 4\pi m_{23}}{L^2}.
\end{eqnarray}
With this link field, the hermitian Wilson-Dirac operator
is diagonalized numerically and the normalized eigenvectors 
with the negative eigenvalues are computed to form the chiral basis $\{ u_j(x) \}$. We checked that the number of the eigenvectors is $n/2 - 8 Q$ for $L=3,4$ with the periodic b.c. and is consistent with the index theorem.

For the constant configuration of the spin field, 
$E^a_0(x) = \delta^{a,10}$, 
$( u^{\rm T}\, i \gamma_5 C_D {\rm T}^a E^a_0 u )_{jk} \, 
(={\cal C}_{jk})$ is a unitary matrix and $\tilde \lambda_i$'s are pure complex phases, while 
$\big( u^\dagger \Gamma^{10}\Gamma^{a} E^{a}_0 u \big)_{jk}$
remains the unit matrix
and $\lambda_i = + 1$.
Then ${\rm pf} ( u^{\rm T}\, i \gamma_5 C_D {\rm T}^a E^a_0 u )$
is a pure complex phase.

\begin{figure}[tbh]
\includegraphics[width =75mm]{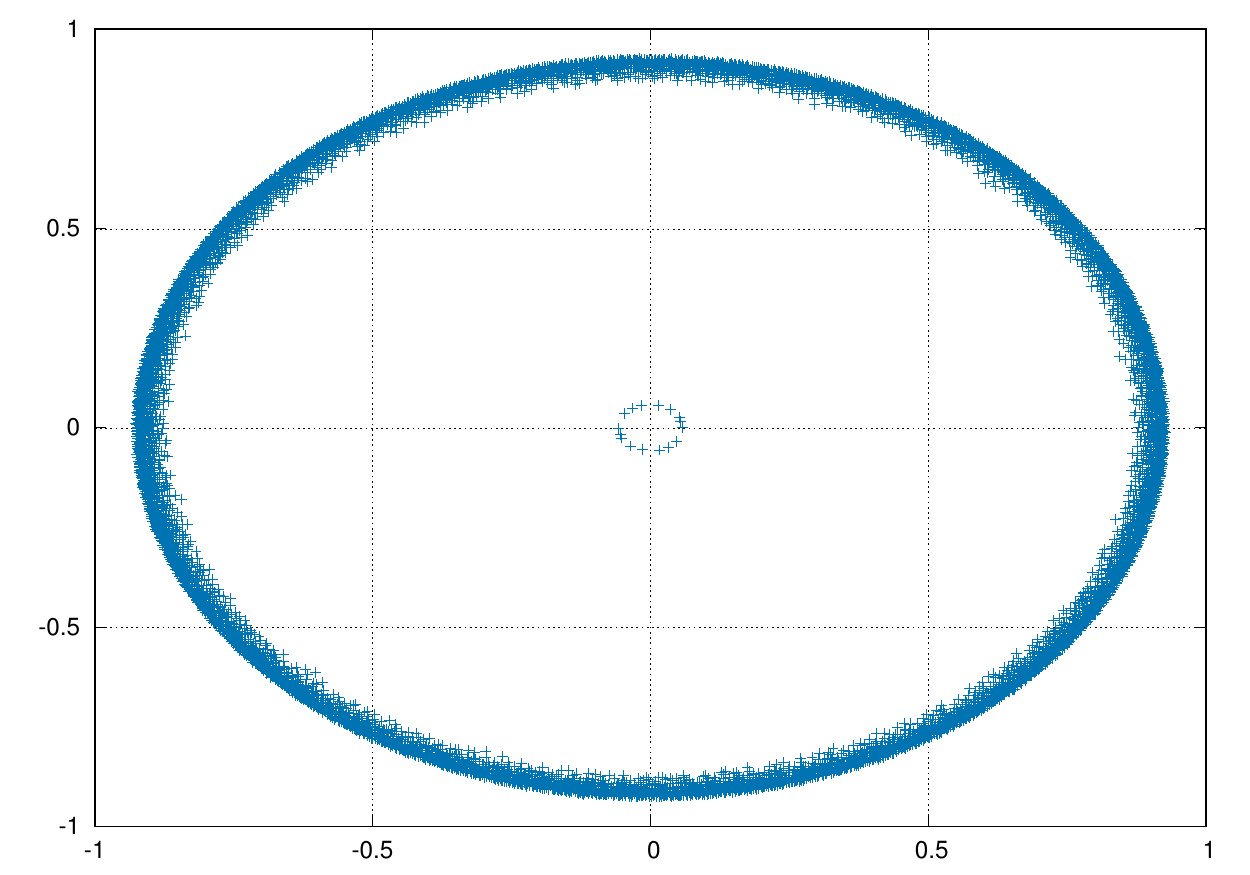} 
\includegraphics[width =75mm]{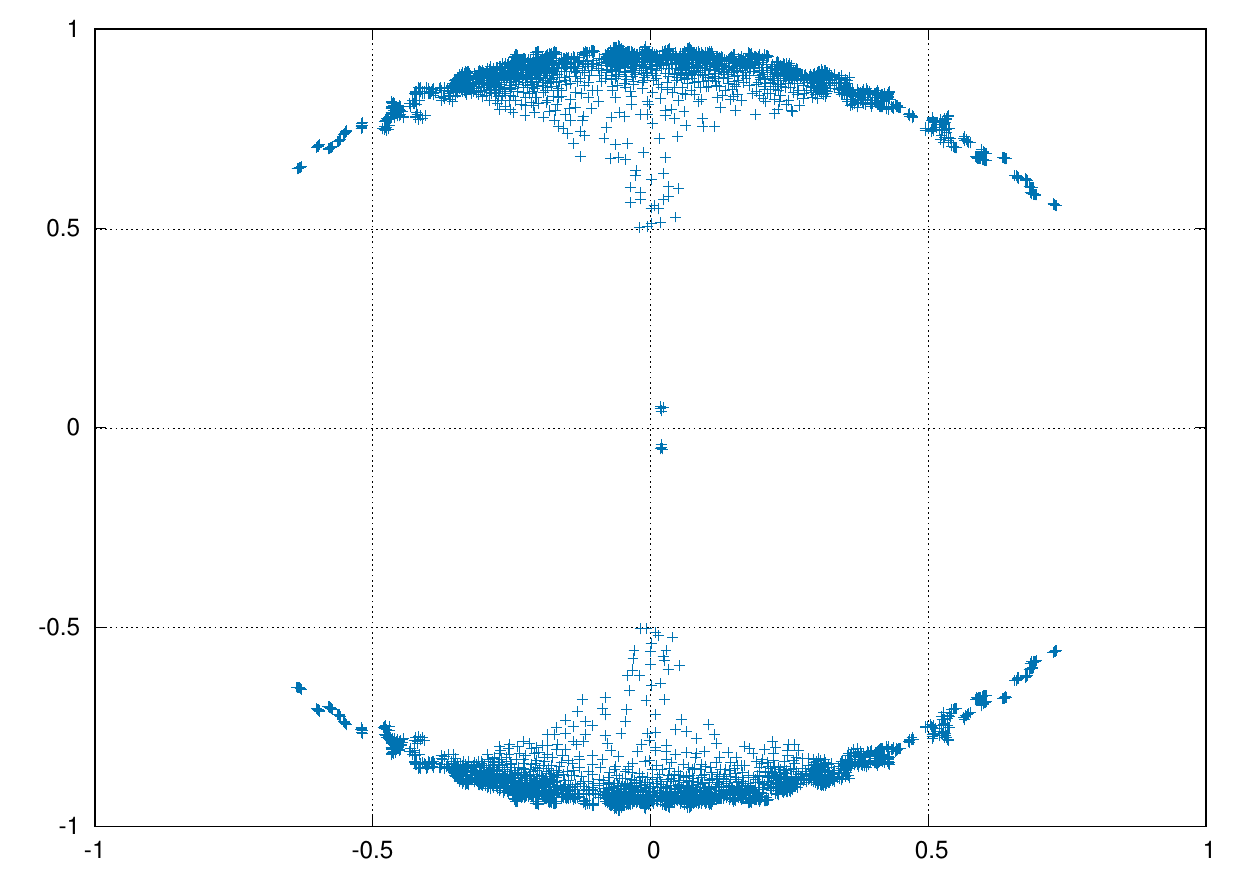} 
\caption{The eigenvalue spectra 
of the matrices
$ \big(u^{\rm T} \,  i \gamma_5 C_D {\rm T}^a E^a u \big) $ and
$ \big( u^\dagger \Gamma^{10} \Gamma^{a} E^{a} u \big) $ 
with a randomly generated spin-field configuration
for the case of the representative SU(2) link field
of the topological sector with $Q=-2$. 
The lattice size is $L=4$ and
the boundary condition for the fermion field is periodic.}
\label{fig:eigenvalues-L=4-Q=2E=R}
\end{figure}

For randomly generated spin-field configurations
with the lattice sizes up to $L=4$, 
we again checked that the eigenvalue spectra of the matrices 
$ \big(u^{\rm T} \,  i \gamma_5 C_D {\rm T}^a E^a u \big) $ and
$ \big( u^\dagger \Gamma^{10} \Gamma^{a} E^{a} u \big) $ 
have the structures of 
$ \{ (\tilde \lambda_i, - \tilde \lambda_i) \, \vert \,  i=1,\cdots, n/4-4Q\}$
and
$\{ (\lambda_i, \lambda_i) \, \vert \, i=1,\cdots, n/4-4Q \}$,
respectively.
All eigenvalues turn out to be non-zero.
But there appear again relatively small eigenvalues for $Q < 0$, the number of those counts to $\vert -8Q \vert$. 
We found that the complex phase of 
${\rm pf}( u^{\rm T}\, i \gamma_5 C_D {\rm T}^a E^a u )$ 
stays constant and equal to that of ${\rm pf} ( u^{\rm T}\, i \gamma_5 C_D {\rm T}^a E^a_0 u )$, while
the half product
${\prod_{i=1}^{n/4-4Q}} \lambda_i$
stays real-positive.
The typical examples of the eigenvalue spectra
are shown in fig.~\ref{fig:eigenvalues-L=4-Q=2E=R} for 
$Q=-2$ and $L=4$ with the periodic boundary condition.

In this case, the relatively small eigenvalues 
can be attributed to the chiral zero modes due to the topologically non-trivial link field.
This is because the number of these small eigenvalues always counts to $\vert -8Q \vert$ consistently with the index theorem.

Based on 
the analytical results in the previous subsection
and
the above numerical observations, 
we again assume that the half product
${\prod_{i=1}^{n/4-4Q}} \lambda_i$ 
stays real 
independently of the spin field $E^a(x)$.
In this case 
the multiplicity of the zero eigenvalues should be $\vert -8Q \vert$
and 
the would-be zero eigenvalues should 
have the approximate structure 
$\{ (\lambda_i, \lambda_i, \lambda_i^\ast, \lambda_i^\ast) \, \vert \,  i=1,\cdots, \vert -2 Q \vert \}$, 
and then
the signature(phase) of the half product 
does not change.
Therefore 
the complex phase of the pfaffian 
is stationary.
%
It then follows that the
path-integration of the pfaffian 
is a non-vanishing complex number
for the representative $SU(2)$ link fields 
of topologically non-trivial sectors  $\mathfrak{U}[Q]$:
\begin{eqnarray}
\big\langle
1
\big\rangle_{E}
&=&
\int {\cal D}[E^a] \,
\det (u^T \, i\gamma_5 C_D \check{\rm T}^a E^a u) 
\nonumber\\
&=&
c[U] \,\, ( \not = 0 ) 
\qquad\quad
( \, U(x,\mu)= {\rm e}^{i \theta_{12}(x,\mu) \Sigma^{12}} \in \mathfrak{U}[Q] \, ; \, g \rightarrow 0) .
\end{eqnarray}

\subsection{Continuity across the mass singularity: $m_0 \rightarrow +0 ; +0 \rightarrow -0 ; -0 \rightarrow - \infty$}

\begin{figure}[tbh]
\begin{center}
\includegraphics[width =70mm]{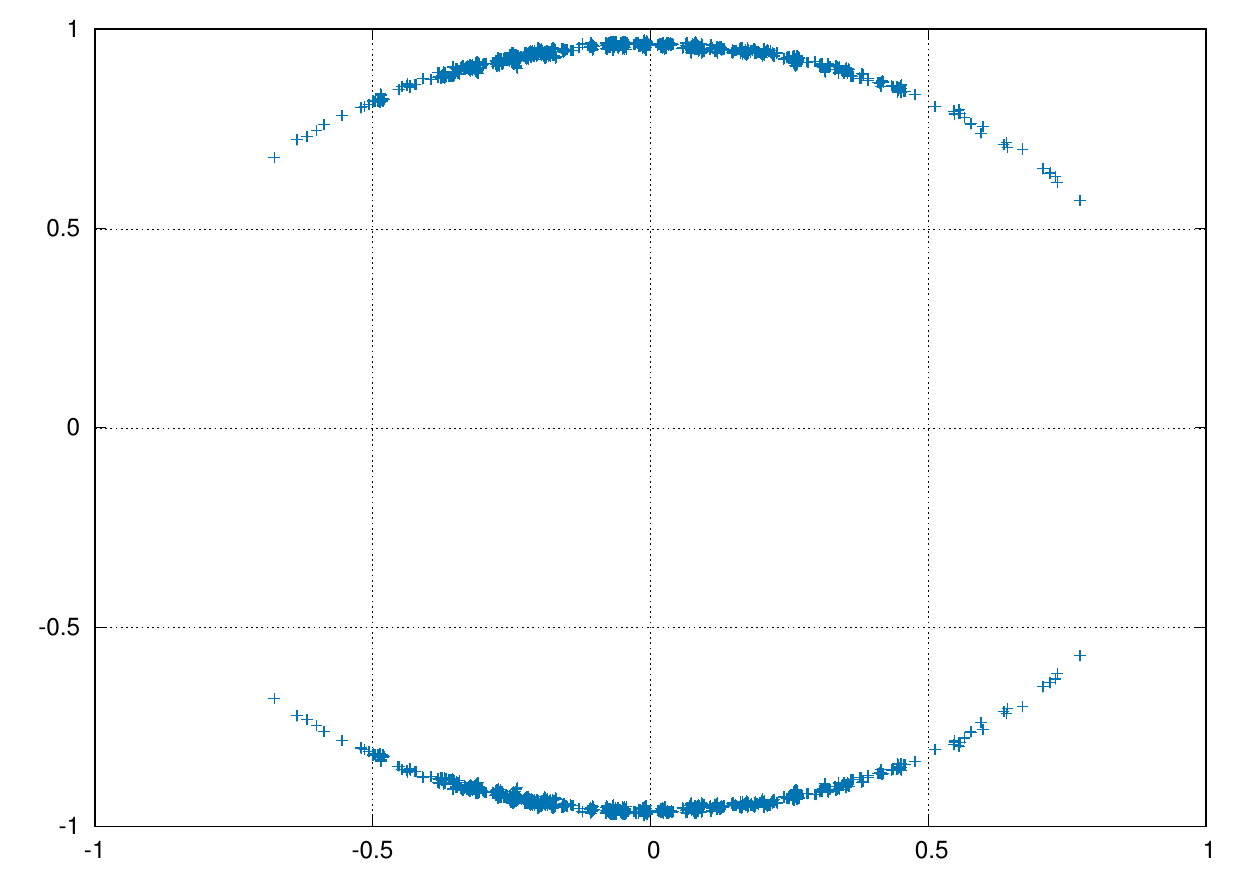} 
\hspace{2em}
\includegraphics[width =70mm]{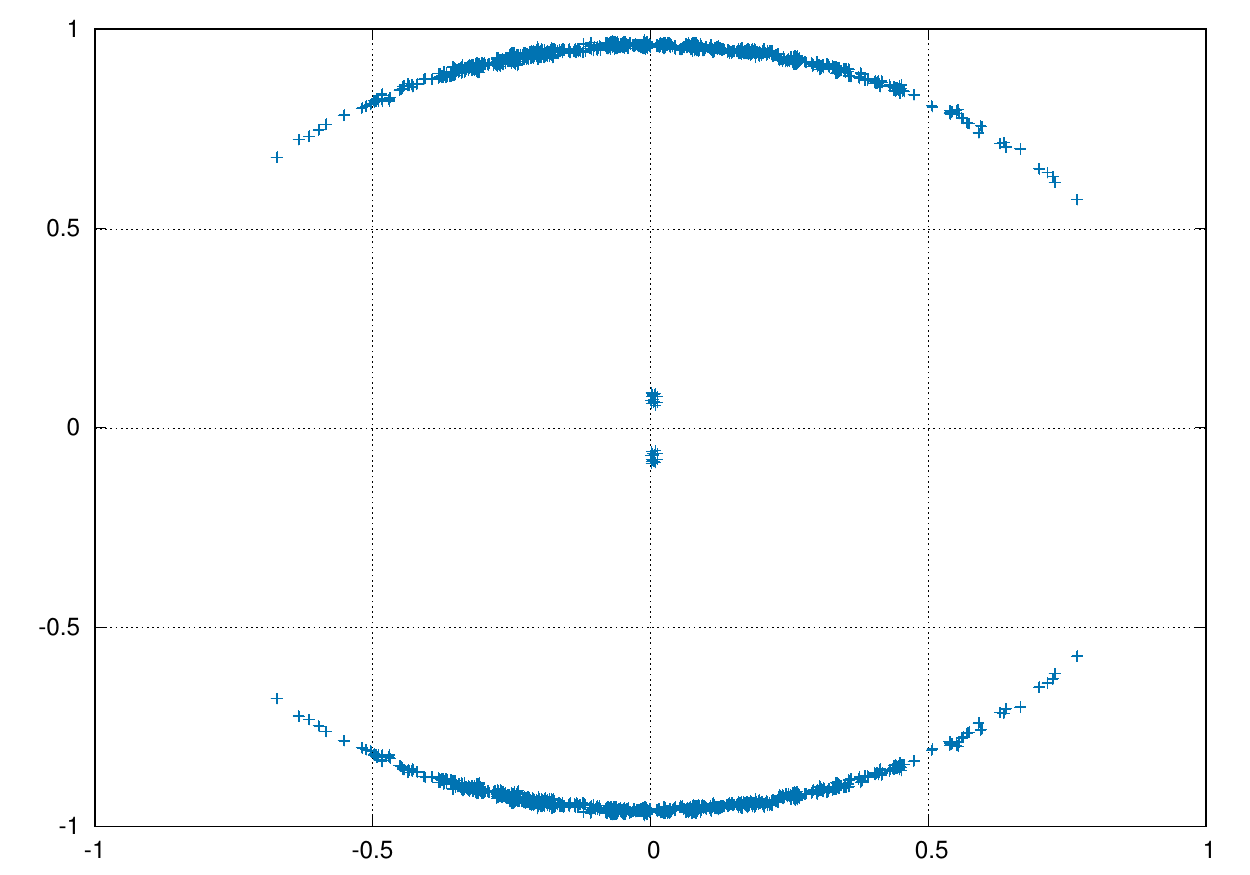} 
\end{center}

\begin{center}
\includegraphics[width =45mm]{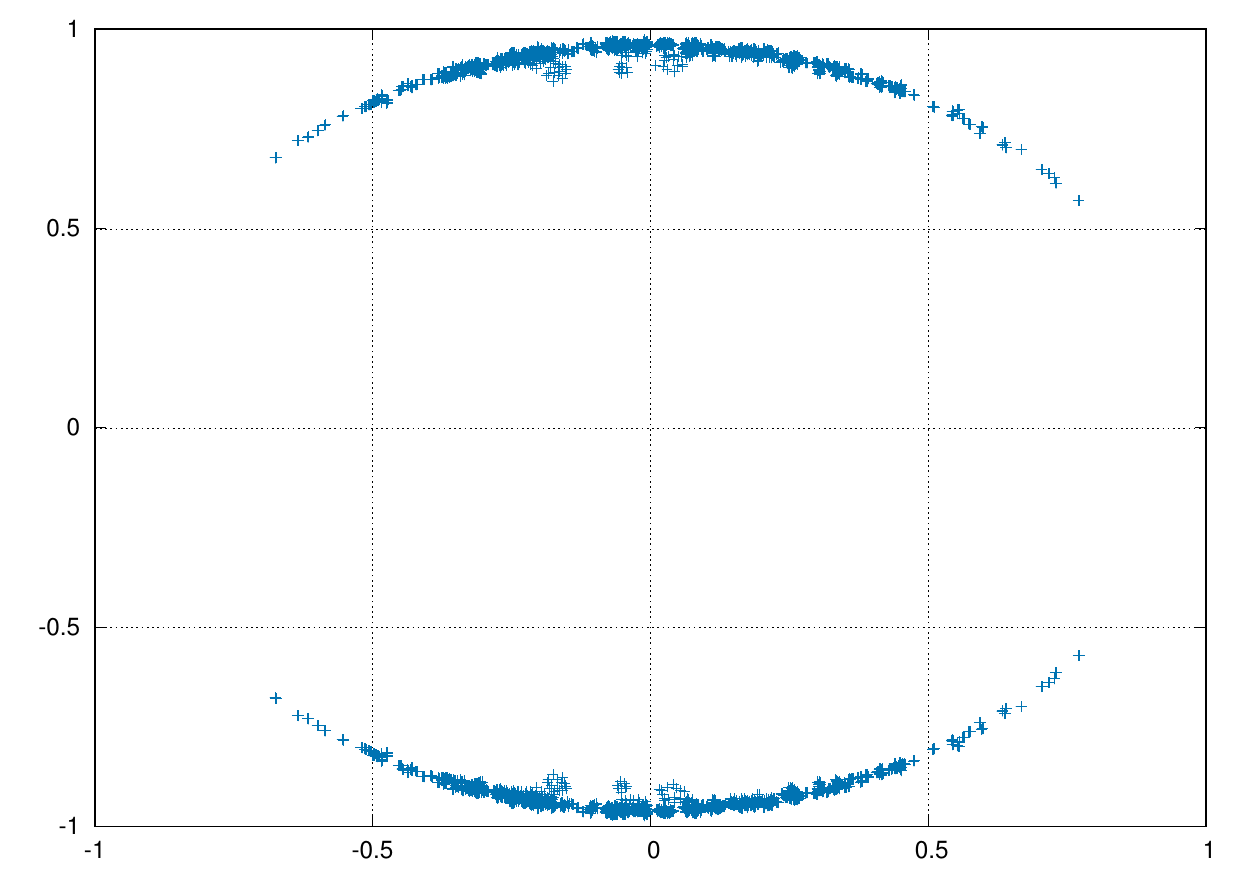} 
\includegraphics[width =45mm]{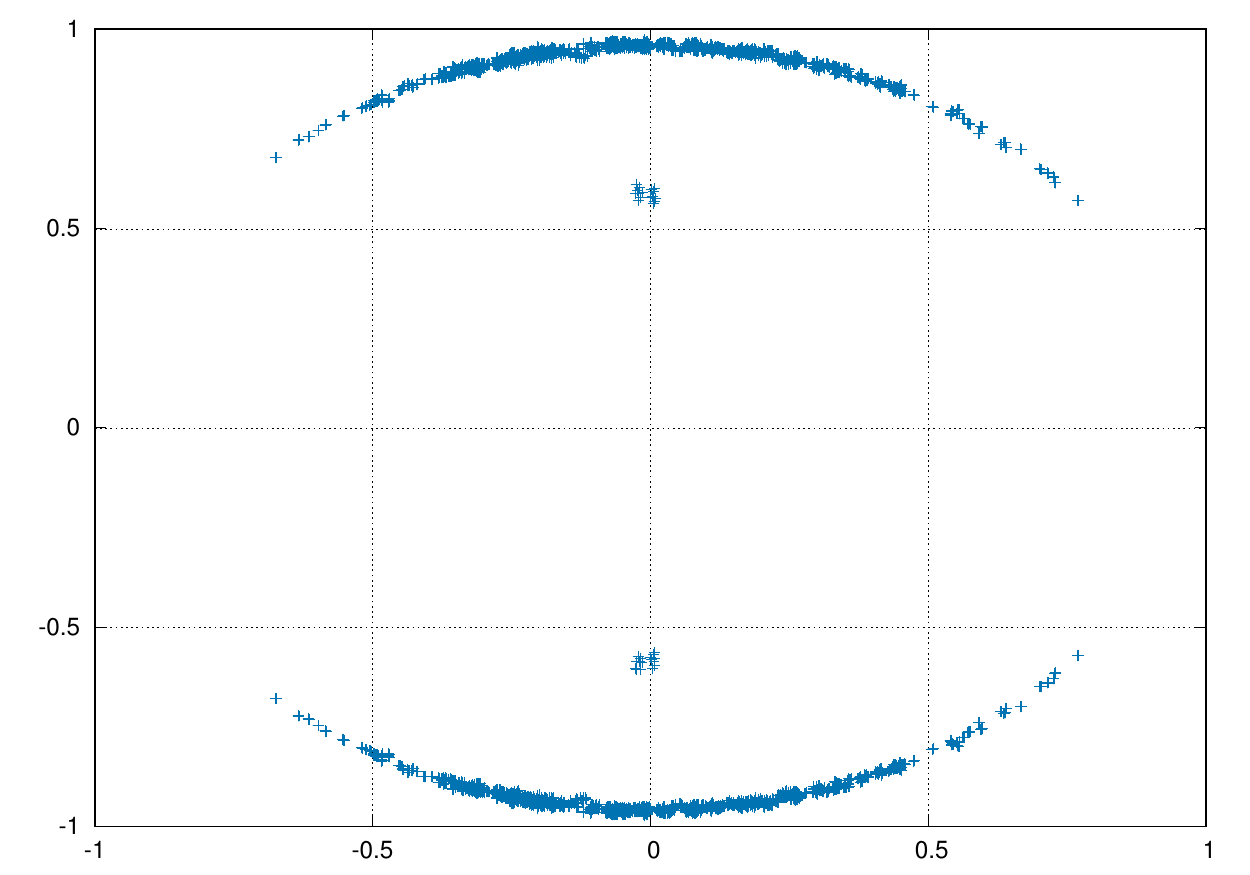} 
\includegraphics[width =45mm]{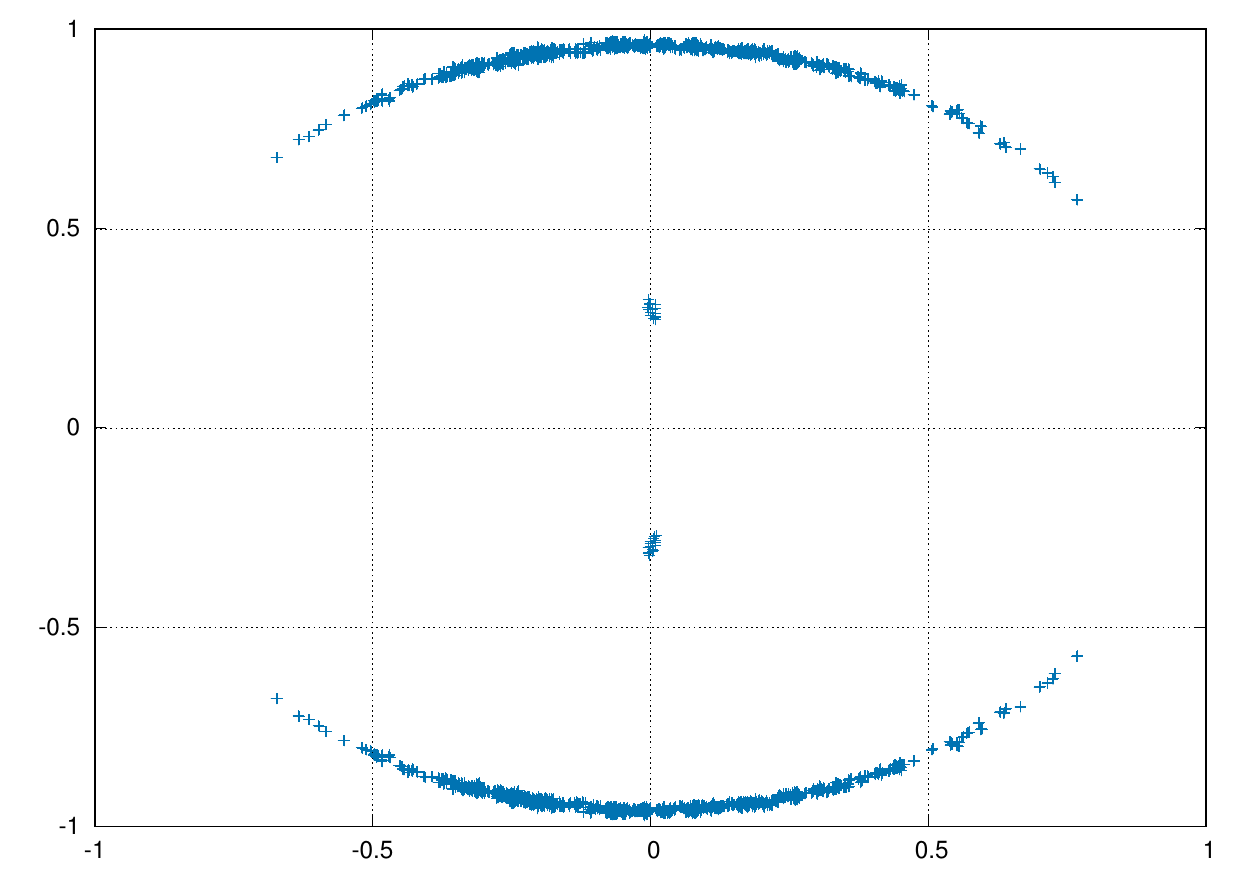} 
\end{center}

\caption{
The eigenvalue spectra of 
$ \big( u^\dagger \Gamma^{10} \Gamma^{a} E^{a} u \big) $
in the limit $m_0 \rightarrow \mp 0$ 
with a randomly generated spin configuration for the
trivial link field.
The interpolation parameter $\theta_\alpha$ is chosen as
$\theta_\alpha = 0, 3\pi/12, 4\pi/12, 5\pi/12, \pi/2$
for the top-left, bottom-left, bottom-middle, bottom-right,
top-right figures, respectively.
The lattice size is $L=4$ and
the boundary condition for the fermion field is periodic.
}
\label{fig:eigenvalues-UTEU-L=4-Q=0E=R-massless}
\end{figure}

\begin{figure}[bth]
\begin{center}
\includegraphics[width =70mm]{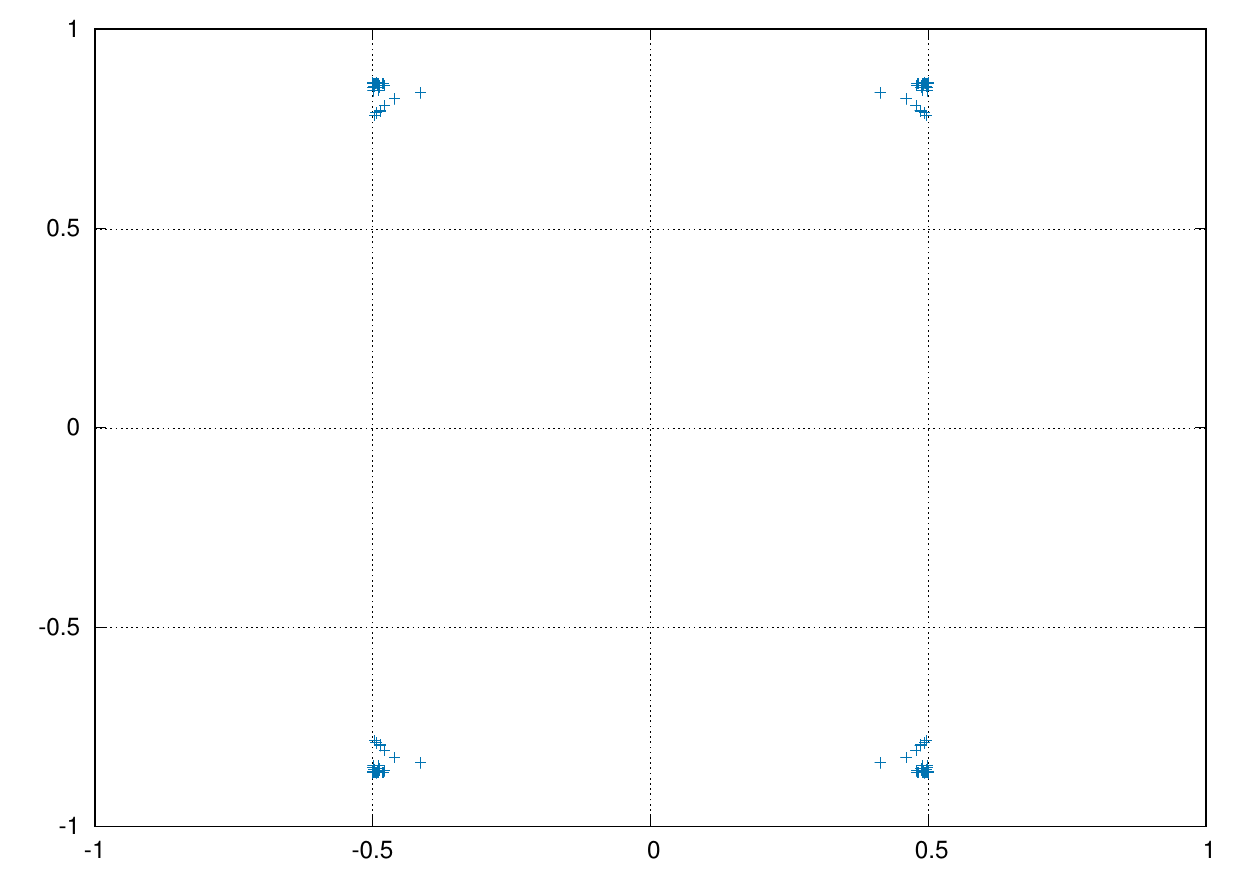} 
\hspace{2em}
\includegraphics[width =70mm]{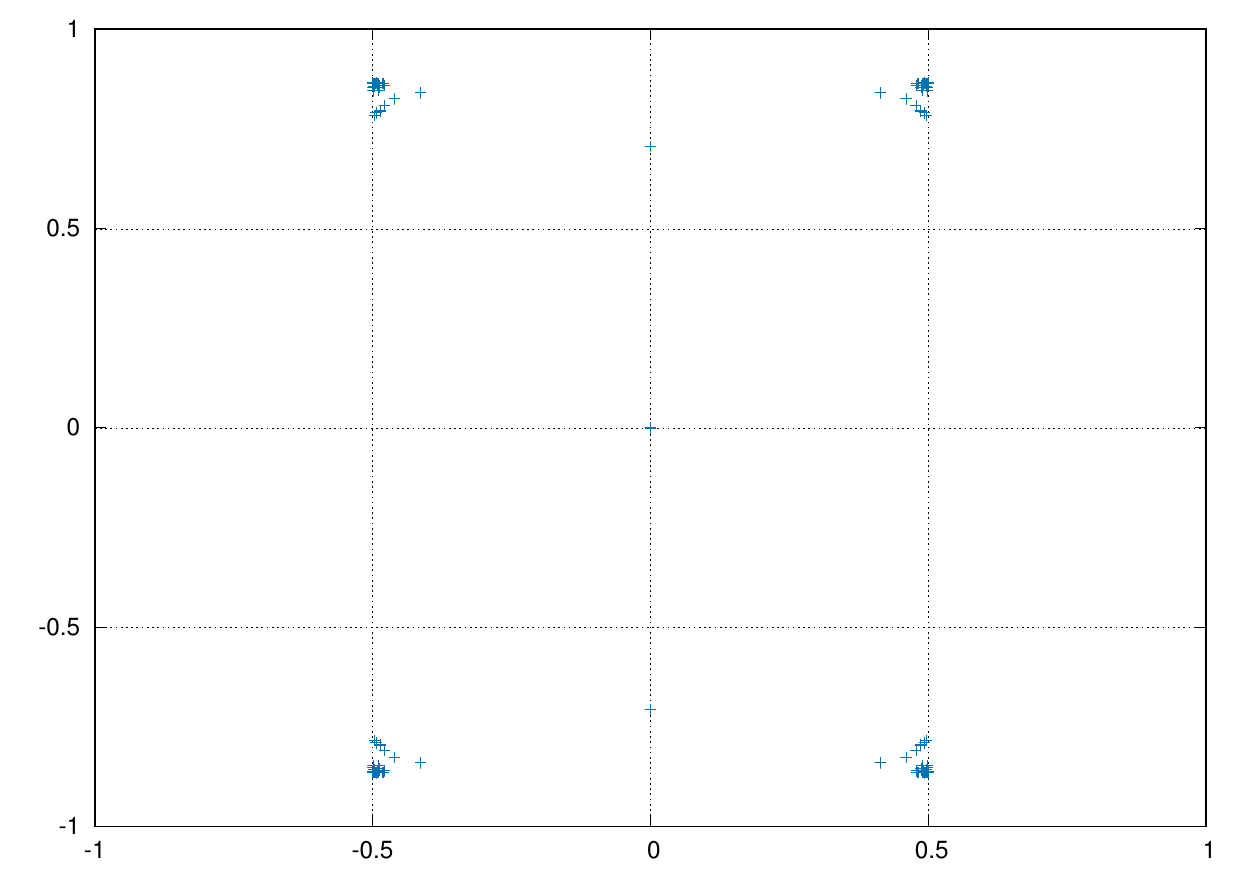} 
\end{center}

\begin{center}
\includegraphics[width =45mm]{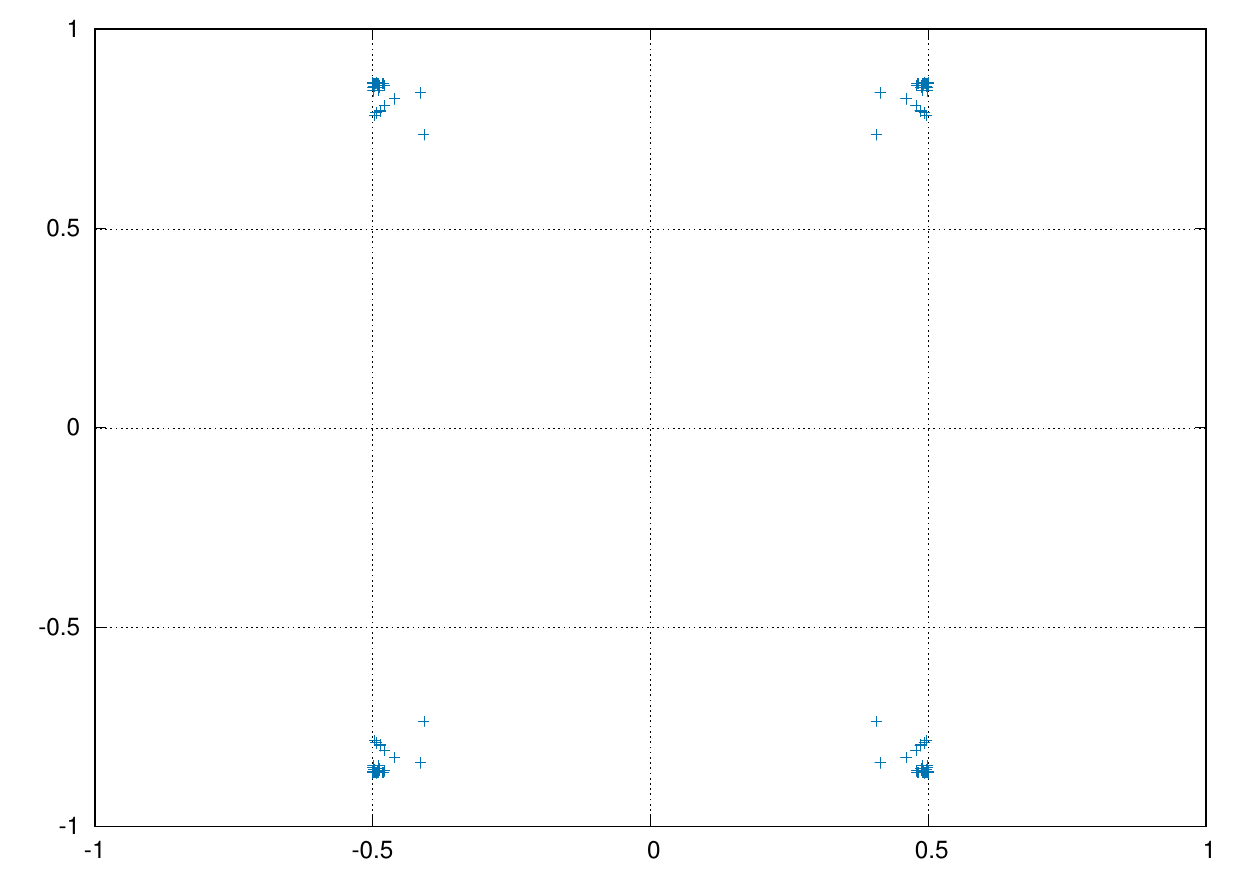} 
\includegraphics[width =45mm]{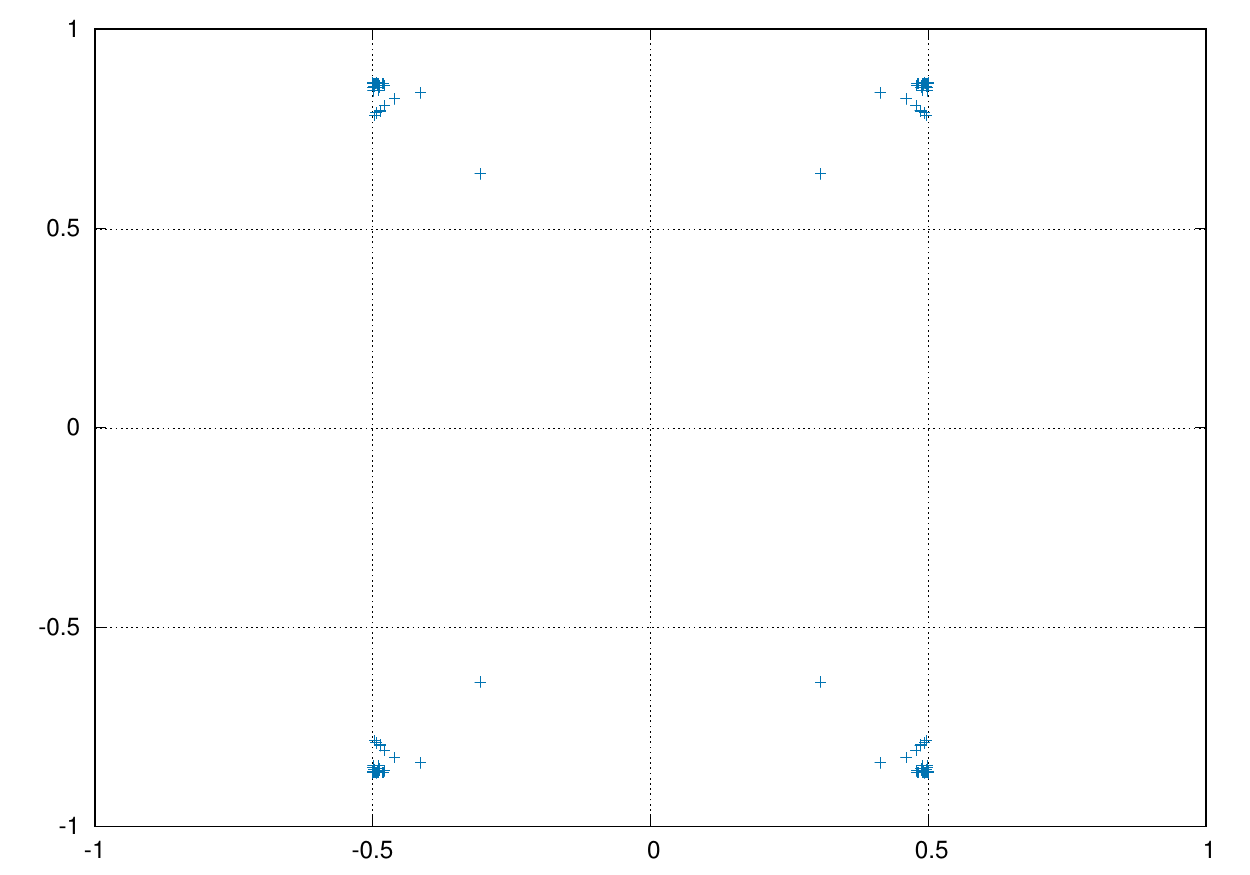} 
\includegraphics[width =45mm]{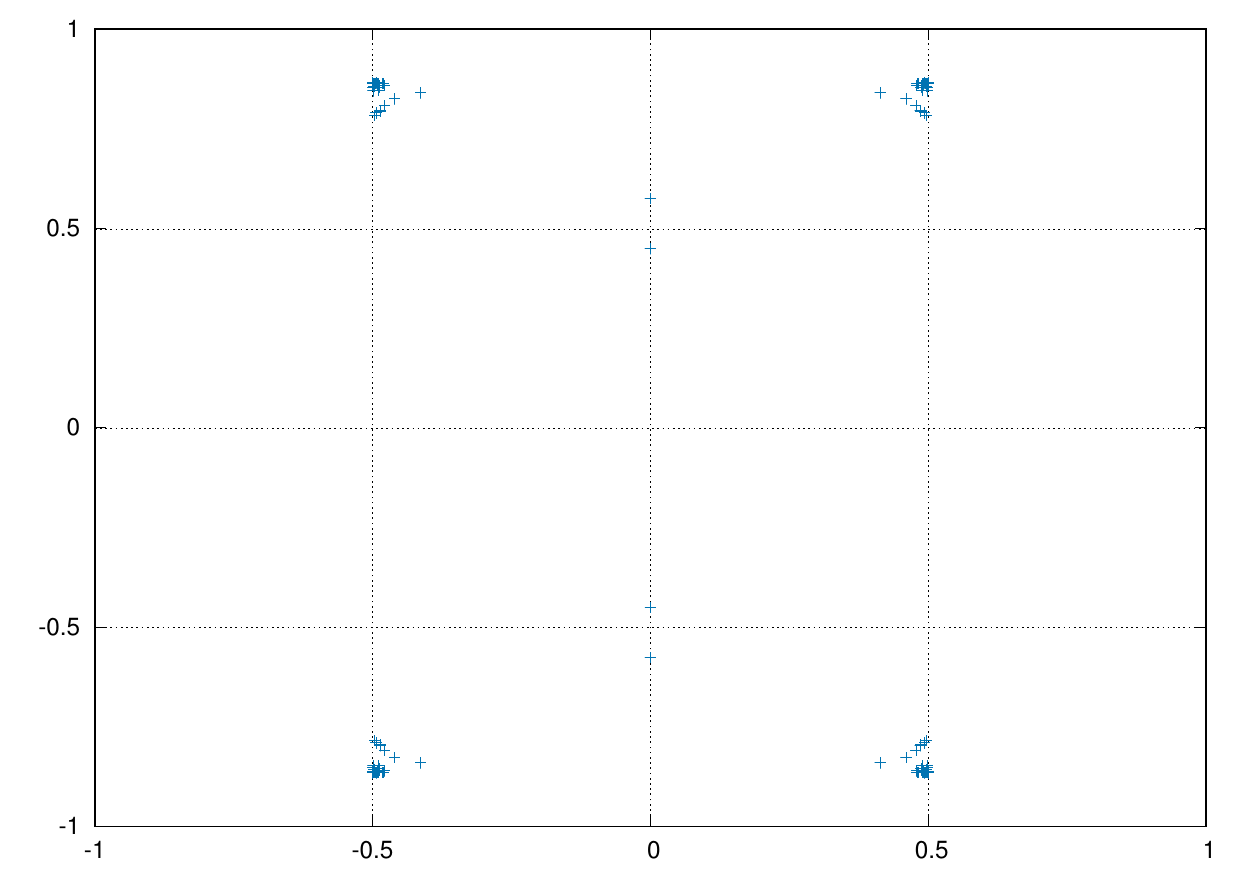} 
\end{center}

\caption{The eigenvalue spectra of 
$ \big( u^\dagger \Gamma^{10} \Gamma^{a} E^{a} u \big) $
in the limit $m_0 \rightarrow \mp 0$ 
with a spin field configuration in the class $E^a_\ast(x)$
for the trivial link field. 
The interpolation parameter $\theta_\alpha$ is chosen as
$\theta_\alpha = 0, 3\pi/12, 4\pi/12, 5\pi/12, \pi/2$
for the top-left, bottom-left, bottom-middle, bottom-right,
top-right figures, respectively.
The lattice size is $L=4$ and
the boundary condition for the fermion field is periodic.}
\label{fig:eigenvalues-UTEU-L=4-Q=0E=E9876equal-massless}
\end{figure}

Another support for the above arguments is followed from the consideration on the continuity in the mass parameter $m_0$
from 
the negative region, $ m_0 < 0$
to
the positive region, $ 0 < m_0 < 2 $.
We note first that 
the chiral basis for the negative mass $ m_0 < 0 $
can be defined by the same formula as
for the positive mass $ 0 < m_0 < 2 $ given by 
eqs.~(\ref{eq:chiral-basis-u-free}) 
and (\ref{eq:chiral-basis-u-planewave}),
except that the vectors of zero modes with $p_\mu = 0$ 
should be flipped in chirality to the left-handed ones 
$(0, \chi_\sigma)^T$ $(\sigma=1,2)$
from the right-handed ones
$(\chi_\sigma, 0)^T$ $(\sigma=1,2)$.
For this case, one can take the limit $m_0 \rightarrow -\infty$
to obtain the trivial basis vectors\footnote{Note that we define
the Wilson Dirac operator $X$ in eq.~(\ref{eq:def-overlap-D-X}) with the negative
signature in front of the mass parameter $m_0$.
The above fact is related to the fact that
one can send to the infinity
the mass parameter $+m_0$ of 
the positive mass region in Kaplan's Domain-wall fermion\cite{Kaplan:1992bt}.
And this is why and how Kaplan's domain-wall fermion (with periodic b.c.) is simplified to Shamir's boundary fermion/vectorlike domain-wall fermion\cite{Shamir:1993zy,Furman:ky}.
} as
\begin{eqnarray}
\label{eq:chiral-basis-u-free-negativemass-infty}
u_j(x) = \frac{1}{\sqrt{L^4}} { \rm e}^{ i p x } \, 
\left( \begin{array}{c} 0 \\ \chi_\sigma \end{array} \right)
 \, \delta_{s,t}    \qquad   ( j=\{ p, \sigma, t \} ) ,
\end{eqnarray}
and one can show that the pfaffian is unity
independent of the spin field $E^a(x)$, 
just like the case of the anti-field $\bar \psi_+(x)$.
We note next that
in the limits $m_0 \rightarrow \mp 0$, the both formula are well-defined as long as $L$ is finite. But, actually at $m_0 = \mp 0$,
the zero modes of both chiralities belong to the same
zero-eigenvalue-sector of the massless hermitian Wilson-Dirac operator $H_w = \gamma_5 D_w$ and degenerate. Then one can interpolate the two regions of $m_0 \rightarrow \mp 0$ smoothly 
by 
the one-parameter family of 
the basis vectors of the zero modes,
\begin{equation}
u_\alpha(0,\sigma)^{(\theta)} = \left( \begin{array}{c} \sin \theta \,  \chi_\sigma \\ \cos \theta \,  \chi_\sigma \end{array} \right) \qquad
\theta \in [ 0, \pi/2 ].
\end{equation}
The examples of the eigenvalue spectra of 
$ \big( u^\dagger \Gamma^{10} \Gamma^{a} E^{a} u \big) $
in the limit $m_0 \rightarrow \mp 0$ are shown
for randomly generated spin configurations
in fig.~\ref{fig:eigenvalues-UTEU-L=4-Q=0E=R-massless} 
and 
for $E^a_\ast(x)$
in fig.~\ref{fig:eigenvalues-UTEU-L=4-Q=0E=E9876equal-massless},
respectively 
both with $L=4$ and the periodic boundary condition.
In both cases,
the two spectra of the limit $m_0 \rightarrow \mp 0$
shown in the upper two panels are interpolated by varying the parameter $\theta$ from $0$ to $\pi/2$, as shown 
in the lower three panels. We found that
in the course of the interpolation, 
the ``half product'' 
${\prod_{i=1}^{n/4-4Q}} \lambda_i$
remains real and positive-definite
and it vanishes only at $\theta=\pi/2$ if $E^a(x)=E^a_\ast(x)$.
This result supports the picture that
the pfaffian for $ 0 < m_0 < 2$ can be zero, but is positive
semi-definite, while the pfaffian for $m_0 < 0$ is positive definite 
all the way down to the limit $m_0 \rightarrow - \infty$, 
where it is unity independently of the spin field $E^a(x)$.
And the positive semi-definite
pfaffian for $0 < m_0 < 2$ can be smoothly connected 
to the positive-definite one for $ m_0 < 0$ at $m_0 = \pm 0 $.
without any singularity.
And it implies that
the pfaffian integrals are both positive definite
for $0 \le m_0 < 2$ and $ m_0 \le 0$, and
there is no massless singularity at the limit $m_0 = \pm 0$.



\subsection{Disorder nature of the auxiliary 
spin-field path integrations} 
\label{subsec:disorder-nature-spin-field}

The path-integrations over the SO(10)-vector real spin fields with unit length, $E^a(x)$ and $\bar E^a(x)$, 
in eq.~(\ref{eq:def-weyl-measure-in-chiral-basis-pfaffian})
define two kind of 
the four-dimensional spin models with the 
partition functions,
\begin{eqnarray}
\label{eq:z-E}
\big\langle 1 \big\rangle_E 
&=& \int {\cal D}[E] \, 
{\rm pf}  \big(u^{\rm T} \,  i \gamma_5 C_D {\rm T}^a E^a u \big) ,
\\
\big\langle 1 \big\rangle_{\bar E}
&=& \int {\cal D}[\bar E] \,
{\rm pf} \big(  \bar u \,  i \gamma_5 C_D {{\rm T}^a}^\dagger \bar E^a\bar u^{\rm T} \big) .
\end{eqnarray}
It is important and useful to understand the dynamical nature
of the path-integrations in these spin models. 

The second model for $\bar E^a(x)$ is trivial.
This is because 
the pfaffian weight is unity,
${\rm pf} \big(  \bar u \,  i \gamma_5 C_D {{\rm T}^a}^\dagger \bar E^a\bar u^{\rm T} \big) =1$. Then
the two-point correlation function is given by
\begin{equation}
\big\langle \bar E^a(x) \bar E^b(y) \big\rangle_{\bar E} 
= \frac{1}{10} \delta_{xy} \delta^{ab} 
\big\langle 1 \big\rangle_{\bar E} 
\qquad ( \big\langle 1 \big\rangle_{\bar E} =1)
\end{equation}
and the spin field is completely disordered.

The first model for $E^a(x)$ is quite non-trivial. 
The pfaffian weight is the rather complicated (non-local) function of the spin field variables, which can be chosen to 
be real and positive semi-definite for the background link fields in the SO(9) subgroup, as we have argued, 
but is complex in general.
The mass parameter $m_0$ is the only parameter
to control the strength of the coupling of the spin field.\footnote{The kinetic term for the spin field $E^a(x)$ such as
$- K \sum_{x,\mu} E^a(x) E^a(x + \hat \mu) $ can be added for the analysis. We omit this term for simplicity.}
One may  
regard the number of the spin components $N(=10)$ as another parameter and consider the large $N$ method. 
In order to get insights into the dynamical nature of this spin model, one needs to apply the methods such as
Monte Carlo simulations and 
the saddle point analysis in the large $N$ expansion.\cite{zinn-justin, itzykson-drouffe}

In the following,
we apply 
the saddle point analysis in the spirit of the large $N$ expansion 
to the model with the trivial link field background (in the weak gauge-coupling limit).
For this purpose,
we introduce the unconstraint (linearized) field $X^a(x)$ and the Lagrange-multiplier field $\lambda(x)$ to impose
the constraint $X^a(x) X^a(x) =1$, and rewrite the
original path integration eq.~(\ref{eq:z-E}) as follows,
\begin{eqnarray}
\big\langle 1 \big\rangle_E 
&=& \int {\cal D}[X] {\cal D}[\lambda] \, 
{\rm pf}  \big(u^{\rm T} \,  i \gamma_5 C_D {\rm T}^a X^a u \big)
\,
{\rm e}^{i \sum_x \lambda(x)(X^a(x) X^a(x) - 1)} .
\end{eqnarray}
Then the field variables are decomposed into
the modes with zero-momentum and other modes of fluctuation as
\begin{eqnarray}
X^a(x) &=& X^a_0 + \tilde X^a(x), \qquad  \sum_x \tilde X^a(x) \, =0 , \\
\lambda(x) &=& \lambda_0 + \tilde \lambda(x), \,\,\quad\qquad \sum_x \tilde \lambda(x) =0 ,
\end{eqnarray}
and the action of $X^a(x)$ and $\lambda(x)$ is
expanded in terms of  $\tilde X^a(x)$ and $\tilde \lambda(x)$ upto the second order. The result reads
\begin{eqnarray}
S[ X^a, \lambda] 
&\equiv& 
- \frac{1}{2} 
\ln \det  \big(u^{\rm T} \,  i \gamma_5 C_D {\rm T}^a X^a u \big) - i \sum_x \lambda(x)(X^a(x) X^a(x) - 1) 
\\
&=&
\big\{ 
- 16 \ln( X^a_0 X^a_0 )^{1/2}  - i  \lambda_0 (X^a_0 X^a_0 - 1)
\big\} V 
-i \sum_x 2 \tilde \lambda(x)  X^a_0 \tilde X^a(x) 
\nonumber\\
&&
+ \sum_{x,y} \tilde X^a(x) \big\{
 4 D(x-y) ( 2 X^a_0 X^b_0 - X^c_0 X^c_0 \, \delta^{ab})/ 
(X^d_0 X^d_0)^2 
-i \lambda_0 \delta_{xy} \delta^{ab}\big\}  \tilde X^b(x) 
\nonumber\\
&& + \cdots ,
\end{eqnarray}
where $D(x-y)$ is the kinetic operator definend 
through the chiral projector
$\hat P_{0 + }(x,y)$ of the single free overlap Dirac fermion 
as
\begin{eqnarray}
D(x-y)
&=&
{\rm tr}[\hat P_{0 +}(x-y) \hat P_{0 +}(y-x)]
\\
&=&
\frac{1}{V} \sum_k {\rm e}^{i k x} \,
\frac{1}{V} \sum_q
\Big\{
1 + \frac{\sin(q+k)_\mu \sin q_\mu + b(q+k) b(q)}{\omega(q+k) \omega(q)}
\Big\} ,
\\
\hat P_{0 +}(x-y) &=&
\frac{1}{V} \sum_q  {\rm e}^{i q x}
\left\{ 
\frac{1}{2} - 
\frac{1}{2} 
\gamma_5 \frac{i \gamma_\mu \sin q_\mu + b(q)}{\omega(q)}
\right\}.
\end{eqnarray}
The path-integration of  $\tilde \lambda(x)$ and $\tilde X^a(x)$ (in this order) 
gives the effective action of $\tilde X^a_0$ and 
$\tilde \lambda_0$ as follows,
\begin{eqnarray}
\label{eq:effective-action-X0-lambda0}
S_{\rm eff}[ X^a_0, \lambda_0] 
&=& 
\big\{ 
- 16 \ln( X^a_0 X^a_0 )^{1/2}  - i  \lambda_0 (X^a_0 X^a_0 - 1)
\big\} V 
\nonumber\\
&& \qquad
+ \frac{(10-1)}{2} \sum_{ k \not = 0}\ln \big\{-4 \tilde D(k)/  X^a_0 X^a_0
-i \lambda_0  \big\} ,
\end{eqnarray}
where $\tilde D(k)$ is the fourier transform of the kinetic
oerator $D(x-y)$.

In eq.~(\ref{eq:effective-action-X0-lambda0}),
we observe that the first term scales as $2^{\frac{N}{2}} (= 32)$ in terms of $N(=10)$ , while the second term scales as $N$. Then, we can assume that $\lambda_0$ scales as $2^{\frac{N}{2}}/N$
so that the first two terms of the classical contribution
both scale as $2^{\frac{N}{2}}$.
The third term of the one-loop correction scales as $N$ and is suppressed by the factor $N / 2^{\frac{N}{2}} (\simeq 9/32)$.

\begin{figure}[t]
\begin{center}
\includegraphics[width =75mm]{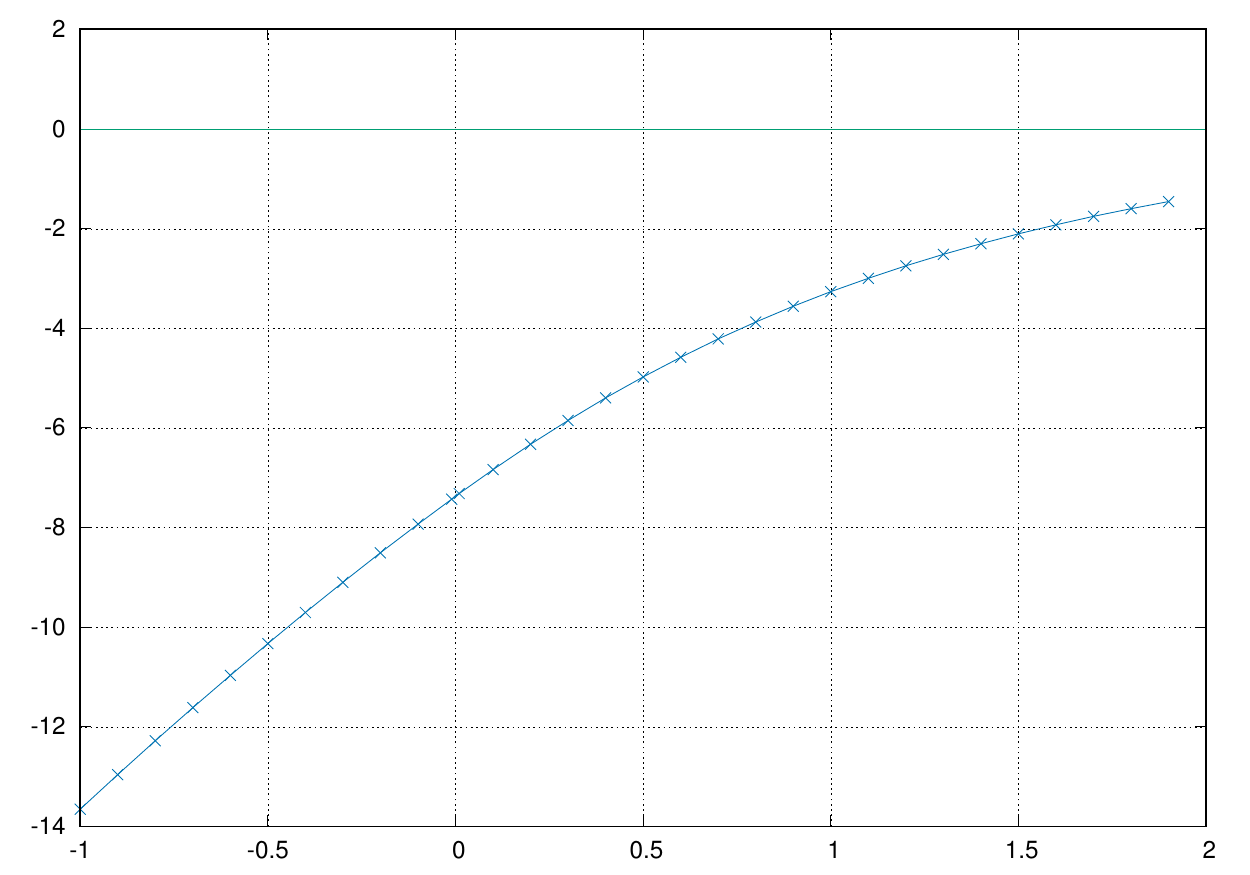} 
\caption{
$f(m_0)$ vs. $m_0$ :
The consistency condition for the 
SO(10) symmetry breaking in the spin model of $E^a(x)$
within the saddle point analysis in the spirit of the large N
expansion.
}
\label{fig:GapEquation}
\end{center}
\end{figure}

The stationary conditions for $X^a_0$ and $\lambda_0$ are given by
\begin{eqnarray}
0 &=& 2 \, X^a_0 
\left\{ 
- \frac{8}{X^c_0 X^c_0} -i \lambda_0
+
\frac{9}{2}\frac{1}{V} \sum_{k \not = 0} 
\Biggl(
-\frac{1}{X^c_0 X^c_0}
-i \lambda_0 \frac{1}{-4 \tilde D(k)  -i \lambda_0  X^d_0 X^d_0} 
\Biggr) 
\right\}  , \\
0 &=& \big( X^c_0 X^c_0 - 1 \big)
+ \frac{9}{2}\frac{1}{V} \sum_{k \not = 0} \frac{1}{-4 \tilde D(k) / X^d_0 X^d_0 -i \lambda_0 } .
\end{eqnarray}
Assuming $X^a_0 \not = 0$,
the above conditions imply that
\begin{eqnarray}
&& - i \lambda_0 = 8 / X^c_0 X^c_0 
-\frac{9}{32}\frac{1}{V} \sum_{k \not = 0} 
\Biggl(
-16
+ \frac{32}{- \tilde D(k) + 2} 
\Biggr) , \\
\label{eq:condition-X0X0}
&& 
X^c_0 X^c_0 
= 1 - \frac{9}{32}\frac{1}{V} \sum_{k \not = 0} \frac{4}{-\tilde D(k) + 2 } ,
\end{eqnarray}
where
the leading results, $- i \lambda_0 = 8 / X^c_0 X^c_0 $ and $X^c_0 X^c_0 = 1$, are substituted 
in the terms suppressed by the factor $N /2^{\frac{N}{2}} (\simeq 9/32  )$.
The r.h.s. of the condition eq.~(\ref{eq:condition-X0X0}) is required to be positive for $X^a_0 \not = 0$.
It is plotted  in fig.~\ref{fig:GapEquation}
as the function of $m_0$,
\begin{equation}
f(m_0) \equiv 1 - \frac{9}{32}\frac{1}{V} \sum_{k \not = 0} \frac{4}{-\tilde D(k) + 2 } .
\end{equation}
One can see that 
$f(m_0) \le 0 $ 
for $m_0 < 2$ 
and it is in contradiction with the assumption $X^a_0 \not = 0$.
In this region of the mass prameter $m_0$,
the fluctuation of the spin field $E^a(x)$
is too large to maintain the non-zero expectation value
of the spin field  $ \langle E^a(x) \rangle $.
The region includes
the positive region 
$ 0 \le m_0 < 2$
and
it also extends to the negative region $ m_0 \le 0$ all the way down to $m_0 \rightarrow -\infty$.

The above result supports the following picture on
the dynamical nature of the spin model. 
The spin model for $E^a(x)$ is well-defined for all values of $m_0$
in the region $[- \infty, 2)$.
For $m_0 \in [ -\infty , 2 )$, 
the spin model 
is in the single disordered phase.
In the limit $m_0 \rightarrow - \infty$, in particular, 
the pfaffian is unity 
and
the spin field is completely disordered, having 
the vanishing correlation length, $\xi_E = 0$.
The correlation length $\xi_E$ is
a monotonically increasing function of $m_0$.
And 
SO(10) global symmetry in the weak gauge-coupling limit
as well as $Z_4 \times Z_4$ discrete symmetries
are not broken spontaneously in the thermodynamic limit $L \rightarrow \infty$.

For our purpose to formulate the Weyl field measure, 
the spin model for $E^a(x)$ should be in the positive
disordered region 
$m_0 \in (0 , 2)$, 
while 
the spin model for $\bar E^a(x)$ is equivalent
to the model in the limit $m_0 \rightarrow -\infty$. Thus the both spin models
have the disorder nature, which are actually in the same disordered phase. 

\subsection{A summary}

Based on the above analytical and numerical results, 
we argue that
in these two cases of the trivial link field
and of the SU(2) link fields with $Q (\not = 0)$, 
the path-integration of the pfaffian 
${\rm pf}( u^{\rm T}\, i \gamma_5 C_D {\rm T}^a E^a u )$ over the spin field $E^a(x)$ gives a non-zero result,
\begin{equation}
\label{eq:path-integral-pfaffian-nonzero-for-field}
\int {\cal D}[E] \, 
{\rm pf}  \big(u^{\rm T} \,  i \gamma_5 C_D {\rm T}^a E^a u \big)
= c \,[ U(x,\mu) ] \not = 0 
\end{equation}
and that
the measure of the right-handed field, 
${\cal D}_\star[\psi_+]$, is indeed saturated 
completely by inserting the product of the 't Hooft vertex 
$T_+(x) [\psi_+]$,
while
the SO(10) symmetry does not break spontaneously in the thermodynamic limit.
Furthermore, we also argue that for the case of 
the trivial link field, 
the path-integral of the pfaffian does not vanish and remains positive-definite in the course of the interpolation
of the mass parameter $m_0$ from the negative region $m_0 < 0$
to the positive region $0< m_0 < 2$.

\section{Other anomalous/anomaly-free chiral gauge theories}
\label{sec:other-theories}

\subsection{Fate of the anomalous SU(2) chiral gauge theory}

Although it is known to be inconsistent due to the global gauge anomaly\cite{Neuberger:1998rn, Bar:2002sa, Bar:2000qa},
it is instructive to try 
to formulate the SU(2) chiral gauge theory
with the Weyl field in the doublet $\underbar{2}$ in the similar manner as the SO(10) model.
This is because the doublet $\underbar{2}$ of SU(2) is the irreducible spinor representation of SO(3) and the (pseudo) scalar bilinear of the Weyl field transforms as the triplet of SO(3), while the coefficient of the perturbative gauge anomaly
vanishes identically as ${\rm Tr} \{ \tau^{a'} 
( \tau^{b'} \tau^{c'} + \tau^{c'} \tau^{b'} ) \} = 0$.
Then one may try to saturate the path-integral measure of
the right-handed Weyl fields by the product of the following
gauge-invariant quartic operators,
\begin{eqnarray}
\frac{1}{2^3}
\big[ 
\psi_+^{\rm T}(x) i \gamma_5 C_D  (i \tau_2 \tau^{a'} ) \psi_+(x) 
\big]^2 ,
\qquad
\frac{1}{2^3}
\big[
\bar \psi_+(x) i \gamma_5 C_D {(i \tau_2 \tau^{a'} )}^\dagger \bar\psi_+ (x)^T
\big]^2 .
\end{eqnarray}

However, this does not work in the topologically nontrivial
sectors $\mathfrak{U}[Q]$, because the index theorem is given by $n_+ - n_- = -Q$ in the SU(2) theory and the number of the zero modes is not necessarily a multiple of four.
In particular, when the topological charge $Q$ is an odd integer,
the dimension of the anti-symmetric matrix $(u_j^T  i \gamma_5 C_D  (i \tau_2 \tau^{a'} ) E^{a'} u_k)$ is odd and 
its pfaffian vanishes identically.
Therefore, the above operators can not always saturate
the right-handed measure. 
%
%
Thus the SU(2) chiral gauge theory with the single Weyl field in 
the doublet $\underbar{2}$ is ill-defined in our formulation, as it should be.

\subsection{Anomaly-free chiral gauge theories descent from SO(10)}

Once the lattice model for the SO(10) chiral gauge theory with the Weyl field in $\underbar{16}$ is formulated, it is straightforward to obtain 
the lattice models for 
the SU(5) chiral gauge theory with $\{\underbar{10}, \underbar{5}^\ast \}$ (the Georgi-Glashow model), 
the SU(4)$\times$SU(2)$\times$SU(2) chiral gauge theory with 
$\{ (\underbar{4},\underbar{2},\underbar{1}), 
    (\underbar{4}^\ast,\underbar{1},\underbar{2}) \}$ (the Pati-Salam model),
the SU(3)$\times$SU(2)$\times$U(1) chiral gauge theory
with
$\{ (\underbar{3},\underbar{2})_{1/6},  (\underbar{3}^\ast,\underbar{1})_{-2/3}, (\underbar{3}^\ast,\underbar{1})_{1/3}, (\underbar{1},\underbar{2})_{-1/2}, (\underbar{1},\underbar{1})_{1}, (\underbar{1},\underbar{1})_{0}\}$ (the standard model)
by reducing the link field in SO(10) to 
the subgroups, SU(5), SU(4)$\times$SU(2)$\times$SU(2), SU(3)$\times$SU(2)$\times$U(1), respectively. 
The spin fields may be kept the SO(10) real vectors so that 
the SO(10) global symmetry is maintained in the weak gauge-coupling limit. The continuous  global symmetries remained in these models are all ``ready to be gauged''
without encountering gauge anomalies, thus anomaly-matched, and  therefore free from the infra-red singularities in the correlation functions of the associated conserved currents 
due to chiral(gauge) anomalies.

\subsection{The standard model plus the right-handed neutrinos}

As to the case of the SU(3)$\times$SU(2)$\times$U(1) chiral gauge theory,
in particular, we can incorporate the three generations of quarks
and leptons plus right-handed neutrinos
by simply preparing the three copies of the Weyl fields in 
$\underbar{16}$ in our formulation.
Then the left-handed part of the fermion measure respects the global flavor SU(3) symmetry,
while the right-handed part respects only the global permutation symmetry, S$_3$, because of the products of the
't Hooft vertices of each right-handed fields.
It is straightforward to incorporate
the Higgs boson field and its Yukawa couplings to
the left-handed fields consisting the three generations of quark
and leptons plus right-handed neutrinos.
In doing this, one can include the chiral projector $P_\pm$
suitably so that the Yukawa coupling terms themselves 
respect the CP symmetry\cite{Suzuki:2000ku, Fujikawa:2002vj, Fujikawa:2002up}
and that 
the sources
of 
the CP violation are given by
the complex phases in the flavor mixing matrices
of quarks (Cabibbo-Kobayashi-Maskawa matrix\cite{Cabibbo:1963yz,Kobayashi:1973fv}),
charged leptons and neutrinos (Pontecorvo-Maki-Nakagawa-Sakata matrix\cite{Pontecorvo:1957qd,Maki:1962mu}) beside the theta terms.
Thus we can formulate the standard model plus the right-handed neutrinos on the lattice with the manifest exact gauge invariance
and the required symmetry properties.
The detail of the formulation will be discussed elsewhere.

\section{Relations with other approaches/proposals}
\label{sec:rel-other-approaches}

The characteristic features of our lattice model for
the SO(10) chiral gauge theory with the Weyl fermions in 
$\underbar{16}$ can be summarised in the formula
of the fermion partition function eq.~(\ref{eq:def-effective-action}):
\begin{eqnarray}
\label{eq:def-effective-action-again}
\big\langle 1 \big\rangle_F
& \equiv & \int {\cal D}[\psi_-] {\cal D}[\bar \psi_-]  \,  {\rm e}^{-S_W[\psi_-, \bar\psi_-]}    \nonumber\\
&=&\int {\cal D}[\psi] {\cal D}[\bar \psi] \,
\, \prod_{x \in \Lambda} F( T_+(x) )
\, \prod_{x \in \Lambda}  F( \bar T_+(x) ) \, 
 {\rm e}^{- S_W[\psi_-, \bar \psi_-] } \nonumber\\
&=&
\int 
{\cal D}[\psi] {\cal D}[\bar \psi] \,
{\cal D}[E] {\cal D}[\bar E] \,
\, {\rm e}^{- S_W[\psi_-, \bar \psi_-] + 
\sum_{x \in \Lambda} \{ 
E^a(x) 
V_+^a(x) 
+ \bar E^a(x)
\bar V_+^a(x)
\}[\psi_+, \bar\psi_+]
} . \nonumber\\
\end{eqnarray}
In this formula eq.~(\ref{eq:def-effective-action-again}), the total action of the model,  including the 't Hooft vertex terms, can be defined as
\begin{eqnarray}
\label{eq:action-SO10}
S_{\rm Ov}[\psi, \bar \psi, E^a, \bar E^a]
&=&
 \sum_{x\in\Lambda} 
\bar \psi_- (x) 
D \psi_-(x)
\nonumber\\
&-&
\sum_{x \in \Lambda} \frac{1}{2}\,\{ 
E^a(x) 
\psi_+^{\rm T}(x) i \gamma_5 C_D {\rm T}^a \psi_+(x)
+ \bar E^a(x)
\bar \psi_+(x) i \gamma_5 C_D {{\rm T}^a}^\dagger \bar\psi_+ (x)^T
\} . \nonumber\\
\end{eqnarray}
Here the right-handed Weyl fields are introduced explicitly,
trying to make the path-integral measure of the left-handed Weyl fields in $\underbar{16}$ simplified and manifestly gauge-invariant.
The SO(10) invariant 't Hooft vertex operators of the right-handed fields are used to saturate completely the right-handed
part of the fermion measure.
The short range correlations of order the lattice spacing
are required for the the right-handed Weyl fields 
and the auxiliary spin fields
so that they 
are decoupled from  physical degrees of freedom, 
preserving the symmetries and 
leaving only the smooth and local terms of the link fields.
These features/requirements are actually 
shared
with
other various approaches and proposals
to decouple the species doubling or mirror modes of models.

An important technical difference lies on the fact that
the path-integral measure of the right-handed Weyl fields, i.e.
the right-handed part of the chiral decomposition of Dirac field measure, 
%
%
are formulated with the non-trivial
chiral basis 
$\{u_i(x) \, \vert \, {\rm P}_+ \otimes \hat P_+ u_i = u_i , i=1, \cdots , n/2-8Q \, \}$,
$\{\bar u_k(x) \, \vert \, \bar u_k  P_- \otimes {\rm P}_+ = u_k , k=1, \cdots , n/2 \, \}$,
which depends on the gauge field,
as given by 
eq.~(\ref{eq:fermion-measure-left-right-in-chiral-basis}),
\begin{eqnarray}
\label{eq:fermion-measure-right-in-chiral-basis}
&&
\psi_+(x) = \sum_i u_i(x) b_i , \quad 
\bar \psi_+(x) = \sum_k \bar b_k \bar u_k(x) , 
\\
&& {\cal D}_\star[\psi_+] {\cal D}_\star[\bar \psi_+]
=
\prod_{j=1}^{n/2-8Q} d b_j  
\prod_{k=1}^{n/2} d \bar b_k .
\end{eqnarray}
We need to make sure
the locality of this right-handed-measure contribution 
to the induced effective action. 

Another important technical difference is that
we choose the product function for the 't Hooft vertices $F(\omega)$ as given by eq.~(\ref{eq:def-of-F}) and therefore
use the unit SO(10)-vector spin fields, 
$E^a(x)$ and $\bar E^a(x)$ with the constraints
$E^a(x)E^a(x)=1$ and $\bar E^a(x) \bar E^a(x) =1$, 
omitting their kinetic(hopping) terms. 
This choice 
allows us to prove the CP symmetry.
It is also relevant for preserving the (global) SO(10) symmetry
in the thermodynamic limit.

The above action $S_{\rm Ov}$ 
can be regarded as a certain limit of the following 
action of the SO(10)-invariant chiral Yukawa model in the framework of the Ginsparg-Wilson fermion,
\begin{eqnarray}
\label{eq:action-GW-Mirror-SO10}
S_{\rm Ov/Mi}[\psi, \bar \psi, X^a, \bar X^a]
&=&
 \sum_{x\in\Lambda} 
\big\{
\bar \psi_- (x) 
D \psi_-(x)
+
z_+ 
\bar \psi_+ (x) 
D \psi_+(x)
\big\}
\nonumber\\
&-&
\sum_{x \in \Lambda}  \{ 
y \,
X^a(x) 
\psi_+^{\rm T}(x) i \gamma_5 C_D {\rm T}^a \psi_+(x)
+ 
\bar y \,
\bar X^a(x)
\bar \psi_+(x) i \gamma_5 C_D {{\rm T}^a}^\dagger \bar\psi_+ (x)^T
\} 
\nonumber\\
&+& 
S_{X}[X^a] ,
\end{eqnarray}
where
\begin{eqnarray}
S_{X}[X^a] &=&
\sum_{x \in \Lambda}  
\left\{ 
- \kappa  \sum_\mu X^a(x)X^a(x +\hat\mu)
+\frac{1}{2} X^a(x)X^a(x)
+ \frac{\lambda'}{2} (X^a(x)X^a(x) - v^2 )^2 
\right.
\nonumber\\
&&
\left. \qquad\,\,\,
- \bar \kappa  \sum_\mu \bar X^a(x) \bar X^a(x +\hat\mu)
+\frac{1}{2} \bar X^a(x) \bar X^a(x)
+ \frac{\bar \lambda'}{2} (\bar X^a(x) \bar X^a(x) - \bar v^2)^2 
\right\} . \nonumber\\
\end{eqnarray}
The limit to the original action $S_{\rm Ov}$ is achieved
by
\begin{eqnarray}
\label{eq:limit-S10-y}
&& y = \bar y, \quad \frac{z_+}{\sqrt{y \bar y}} \rightarrow 0 , \\
\label{eq:limit-S10-lambda}
&& v = \bar v =1, \quad \lambda'= \bar \lambda' \rightarrow \infty,\\
\label{eq:limit-S10-kappa}
&& \kappa= \bar \kappa \rightarrow 0 .
\end{eqnarray}
In the lattice model defined with the action, 
$S_{\rm Ov/Mi}$, 
the global U(1) symmetry of the right-handed
fields is broken to $Z_4$ by the Yukawa couplings 
$y$ and $\bar y$.
But the proof of the CP symmetry is not successful so far.\footnote{
In the other limit as
\begin{eqnarray}
\label{eq:limit-EP}
&& \lambda'= \bar \lambda' \rightarrow 0,\\
&& \kappa= \bar \kappa \rightarrow 0 ,
\end{eqnarray}
it reduces to the model with quartic interaction of 
the 't Hooft vertices,
\begin{eqnarray}
S_{\rm Ov/EP}[\psi, \bar \psi] 
&=&
 \sum_{x\in\Lambda} 
\big\{
\bar \psi_- (x) 
D \psi_-(x)
+
z_+
\bar \psi_+ (x) 
D \psi_+(x)
\big\}
\nonumber\\
&-&
\sum_{x \in \Lambda}  \{ 
y^2 \,
\frac{1}{2}
\big[ 
\psi_+^{\rm T}(x) i \gamma_5 C_D {\rm T}^a \psi_+(x) 
\big]^2
+ 
\bar y^2 \,
\frac{1}{2}
\big[
\bar \psi_+(x) i \gamma_5 C_D {{\rm T}^a}^\dagger \bar\psi_+ (x)^T
\big]^2
\} .
\end{eqnarray}
This action (in the limit $z_+ \rightarrow 0$) corresponds
to the other choice of the product function 
$F(\omega)$ as $F(\omega)=e^{\omega}$.
}

In the following, 
we discuss the relations to 
Eichten-Preskill model, 
Ginsparg-Wilson Mirror-fermion model, 
Domain wall fermion model with the boundary Eichten-Preskill term,  
4D Topological Insurators/Superconductors
with gapped boundary phases,
and  
the recent studies on the Paramagnetic Strong-coupling (PMS) phase/Mass without symmetry breaking, 
trying to clarify the similarity and
the difference in technical detail and to show 
that our proposal is a well-defined testing ground for that basic question.
%
%

\subsection{cf. Eichten-Preskill model  
}
\label{subsec:cf-EP}

The SO(10) invariant interaction terms of the 't Hooft vertex
were first used 
by Eichten and Preskill\cite{Eichten:1985ft}
to decouple the species doublers 
in their formulation of chiral lattice gauge theories
based on the generalized Wilson term:
\begin{eqnarray}
S_{\rm EP} 
&=& 
 \sum_{x\in\Lambda} 
\big\{
\bar \psi (x) 
  \gamma_\mu P_-  ([\nabla_\mu - \nabla_\mu^\dagger]/2)
\psi(x)
+
z_+ 
\bar \psi (x) 
 \gamma_\mu P_+  ([\nabla_\mu - \nabla_\mu^\dagger]/2)
 \psi(x)
\big\}
\nonumber\\
&-&
\sum_{x \in \Lambda}  \{ 
\frac{\lambda}{24}
\big[ 
\psi_+^{\rm T}(x) i \gamma_5 C_D {\rm T}^a \psi_+(x) 
\big]^2
+ 
\frac{\lambda}{24}
\big[
\bar \psi_+(x) i \gamma_5 C_D {{\rm T}^a}^\dagger \bar\psi_+ (x)^T
\big]^2
\} 
\nonumber\\
&-&
\sum_{x \in \Lambda}  \{ 
\frac{r}{48}
\Delta
\big[ 
\psi^{\rm T}(x) i \gamma_5 C_D {\rm T}^a P_+ \psi(x) 
\big]^2
+ 
\frac{r}{48}
\Delta
\big[
\bar \psi(x) P_- i \gamma_5 C_D {{\rm T}^a}^\dagger  \bar\psi (x)^T
\big]^2
\} ,
\nonumber\\
\end{eqnarray}
where 
\begin{eqnarray}
&&\Delta\{ A(x)B(x)C(x)D(x)\}
\nonumber\\
&&\qquad
\equiv 
+\frac{1}{2}
\sum_{\mu}
\Big\{ 
\big(\nabla_\mu \nabla_\mu^\dagger A(x)\big)B(x)C(x)D(x)
+A(x) \big(\nabla_\mu \nabla_\mu^\dagger B(x)\big)C(x)D(x)
\nonumber\\
&&\qquad\qquad\quad
+A(x)B(x)\big(\nabla_\mu \nabla_\mu^\dagger C(x)\big)D(x)
+A(x)B(x)C(x)\big(\nabla_\mu \nabla_\mu^\dagger D(x)\big)
\Big\}.
\end{eqnarray}
In this action, the right(left)-handed Weyl fields are formulated
by the naive chiral projectors as 
$P_+ \psi(x)$, $\bar \psi(x) P_-$ ($P_- \psi(x)$, $\bar \psi(x) P_+$).
The global U(1) symmetry of the right-handed
fields is broken to $Z_4$ by the quartic couplings 
$\lambda$, $\bar \lambda$ and $r$, $\bar r$, 
and the CP symmetry is manifest
thanks to the naive definition of chirality.
On the other hand,  
in their analysis of the Eichten-Preskill model, Golterman, Petcher and Rivas\cite{Golterman:1992yha} have considered the same type of chiral Yukawa model as $S_{\rm Ov/Mi}$, but with the naive Dirac operator, the naive chiral projectors, 
and the additional Wilson-Yukawa coupling with the lattice laplacian included:
\begin{eqnarray}
S_{\rm EP/WY}
&=&
 \sum_{x\in\Lambda} 
\big\{
\bar \psi (x) 
  \gamma_\mu P_-  ([\nabla_\mu - \nabla_\mu^\dagger]/2)
\psi(x)
+
z_+ 
\bar \psi (x) 
 \gamma_\mu P_+  ([\nabla_\mu - \nabla_\mu^\dagger]/2)
 \psi(x)
\big\}
\nonumber\\
&-&
\sum_{x \in \Lambda}  \{ 
y \,
X^a(x) 
\psi^{\rm T}(x) i \gamma_5 C_D {\rm T}^a P_+ \psi(x)
+ 
y \,
\bar X^a(x)
\bar \psi(x) P_- i \gamma_5 C_D {{\rm T}^a}^\dagger \bar\psi (x)^T
\} 
\nonumber\\
&-&
\sum_{x \in \Lambda}  \{ 
w \,
X^a(x) 
\psi^{\rm T}(x) i \gamma_5 C_D {\rm T}^a 
(\nabla_\mu\nabla_\mu^\dagger/2) 
P_+ \psi(x)
\nonumber\\
&&
\qquad
+ 
w \,
\bar X^a(x)
\bar \psi(x) P_- i \gamma_5 C_D {{\rm T}^a}^\dagger
(\nabla_\mu\nabla_\mu^\dagger/2) 
 \bar\psi (x)^T
\} 
\nonumber\\
&+& 
S_{X}[X^a] \big\vert_{\lambda'=\bar\lambda'} .
\end{eqnarray}
%
The authors' intention here is to consider the right-handed sector of the above models and to leave only
the physical mode (with $p_\mu \simeq 0$) of the right-handed Weyl fields 
$P_+ \psi(x)$, $\bar \psi(x) P_-$
and split/gap/decouple the speices doubling modes (with $p_\mu \simeq \pi^{(A)}$, $A=1,\cdots,15$) of the fields by a suitable choice of the couplings, $\lambda$, $r$
or $y$, $w$ 
just like the Wilson term does 
for the naive Dirac field.

As Eichten and Preskill
showed, in the strong quartic-coupling limit
$z_+ /\sqrt{\lambda} \rightarrow 0$ and
$r/\lambda \rightarrow 0$, 
the path-integral measure of the right-handed Weyl fields,
\begin{equation}
\label{eq:right-handed-measure-WYEP}
{\cal D}_\diamond [\psi_+] 
{\cal D}_\diamond [\bar \psi_+]
=
\prod_{x \in \Lambda} \prod_{\alpha=1}^2\prod_{s=1}^{16} 
 d \psi_{\alpha s}(x) 
\prod_{x \in \Lambda} \prod_{\alpha=3}^4\prod_{s=1}^{16} 
 d\bar\psi_{\alpha s}(x),
\end{equation}
are saturated completely by the 't Hooft vertices in terms of the right-handed fields, as reproduced in eq.(\ref{eq:path-integral-pfaffian-nonzero-for-antifield-2}):
\begin{eqnarray}
&&
\int \,
%
{\cal D}_\diamond [\psi_+]
\prod_{x \in \Lambda} 
\frac{4!}{8! 12!}
\left\{
\frac{1}{2^3}\,
\psi(x)^T P_+  i \gamma_5 C_D {\rm T}^a \psi(x) \, 
\psi(x)^T P_+ i \gamma_5 C_D {\rm T}^a \psi (x)
\right\}^{8}   
=1,
\nonumber\\
&&\\
&&
\int \,
%
{\cal D}_\diamond [\bar \psi_+]
\prod_{x \in \Lambda} 
\frac{4!}{8! 12!}
\left\{
\frac{1}{2^3}\,
\bar \psi(x) P_-  i \gamma_5 C_D {\rm T}^{a \dagger} \bar\psi(x)^{\rm T} \, 
\bar \psi(x) P_- i \gamma_5 C_D {\rm T}^{a \dagger} \bar\psi (x)^{\rm T}
\right\}^{8}   
=1.
\nonumber\\
\end{eqnarray}
Moreover,
in the hopping parameter expansion w.r.t. 
$z_+ / \sqrt{\lambda}$, $ r / \lambda$, 
all the modes of the right-handed Weyl fields
$P_+ \psi(x)$, $\bar \psi(x) P_-$ acquire masses
without breaking the SO(10) symmetry
through
the mixings 
with the modes of the composite fields 
\begin{eqnarray}
B_-(x) &=&  
P_- i \gamma_5 C_D  {{\rm T}^a}^\dagger  \bar\psi (x)^T \, 
\bar \psi(x) P_- i \gamma_5 C_D {{\rm T}^a}^\dagger 
  \bar\psi (x)^T, \\
\bar B_-(x) &=& 
\psi^{\rm T}(x) i \gamma_5 C_D P_+ {\rm T}^a \psi(x) \, 
\psi^{\rm T}(x) i \gamma_5 C_D P_+ {\rm T}^a ,
\end{eqnarray}
as chiral partners.

Golterman, Petcher and Rivas applied a method of large $N_f$ expansion to the model $S_{\rm EP/WY}$, where
$N_f$ is the number of copies of the right(left)-handed Weyl fields
in $\underbar{16}$.
In the region of the strong Yukawa/Wilson-Yukawa couplings, 
where the couplings are scaled as
$y=\sqrt{N_f} \tilde y$, $w=\sqrt{N_f} \tilde w$ ($\lambda'=0$, $z_+=1$), 
the effective hopping parameter for the spin fields 
$X^a(x)$ 
is given by
\begin{eqnarray}
\kappa_{\rm eff} 
= 
\kappa + 
\frac{1}{8}\int_{-\pi}^{\pi} \frac{d^4 p}{(2\pi)^4} \frac{\sum_\mu \sin^2 p_\mu}{\big(\tilde y + \tilde w \sum_\mu(1-\cos p_\mu) \big)^2} .
\end{eqnarray}
For larger $\tilde y$ and $\tilde w$, $\kappa_{\rm eff}$
is less than  the mean-field value of the critical hopping parameter $\kappa_c = 5/2$. Then the SO(10) symmetry is not broken spontaneously and the model is in the PMS phase.
For smaller $\tilde y$ and $\tilde w$, 
$\kappa_{\rm eff}$ exceeds the critical value.
Then the SO(10) symmetry is broken spontaneously and the model is in the FM phase.
All the modes of the right-handed Weyl fields
$P_+ \psi(x)$, $\bar \psi(x) P_-$ 
form massive Wilson-Dirac fermions
with the modes of the composite fields 
\begin{eqnarray}
B'_-(x) &=&
 P_- i \gamma_5 C_D  
\bar\psi (x)^T {{\rm T}^a}^\dagger \bar X^a(x) ,  \\
\bar B'_-(x) &=& 
\psi^{\rm T}(x) i \gamma_5 C_D P_+ {\rm T}^a X^a(x) , 
\end{eqnarray}
as chiral partners.
The inverse propagator is given by
\begin{eqnarray}
S(p)^{-1} &=& (P_+ z^2 + P_- ) i \gamma_\mu \sin p_\mu +
y + w  \sum_\mu(1-\cos p_\mu) , 
\end{eqnarray}
where $z^2 = \frac{1}{32} \langle X^a(x)X^a(x\pm\hat\mu) 
+ \bar X^a(x) \bar X^a(x\pm\hat\mu) \rangle$
up to ${\cal O}(1/N_f)$ corrections.
Through the transition from the PMS phase to the FM phase, 
it remains that $z^2 \not = 0$ and 
the right-handed Weyl fields
keep to form the Wilson-Dirac fermions
with the composite-field chiral partners.

On the other hand, 
in the region of the weak Yukawa/Wilson-Yukawa couplings, 
where the couplings are scaled as
$y=\tilde y / \sqrt{N_f}$, $w=\tilde w /\sqrt{N_f}$, 
$\lambda' = \tilde \lambda / N_f$ ($z_+ = 1$), 
the SO(10) symmetry is broken spontaneously
for larger $\tilde y$ and $\tilde w$ and the model is in the FM phase.
The non-zero vacuum expectation value,
$\langle X^a(x) \rangle = \delta^{a 10} \tilde v \, \sqrt{N_f}$,
is determined through the gap equation of the fermion self-energy,
\begin{eqnarray}
\Sigma(p) &=&
i \gamma_5 C_D P_+ \check {\rm C} \,  \tilde v 
\big(\tilde y + \tilde w \sum_\mu (1-\cos p_\mu) \big), 
\end{eqnarray} 
by 
\begin{eqnarray}
1 + 2 \tilde \lambda \tilde v^2 - 8 \kappa 
= 32 
\int_{-\pi}^{\pi} \frac{d^4 p}{(2\pi)^4} 
\frac{
\big(\tilde y + \tilde w \sum_\mu(1-\cos p_\mu) \big)^2
}{
\sum_\mu \sin^2 p_\mu 
+
\tilde v^2 
\big(\tilde y + \tilde w \sum_\mu(1-\cos p_\mu) \big)^2
} .
\end{eqnarray}
The right-handed Weyl fields acquire 
the non-vanishing Majorana-Wilson masses.
For smaller $\tilde y$ and $\tilde w$,  $\tilde v$ vanishes
and the model is in the PMW phase.
The Majorana-Wilson masses vanish identically
and all the modes of the right-handed Weyl fields
become massless.

Thus both in the PMS and PMW phases
the massless fermion spectrum
consist of Dirac fermions in $\underbar{16}$.
Then it was argued by the authors
that the existence of the FM phase which 
separates the PMS and PMW phases
is the crucial ingredient for the failure of the proposal.



For our purpose, 
the above results in the strong coupling limits are indeed the ideal situation. 
Our intention here is actually to show the same situation occurs
for the Weyl fields $\hat P_+ \psi(x)$, $\bar \psi(x) P_-$
in the framework of the overalp fermion/the Ginsparg-Wilson relation as defined in 
the model $S_{\rm Ov/Mi}$ 
and 
the model $S_{\rm Ov}$.  
As we mentioned above, 
an important technical difference here
lies on the fact that
the path-integral measure of the right-handed Weyl fields, i.e.
the right-handed part of the chiral decomposition of Dirac field measure, are simple and gauge-invariant
for the model $S_{\rm WY/EP}$ as given by eq.~(\ref{eq:right-handed-measure-WYEP}), 
but are non-trivial and 
gauge-field dependent for the models
$S_{\rm Ov/Mi}$ and $S_{\rm Ov}$ as given by eq.~(\ref{eq:fermion-measure-left-right-in-chiral-basis}).
This is why we need to make sure
the saturation of the right-handed-measure
by the 't Hooft vertices and 
the locality of the right-handed-measure contribution to the induced effective action for our case.

Another important technical difference is that
we 
choose the product function %
$F(\omega)$ as given by eq.~(\ref{eq:def-of-F}) and therefore
use the unit SO(10)-vector spin fields, 
$E^a(x)$ and $\bar E^a(x)$ with the constraints
$E^a(x)E^a(x)=1$ and $\bar E^a(x) \bar E^a(x) =1$, 
omitting their kinetic(hopping) terms. 
This choice allows us to prove the CP symmetry.
But 
it is also relevant for preserving the (global) SO(10) symmetry.
This corresponds to 
taking the limits eqs.~(\ref{eq:limit-S10-lambda}) 
and (\ref{eq:limit-S10-kappa}) in $S_{\rm EP/WY}$ as 
\begin{eqnarray}
\label{eq:EP-WY-with-spin-field-E}
S_{\rm EP/WY}
\big\vert^{v=1, \lambda' \rightarrow \infty}_{\kappa=\bar \kappa=0}
&=&
 \sum_{x\in\Lambda} 
\big\{
\bar \psi (x) 
  \gamma_\mu P_-  ([\nabla_\mu - \nabla_\mu^\dagger]/2)
\psi(x)
+
z_+ 
\bar \psi (x) 
 \gamma_\mu P_+  ([\nabla_\mu - \nabla_\mu^\dagger]/2)
 \psi(x)
\big\}
\nonumber\\
&-&
\sum_{x \in \Lambda}  \{ 
y \,
E^a(x) 
\psi^{\rm T}(x) i \gamma_5 C_D {\rm T}^a P_+ \psi(x)
+ 
y \,
\bar E^a(x)
\bar \psi(x) P_- i \gamma_5 C_D {{\rm T}^a}^\dagger \bar\psi (x)^T
\} 
\nonumber\\
&-&
\sum_{x \in \Lambda}  \{ 
w \,
E^a(x) 
\psi^{\rm T}(x) i \gamma_5 C_D {\rm T}^a 
(\nabla_\mu\nabla_\mu^\dagger/2) 
P_+ \psi(x)
\nonumber\\
&&
\qquad
+ 
w \,
\bar E^a(x)
\bar \psi(x) P_- i \gamma_5 C_D {{\rm T}^a}^\dagger
(\nabla_\mu\nabla_\mu^\dagger/2) 
 \bar\psi (x)^T
\} .
\end{eqnarray}
This region in the coupling-constant space of $S_{\rm EP/WY}$
has not been explored by Golterman, Petcher and Rivas\cite{Golterman:1992yha}.
And, in fact, we find no phase transition from
the PMS phase to the FM phase towards the weak-coupling limit
$y/z_+, w/z_+ \rightarrow 0$ 
within
the saddle point analysis in the spirit of the large N expansion.
We will come back to this point later
in relation to the discussion about the recent studies on
the PMS phase/``Mass without symmetry breaking''. 

\subsection{cf. Ginsparg-Wilson Mirror-fermion model  
}
\label{subsec:cf-GWMi}

The SO(10) invariant action $S_{\rm Ov/Mi}$, eq.~(\ref{eq:action-GW-Mirror-SO10}), 
defines a Mirror-fermion model 
for the SO(10) chiral gauge theory
in the framework of the Ginsparg-Wilson fermion.
It is formulated in the spirit of 
the series of works by 
Bhattacharya, Chen, Giedt, Poppitz and Shang\cite{Bhattacharya:2006dc,Giedt:2007qg,Poppitz:2007tu,Poppitz:2008au,Poppitz:2009gt,Poppitz:2010at,Chen:2012di,Giedt:2014pha},
although any SO(10) model has not been discussed in the literature.
In the action $S_{\rm Ov/Mi}$, 
the global U(1) symmetry of the right-handed
fields is broken to $Z_4$ by the Yukawa couplings 
$y$ and $\bar y$.
As mentioned above, however, the proof of the CP symmetry is
not successful so far.
%

In this respect, 
we note that 
one can prove the CP invariance 
of the effective action 
if one modifies 
the Yukawa couplings by the insertion of the chiral projectors $P_\pm$ \cite{Suzuki:2000ku, Fujikawa:2002vj, Fujikawa:2002up}
as
\begin{eqnarray}
\label{eq:singular-yukawa-coupling}
&&
-
\sum_{x \in \Lambda}  \{ 
y \,
X^a(x) 
\psi_+^{\rm T}(x) i \gamma_5 C_D {\rm T}^a P_+ \psi_+(x)
+ 
\bar y \,
\bar X^a(x)
\bar \psi_+(x) i \gamma_5 C_D {{\rm T}^a}^\dagger P_- \bar\psi_+(x)^T
\} 
\nonumber\\
&=&
-
\sum_{x \in \Lambda}  \{ 
y \,
\psi^{\rm T}(1-D)^T i \gamma_5 C_D {\rm T}^a X^a P_+ (1-D) \psi(x)
+ 
\bar y \,
\bar \psi i \gamma_5 C_D {{\rm T}^a}^\dagger \bar X^a P_- \bar\psi^T(x)
\} , \nonumber\\
\end{eqnarray} 
and 
to take
the following action,
\begin{eqnarray}
\label{eq:action-GW-Mirror-SO10-naive}
S'_{\rm Ov/Mi}[\psi, \bar \psi, X^a, \bar X^a]
&=&
 \sum_{x\in\Lambda} 
\big\{
\bar \psi_- (x) 
D \psi_-(x)
+
z_+ 
\bar \psi_+ (x) 
D \psi_+(x)
\big\}
\nonumber\\
&-&
\sum_{x \in \Lambda}  \{ 
y \,
X^a(x) 
\psi_+^{\rm T}(x) i \gamma_5 C_D {\rm T}^a P_+ \psi_+(x)
\nonumber\\
&&\qquad
+ 
\bar y \,
\bar X^a(x)
\bar \psi_+(x) P_- i \gamma_5 C_D {{\rm T}^a}^\dagger \bar\psi_+ (x)^T
\} 
\nonumber\\
&+& 
S_{X}[X^a] .
\end{eqnarray} 
But this type of Yukawa coupling is singular in the large limit 
$\frac{z_+}{\sqrt{y \bar y}} \rightarrow 0$:
the saturation of the right-handed part of  the measure
is incomplete, because 
\begin{equation}
P_+  \psi_+(x) = P_+  \hat P_+ \psi(x) = P_+(1-D) \psi(x)
\end{equation}
and the factor $(1-D)$ projects out
the modes with the momenta $\pi_\mu^{(A)}$ ($A=1,\cdots, 15$).

We note that this is the common property of the mass-like terms of the Ginsparg-Wilson fermion.
For the Dirac mass term, it is usually formulated as
\begin{equation}
S_D = \sum_{x \in \Lambda}  \{ 
\bar \psi(x) D \psi(x) + m_D \, \bar \psi (1-D) \psi(x) 
\},
\end{equation}
because the scalar and pseudo scalar operators,
$\bar \psi (1-D) \psi(x)$ and 
$\bar \psi i \gamma_5(1-D) \psi(x) $,
have the good transformation properties
under the chiral transformation, 
$\delta \psi(x)=\gamma_5 (1-2D) \psi(x)$, 
$\delta \bar\psi(x) = \bar\psi(x) \gamma_5$.
However, this choice makes the limit of the large mass parameter
$m_D$ singular by the same reason as above. The maximal value of the mass is given at $m_D=1$, where $D$ cancels out in the action and 
the simple bilinear operator $\bar\psi(x)\psi(x)$ saturates
the path-integral measure of the Dirac fields completely.
To make the limit of the large mass parameter well-defined, we should write the action as
\begin{equation}
S_D = \sum_{x \in \Lambda}  \{ 
 z \, \bar \psi(x) D \psi(x) + m \, \bar \psi(x) \psi(x) 
\},
\end{equation}
where $z= 1-m_D$ and $m=m_D$ and should take the limit
$z/m = (1-m_D)/m_D \rightarrow 0$.

As for the Majorana mass term, 
one often formulates the action as
\begin{eqnarray}
S_M 
&=& \sum_{x \in \Lambda}  \{ 
\bar \psi_+(x)  D  \psi_+(x)
\nonumber\\
&&\qquad
 + m_M \,( 
\psi_+(x)^T 
C_D  \psi_+(x) 
+
\bar \psi_+(x) 
C_D \bar \psi_+(x)^T 
)
\}
\\
&=& \sum_{x \in \Lambda}  \{ 
\bar \psi(x) P_- D  \psi(x)
\nonumber\\
&&\qquad
 + m_M \,( 
\psi^T \hat P_+^T 
C_D \hat P_+ \psi(x) 
+
\bar \psi P_- 
C_D P_-^T \bar \psi^T(x)
) 
\} .
\end{eqnarray}
But this Majorana mass term has the
matrix elements as
\begin{eqnarray}
u^T_j  C_D u_k 
&=&
- \delta_{p+p',0} \, \frac{b(p')}{\omega(p')} \epsilon_{\sigma, \sigma'} \, \quad 
(j=\{p,\sigma\}, k=\{p',\sigma'\}) , \\
\bar u_j  C_D \bar u_k^T
&=&
- \delta_{x,x'} \epsilon_{\sigma, \sigma'} \, \quad \quad \,\,\,\qquad
(j=\{x,\sigma\}, k=\{x',\sigma'\}) ,
\end{eqnarray}
and the pfaffian of the first matrix has the factor
$\prod_{p'} \{ b(p')/\omega(p')\}$, 
while the second one is unity.
Since $b(p')$ can vanish for $ 0 < m_0 < 2$, it is singular in the limit of the large mass parameter $m_M$.
This type of the Majorana-Yukawa couplings are used
in the formulation of the 2D ``10'',  ``3450'' models by Chen, Giedt, Poppitz and Shang.\cite{Bhattacharya:2006dc,Giedt:2007qg,Poppitz:2007tu,Poppitz:2008au,Poppitz:2009gt,Poppitz:2010at,Chen:2012di,Giedt:2014pha} 
Instead, one can formulate the action as
\begin{eqnarray}
S_M 
&=& \sum_{x \in \Lambda}  \{ 
 z \, \bar \psi_+(x)  D  \psi_+(x)
\nonumber\\
&&\qquad
 + M \,( 
\psi_+(x)^T i\gamma_5 C_D  \psi_+(x) 
+
\bar \psi_+(x) i\gamma_5 C_D \bar \psi_+(x)^T 
)
\}
\\
&=& \sum_{x \in \Lambda}  \{ 
 z \, \bar \psi(x) P_- D  \psi(x)
\nonumber\\
&&\qquad
 + M \,( 
\psi^T \hat P_+^T i\gamma_5 C_D \hat P_+ \psi(x) 
+
\bar \psi P_- i\gamma_5 C_D P_-^T \bar \psi^T(x)
) 
\} .
\end{eqnarray}
In the chiral basis, this Majorana mass term has the
matrix elements as
\begin{eqnarray}
u^T_j  i\gamma_5 C_D u_k 
&=&
i \delta_{p+p',0} \epsilon_{\sigma, \sigma'} \, \quad 
(j=\{p,\sigma\}, k=\{p',\sigma'\}) , \\
\bar u_j  i\gamma_5 C_D \bar u_k^T
&=&
i \delta_{x,x'} \epsilon_{\sigma, \sigma'} \, \quad \quad
(j=\{x,\sigma\}, k=\{x',\sigma'\}) ,
\end{eqnarray}
and the pfaffians of these matrices are both unity.
Then the limit $z/M \rightarrow 0$ is well-defined
and the right-handed measure is indeed saturated completely.
The Majorana-Yukawa couplings in $S_{\rm Ov/Mi}$ and 
$S_{\rm Ov}$ have precisely the latter structure.


As argued in
sections~\ref{sec:path-integration-measure} 
and \ref{sec:saturation-right-handed-measure},
in the model $S_{\rm Ov}$
the functional pfaffian is real positive semi-definite 
in the weak gauge-coupling limit, where the link variables are set to unity, $U(x,\mu)=1$, 
and the pfaffian path-integration over the spin fields
is non-vanishing. 
Moreover,
the correlation functions of
the right-handed fields $\psi_+(x), \bar \psi_+(x)$
are short-ranged,
and 
the spin fields $E^a(x), \bar E^a(x)$ are in the disordered phase.
Therefore,  
the limit of large Majorana-Yukawa couplings, 
eqs.~(\ref{eq:limit-S10-y}), (\ref{eq:limit-S10-lambda})
and (\ref{eq:limit-S10-kappa}),
is indeed well-defined
and 
there exists the PMS phase in that region of the coupling-constant space of $S_{\rm Ov/Mi}$.



\subsection{cf. Recent studies on the PMS phase/Mass without Symmetry Breaking
}
\label{subsec:cf-Mass-without-SB}

As to the possible phase transitions from the PMS phase to the FM and PMW phases, 
the recent lattice studies on the PMS phase/``Mass without Symmetry Breaking''
by Ayyar and Chandrasekharan and by Catterall, Schaich and Butt 
are interesting and suggestive\cite{Ayyar:2014eua, Ayyar:2015lrd, Ayyar:2016lxq, Catterall:2015zua, Catterall:2016dzf, Catterall:2017ogi,Schaich:2017czc}.
As to the four-dimensional case, in particular, 
the authors consider the 
reduced staggered fermion model with a certain quartic interaction term, where
there exist SU(4)/SO(4) and Z$_4$ symmetries and
any quadratic mass terms are forbidden due to the symmetries. 
In the classical continuum limit within the weak coupling phase,
the reduced staggered fermion model describes sixteen Majorana fermions (= sixteen Weyl fermions) interacting through 
SU(4) $\times$ Z$_4$ symmetric
quartic (quadratic Yukawa) interaction. 
In this model, the strong-coupling limit is well-defined because the fermion measure is indeed saturated
by the quartic interaction term completely, and 
the strong coupling expansion can be formulated.
One can show in the strong coupling regime that
the fermion field and the auxiliary boson field are both massive
without any symmetry breaking
and the model is indeed in the PMS phase.
On the other hand, the path-integral weight (fermion pfaffian) can be managed to be real positive and Monte Carlo methods are applicable.
Their numerical simulations have confirmed the PMS phase.
Moreover, the authors have found
the numerical evidences for 
the very narrow FM intermediate phase 
between the PMS and PMW phases
in the four-dimensional model. 

For the purpose to decouple the mirror (Overlap/Ginsparg-Wilson) fermions,
the interest lies in the behavior of the model
deep inside the PMS phase off the phase transition.
It is still important and useful to study
the nature of the possible phase transitions from the PMS phase 
to the FM, PMW phases in our case of the SO(10) theories, because
it gives us the understanding of the relation 
between the phase of the massless left-handed (target) Weyl fields
and the phase of the massive/gapped right-handed mirror fields. 

For this purpose, let us consider the following model, 
as the simplest possible extension of $S_{\rm Ov}$, 
which is obtained from $S_{\rm Ov/Mi}$ 
by taking the limit 
(\ref{eq:limit-S10-lambda}) and
(\ref{eq:limit-S10-kappa}) and by further
reducing the degrees of freedom of the spin fields through
the identification $E^{a}(x) = \bar E^{a}$, $y=\bar y$:
\begin{eqnarray}
\label{eq:action-SO10-with-kinetic-term-z}
\tilde S_{\rm Ov}[\psi, \bar \psi, E^a]
&=&
 \sum_{x\in\Lambda} 
\bar \psi_- (x) 
D \psi_-(x)
+
 \sum_{x\in\Lambda} 
z_+ \,
\bar \psi_+ (x) 
D \psi_+(x)
\\
&-&
\sum_{x \in \Lambda} 
y \,
E^a(x) 
\{
\psi_+^{\rm T}(x) i \gamma_5 C_D {\rm T}^a \psi_+(x)
+  
\bar \psi_+(x) i \gamma_5 C_D {{\rm T}^a}^\dagger \bar\psi_+ (x)^T
\} . \nonumber 
\end{eqnarray}
We note that although the degrees of freedom of the spin fields are halved, 
this model still exactly reduces
to $S_{\rm Ov}$ in the limit $z_+/y \rightarrow 0$
because the functional pfaffian factorizes into
those of the field $\psi_+(x)$ and the anti-fields 
$\bar \psi_+(x)$ in this limit
and the latter is unity,
${\rm pf}(\bar u \,  i \gamma_5 C_D T^{a \dagger} E^a \bar u^T) =1$, independently of the spin field $E^a(x)$.

The partition function of the model $\tilde S_{\rm Ov}$
is obtained as follows.
\begin{eqnarray}
\tilde Z_{\rm Ov} &=&
\det ( \bar v D v  ) \,\,
\big\langle 
{\rm pf} \big(u^T i \gamma_5 C_{\rm D}{\rm T}^a E^a 
\big[1+(z_+/2y)^2 P_- \big] u \big) 
\big\rangle'_{E} \, .
\end{eqnarray}
This is because 
the path-integration of the right-handed anti-field 
$\bar \psi_+(x)$
can be performed explicitly and the effective action reads 
\begin{eqnarray}
\label{eq:action-SO10-with-kinetic-term-z-psibar-integrated}
\tilde S_{\rm Ov}
&=&
 \sum_{x\in\Lambda} 
\bar \psi_- (x) 
D \psi_-(x)
\nonumber
\\
&-&
\sum_{x \in \Lambda} 
y \,
E^a(x) 
\{
\psi_+^{\rm T}(x) i \gamma_5 C_D {\rm T}^a 
\big[1+(z_+/2y)^2 P_- \big]  \psi_+(x)
\} ,  
\end{eqnarray}
which has the same form as $S_{\rm Ov}$ except for the factor
$\big[1+(z_+/2y)^2 P_- \big]$. Then, by the similar
reasoning given in section~\ref{sec:saturation-right-handed-measure},
we can argue that the pfaffian 
$
{\rm pf} \big(u^T i \gamma_5 C_{\rm D}{\rm T}^a E^a 
\big[1+(z_+/2y)^2 P_- \big] u \big) 
$
is real and positive semi-definite.
Therefore, 
the path-integration of the pfaffian over the spin field
is positive definite. This holds true as long as 
$ 0 \le (z_+/2y) < \infty$:
\begin{eqnarray}
\big\langle 
{\rm pf} \big(u^T i \gamma_5 C_{\rm D}{\rm T}^a E^a 
\big[1+(z_+/2y)^2 P_- \big] u \big) 
\big\rangle'_{E} \,\,  > \,\,  0 
\qquad (  0 \le z_+/2y < \infty ) .
\end{eqnarray}
Only at the limit of the weak Majorana-Yukawa coupling,
$ z_+/2y = \infty$ or $y/ z_+ =0 \,  (z_+ = 1)$, 
the partition function can show the massless singularity
because 
the pfaffian is evaluated (formally) as 
\begin{eqnarray}
{\rm pf} \Big(u^T i \gamma_5 C_{\rm D}{\rm T}^a E^a 
\big[ P_- \big] u \Big) 
&=&
{\rm pf} \Big\{
(\bar u u)^T 
(\bar u^\dagger i \gamma_5 C_{\rm D}{\rm T}^a E^a \bar u^T)
(\bar u u)
\Big\}
\nonumber\\
&=&
\det (\bar u u) \, 
{\rm pf} 
(\bar u^\dagger i \gamma_5 C_{\rm D}{\rm T}^a E^a \bar u^T)
\nonumber\\
&=&
\det (\bar u D u) .
\end{eqnarray}
This result implies that the PMS phase extends
all the way to the limit of the weak Majorana-Yukawa coupling
$y/z_+ =0 \,  (z_+ = 1)$
and 
the FM phase is  absent
within the coupling-constant space of $\tilde S_{\rm Ov}$.

\begin{figure}[t]
\begin{center}
\includegraphics[width =75mm]{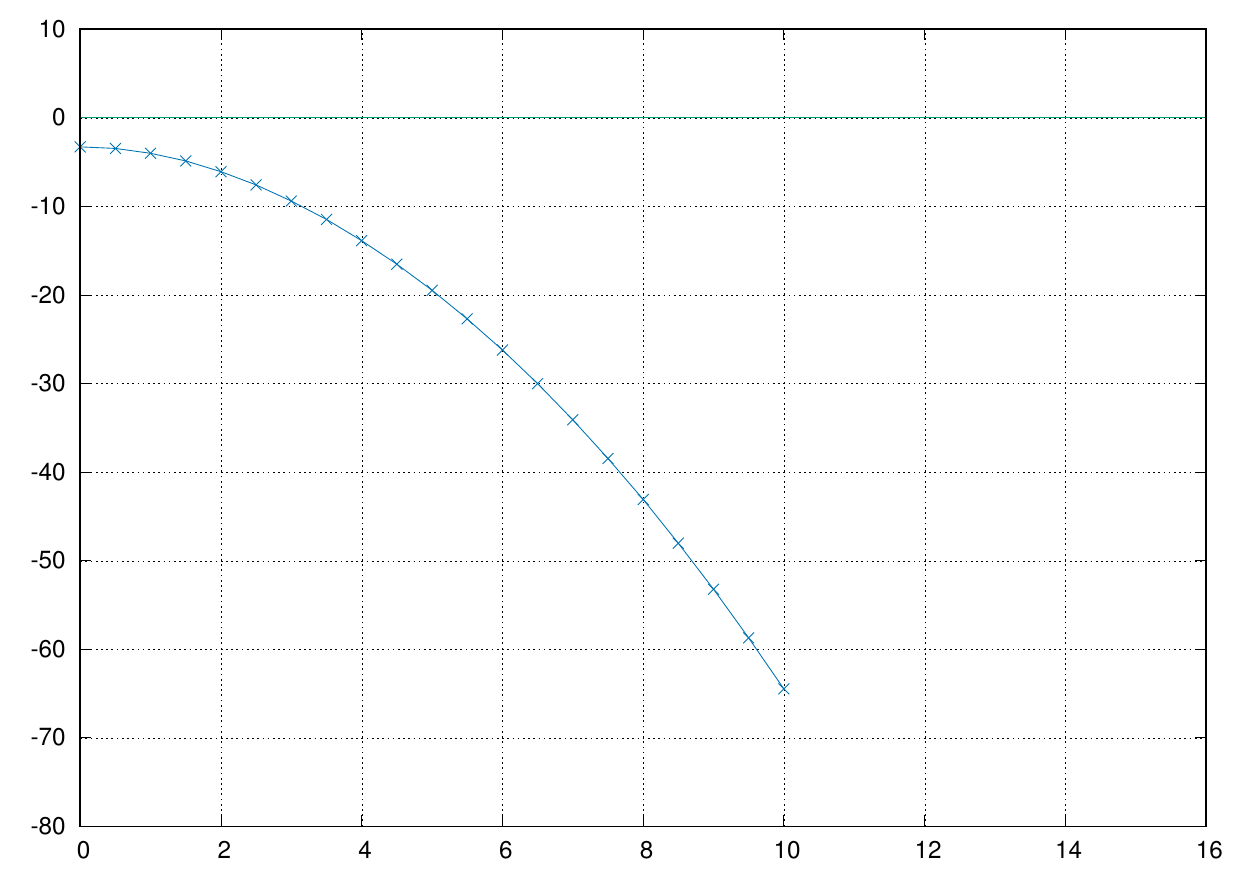} 
\caption{
$f(m_0, z_+, y)$ vs. $z_+$ ($y=1$):
The consistency condition for the 
SO(10) symmetry breaking in the effective spin model of 
$\tilde S_{\rm Ov}$
within the saddle point analysis in the spirit of the large N
expansion.
}
\label{fig:GapEquationSO10_z}
\end{center}
\end{figure}

\begin{figure}[t]
\begin{center}
\includegraphics[width =70mm]{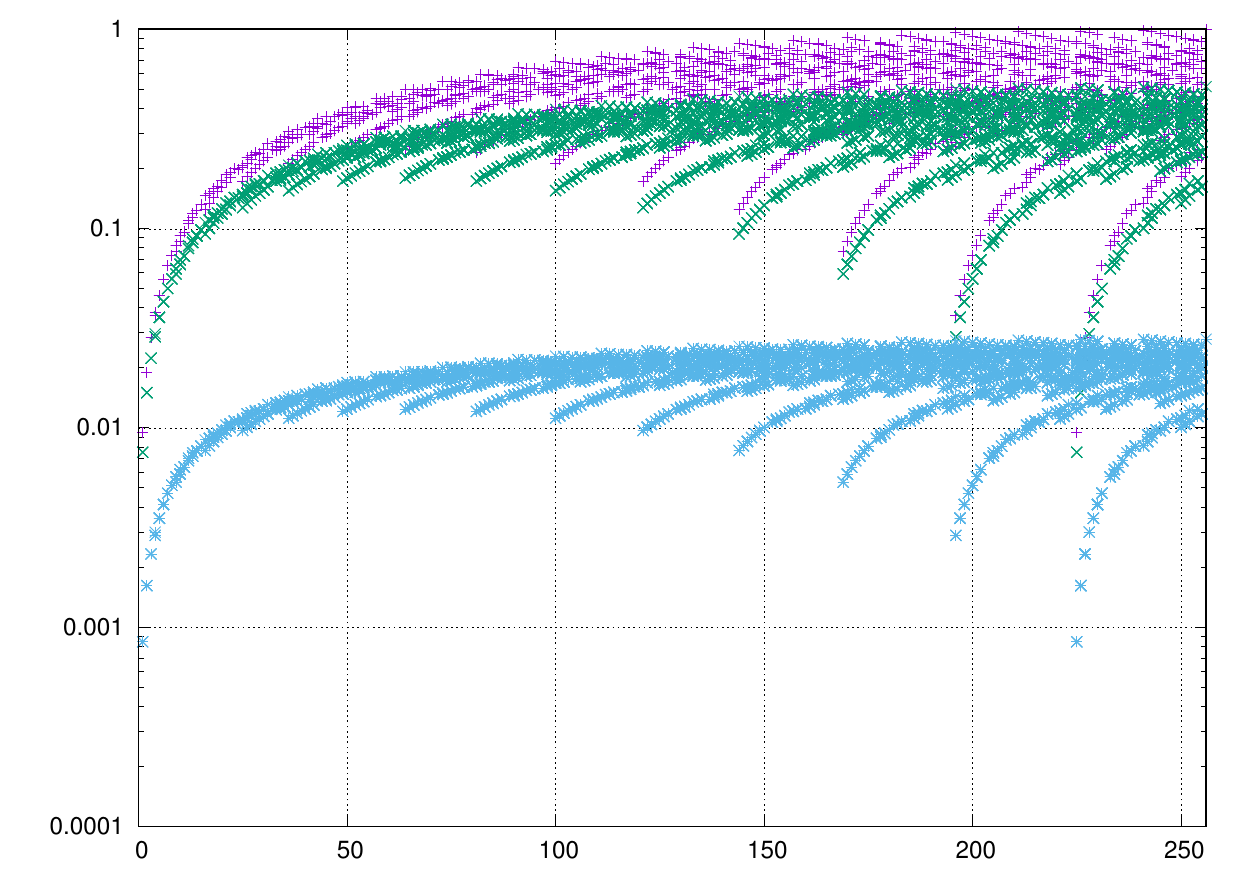} 
\includegraphics[width =70mm]{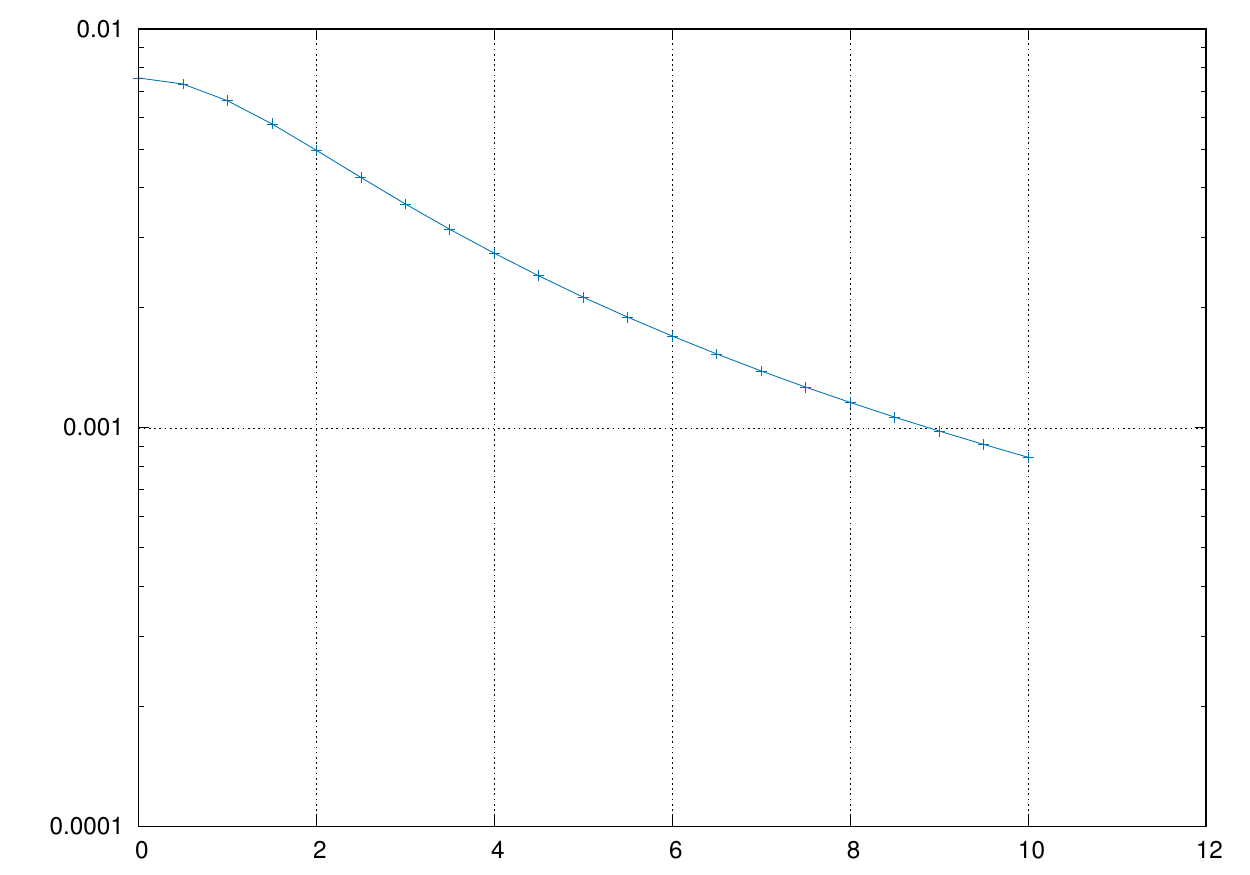} 
\caption{
[left]: 
$\tilde \Gamma(k)$
as the function of $k^2/ (2\pi/L)^2$
for $z_+=0$ (green) and $z_+=10.0$ (blue) \, ($y=1$)
in comparison with
$(1/16) \sum_\mu 4 \sin^2 (k_\mu/2)$ (purple).
[right]: 
$\tilde \Gamma(k) \vert_{ k^2 = (2\pi/L)^2 }$
in logarithmic scale as the function of $z_+$ \,($y=1$).
}
\label{fig:GapEquationSO10_Eprop}
\end{center}
\end{figure}

We can confirm the above result by 
applying 
the saddle point analysis in the spirit of the large $N$ expansion to this model, just as discussed in section~\ref{subsec:disorder-nature-spin-field}.
The effective action of the spin field $X^a(x) (= X^a_0 + \tilde X^a(x))$ and the Lagrange
multiplier field $\lambda(x) (= \lambda^a_0 + \tilde \lambda^a(x))$ in this case is given by
\begin{eqnarray}
S[ X^a, \lambda] 
&\equiv& 
- \frac{1}{2} 
\ln \det \left( 
\begin{array}{cc}
\big(u^{\rm T} \,  i \gamma_5 C_D {\rm T}^a X^a u \big) &
(z_+/2y)(u^T \bar u^T) \\
-(z_+/2y)(\bar u u) &
\big(\bar u \,  i \gamma_5 C_D {\rm T}^{a \dagger} X^a \bar u^T \big) 
\end{array}
\right)
\nonumber\\
&& 
 - i \sum_x \lambda(x)(X^a(x) X^a(x) - 1) 
\\
&=&
- 32 \sum_p \ln \big( 
X^a_0 X^a_0 +(z_+/2y)^2 \tilde \Delta(p)
\big)^{1/2}  
\nonumber\\
&&
- i  \lambda_0 (X^a_0 X^a_0 - 1) V 
-i \sum_x 2 \tilde \lambda(x)  X^a_0 \tilde X^a(x) 
\nonumber\\
&&
+ \sum_{x,y} \tilde X^a(x) \big\{
 4 B(x-y) ( 2 X^a_0 X^b_0 - X^c_0 X^c_0 \, \delta^{ab})
\nonumber\\
&&\qquad\qquad\qquad\qquad
- 4 A(x-y) (z_+/2y)^2 \, \delta^{ab} 
-i \lambda_0 \delta_{xy} \delta^{ab}\big\}  \tilde X^b(x) 
\nonumber\\
&& + \cdots ,
\end{eqnarray}
where $B(x-y)$ and $A(x-y)$ are the kinetic operators defined 
through the chiral projector
$\hat P_{0 + }(x,y)$ of the single free overlap Dirac fermion 
as
\begin{eqnarray}
B(x-y)
&=&
{\rm tr}\Big[
\hat P_{0 +}\frac{1}{X^a_0 X^a_0 +(z_+/2y)^2 \Delta}(x-y) 
\hat P_{0 +}\frac{1}{X^a_0 X^a_0 +(z_+/2y)^2 \Delta}(y-x)
\Big]
\nonumber\\
&+&
{\rm tr}\Big[
 P_{ -}\frac{1}{X^a_0 X^a_0 +(z_+/2y)^2 \Delta}(x-y) 
 P_{ -}\frac{1}{X^a_0 X^a_0 +(z_+/2y)^2 \Delta}(y-x)
\Big]
\nonumber\\
&=&
\frac{1}{V} \sum_k {\rm e}^{i k (x-y)} \,
\frac{1}{V} \sum_q
\Big\{
3 + 
\frac{\sin(q+k)_\mu \sin q_\mu + b(q+k) b(q)}
{\omega(q+k) \omega(q)}
\Big\} \times 
\nonumber\\
&&\qquad\qquad\qquad\qquad\quad
\frac{1}{X^a_0 X^a_0 +(z_+/2y)^2 \tilde \Delta(q+k)}
\frac{1}{X^a_0 X^a_0 +(z_+/2y)^2 \tilde \Delta(q)}   \, \,, 
\nonumber\\
&&\\
A(x-y)
&=&
{\rm tr}\Big[
P_{ -}\frac{1}{X^a_0 X^a_0 +(z_+/2y)^2 \Delta}\hat P_{0 +}(x-y) 
\hat P_{0 +}\frac{1}{X^a_0 X^a_0 +(z_+/2y)^2 \Delta}P_{ -}(y-x)
\Big]
\nonumber\\
&+&
{\rm tr}\Big[
\hat P_{0 +}\frac{1}{X^a_0 X^a_0 +(z_+/2y)^2 \Delta} P_{ -}(x-y) 
 P_{ -}\frac{1}{X^a_0 X^a_0 +(z_+/2y)^2 \Delta}\hat P_{0 +}(y-x)
\Big]
\nonumber\\
&=&
\frac{1}{V} \sum_k {\rm e}^{i k (x-y)} \,
\frac{1}{V} \sum_q
\Big\{
\frac{\sin(q+k)_\mu \sin q_\mu +
(\omega(q+k)+b(q+k))(\omega(q) +b(q))}
{\omega(q+k) \omega(q)}
\Big\} \times 
\nonumber\\
&&\qquad\qquad\qquad\qquad\quad
\frac{1}{X^a_0 X^a_0 +(z_+/2y)^2 \tilde \Delta(q+k)}
\frac{1}{X^a_0 X^a_0 +(z_+/2y)^2 \tilde \Delta(q)}   \, \,, 
\nonumber\\
\end{eqnarray}
and
\begin{eqnarray}
\hat P_{0 +}(x-y) &=&
\frac{1}{V} \sum_q  {\rm e}^{i q (x-y)}
\left\{ 
\frac{1}{2} - 
\frac{1}{2} 
\gamma_5 \frac{i \gamma_\mu \sin q_\mu + b(q)}{\omega(q)}
\right\} ,
\\
\Delta(x-y) &=&
\frac{1}{V} \sum_q  {\rm e}^{i q (x-y)}
\left\{ \frac{\omega(q) + b(q)}{2 \, \omega(q)}
\right\} .
\end{eqnarray}
%
The path-integration of the fluctuations  $\tilde \lambda(x)$ and $\tilde X^a(x)$ (in this order) 
gives the effective action of $X^a_0$ and 
$\lambda_0$ as follows,
\begin{eqnarray}
\label{eq:effective-action-X0-lambda0}
S_{\rm eff}[ X^a_0, \lambda_0] 
&=& 
\big\{ 
- 32 \sum_p \ln \big( 
X^a_0 X^a_0 +(z/2)^2 \tilde \Delta(p)
\big)^{1/2}  
 - i  \lambda_0 (X^a_0 X^a_0 - 1)
\big\} V 
\nonumber\\
&& \qquad
+ \frac{(10-1)}{2} \sum_{ k \not = 0}\ln \big\{-4 \tilde D(k \, ; X_0)
-i \lambda_0  \big\} ,
\end{eqnarray}
where $\tilde D(k \, ; X_0)$ is the fourier transform of the kinetic
operator, 
\begin{eqnarray}
D(x-y \, ; X_0) &=& X_0^c X_0^c B(x-y) + (z_+/2y)^2 A(x-y) .
\end{eqnarray}
%
The stationary conditions for $X^a_0$ and $\lambda_0$ are given by
\begin{eqnarray}
0 &=& 2 \, X^a_0 
\Biggl\{ 
- 16 \frac{1}{V}\sum_p \frac{1}{X^c_0 X^c_0 +(z_+/2y)^2 \tilde \Delta(p)} 
-i \lambda_0
\nonumber\\
&& \qquad\qquad
+
\frac{9}{2}\frac{1}{V} \sum_{k \not = 0} 
\Biggl(
-4 \frac{\partial}{\partial X_0^2} \tilde D(k \, ; X_0)
\frac{1}{-4 \tilde D(k \, ; X_0)  -i \lambda_0} 
\Biggr) 
\Biggr\}  , \nonumber\\
&&\\
0 &=& \big( X^c_0 X^c_0 - 1 \big)
+ \frac{9}{2}\frac{1}{V} \sum_{k \not = 0}
 \frac{1}{-4 \tilde D(k \, ; X_0)  -i \lambda_0 } .
\end{eqnarray}
Assuming $X^a_0 \not = 0$,
the above conditions imply that
\begin{eqnarray}
&& - i \lambda_0 = 
16 \frac{1}{V}\sum_p \frac{1}{X^c_0 X^c_0 +(z_+/2y)^2 \tilde \Delta(p)} 
\nonumber\\
&& \qquad\qquad
-\frac{9}{32}\frac{1}{V} \sum_{k \not = 0} 
\Biggl(
-\frac{\partial}{\partial X_0^2} \tilde D(k\, ; X_0)
\frac{16}{- \tilde D(k \, ; X_0) + \tilde D(0 \, ; X_0) } 
\Biggr) 
\Bigg\vert_{X_0^c X_0^c = 1} , \nonumber\\
&&\\
\label{eq:condition-X0X0-tilde-S-Ov}
&& 
X^c_0 X^c_0 
= 1 - \frac{9}{32}\frac{1}{V} \sum_{k \not = 0} 
\frac{4}{-\tilde D(k \, ; X_0) + \tilde D(0 \,; X_0)}\, \bigg\vert_{X_0^c X_0^c = 1}  ,
\end{eqnarray}
where
the leading results, $- i \lambda_0 =16 \frac{1}{V}\sum_p \frac{1}{X^c_0 X^c_0 +(z_+/2y)^2 \tilde \Delta(p)} $ and $X^c_0 X^c_0 = 1$, are substituted 
in the terms suppressed by the factor $N /2^{\frac{N}{2}} (\simeq 9/32  )$.
The r.h.s. of the condition eq.~(\ref{eq:condition-X0X0-tilde-S-Ov}) is required to be positive for $X^a_0 \not = 0$.
It is plotted  in fig.~\ref{fig:GapEquationSO10_z}
as the function of $z_+$ for $y=1$ and $m_0 = 1$,
\begin{equation}
\label{eq:concistency-condition-X0X0-tilde-S-Ov}
 f(m_0, z_+, y) \equiv 1 - \frac{9}{32}\frac{1}{V} 
\sum_{k \not = 0} 
\frac{4}{-\tilde D(k \, ; X_0) + \tilde D(0 \, ; X_0)}\, \bigg\vert_{X_0^c X_0^c = 1}  .
\end{equation}
One can see that 
$f(m_0, z_+, y) < 0 $ 
for $ z_+ \ge 0 $ up to $z_+ \simeq 10$ \, ($y=1$, $m_0=1$)
and it is in contradiction with the assumption $X^a_0 \not = 0$.
In this region of the coupling $z_+$, 
the fluctuation of the spin field $E^a(x)$
is too large to maintain the non-zero expectation value
of the spin field  $ \langle E^a(x) \rangle $,
and we do not see any evidence for the FM phase.
In fig.~\ref{fig:GapEquationSO10_Eprop}, 
the kinetic term of the fluctuation modes $\tilde X^a(x)$, 
given by $\tilde \Gamma(k) \equiv (1/4)
\big[-\tilde D(k \, ; X_0) + \tilde D(0 \, ; X_0)\big]
\vert_{X_0^2 = 1}$,
is shown as the function of $k^2/ (2\pi/L)^2$
in comparison with 
$(1/16) \sum_\mu 4 \sin^2 (k_\mu/2)$.
One can see that
the kinetic term looks like the canonical form
of the free theory and 
the normalization of the kinetic term, defined by
the value of $\tilde \Gamma(k)$ at $k^2 = (2\pi/L)^2$,
is decreasing monotonically and 
exponentially in $z_+ / y$ \, $(y=1)$.
These results support the picture that the PMS phase extends
all the way to the limit of the weak Majorana-Yukawa coupling
$y/z_+ =0$
and that the FM and PMW phases are absent
within the coupling space of $\tilde S_{\rm Ov}$.

%

\begin{figure}[t]
\begin{center}
\includegraphics[width =75mm]{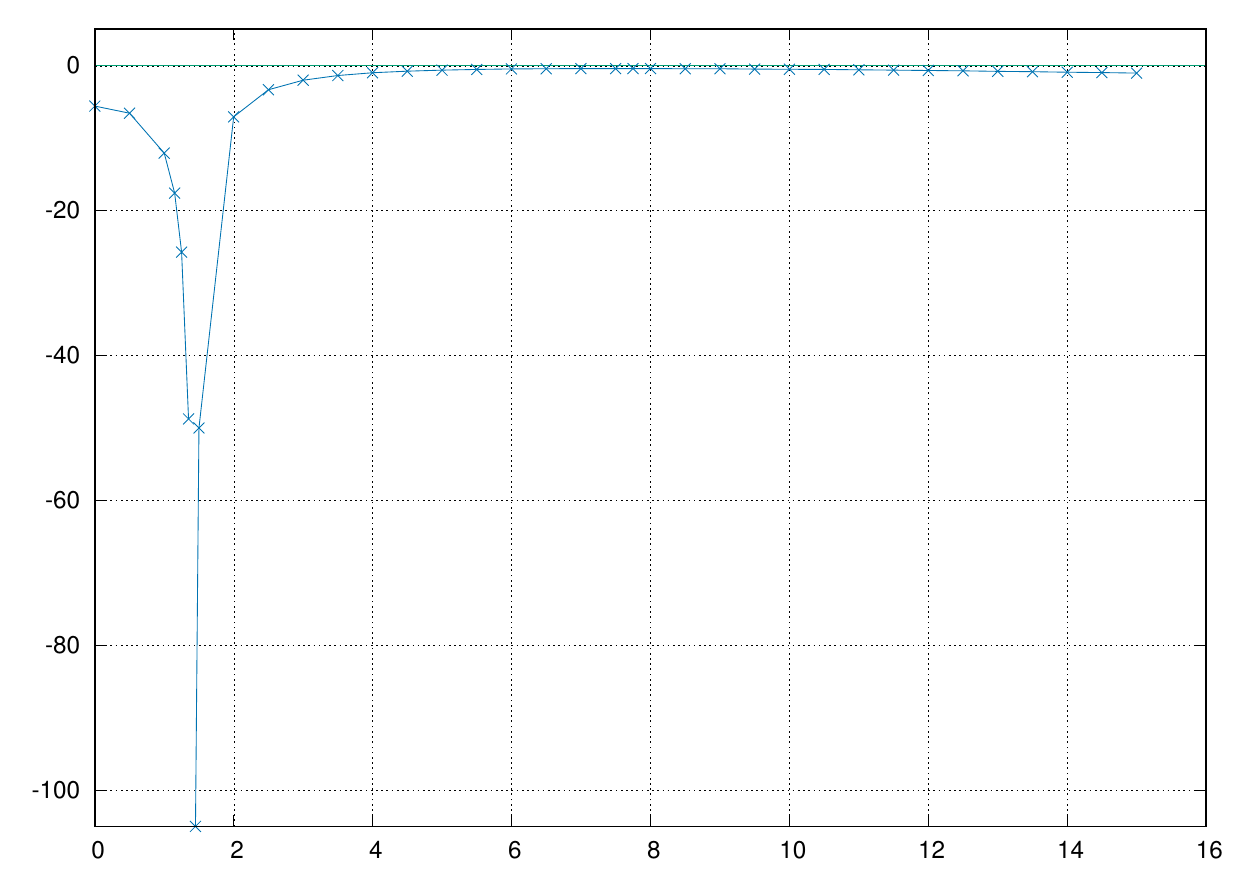} 
\includegraphics[width =75mm]{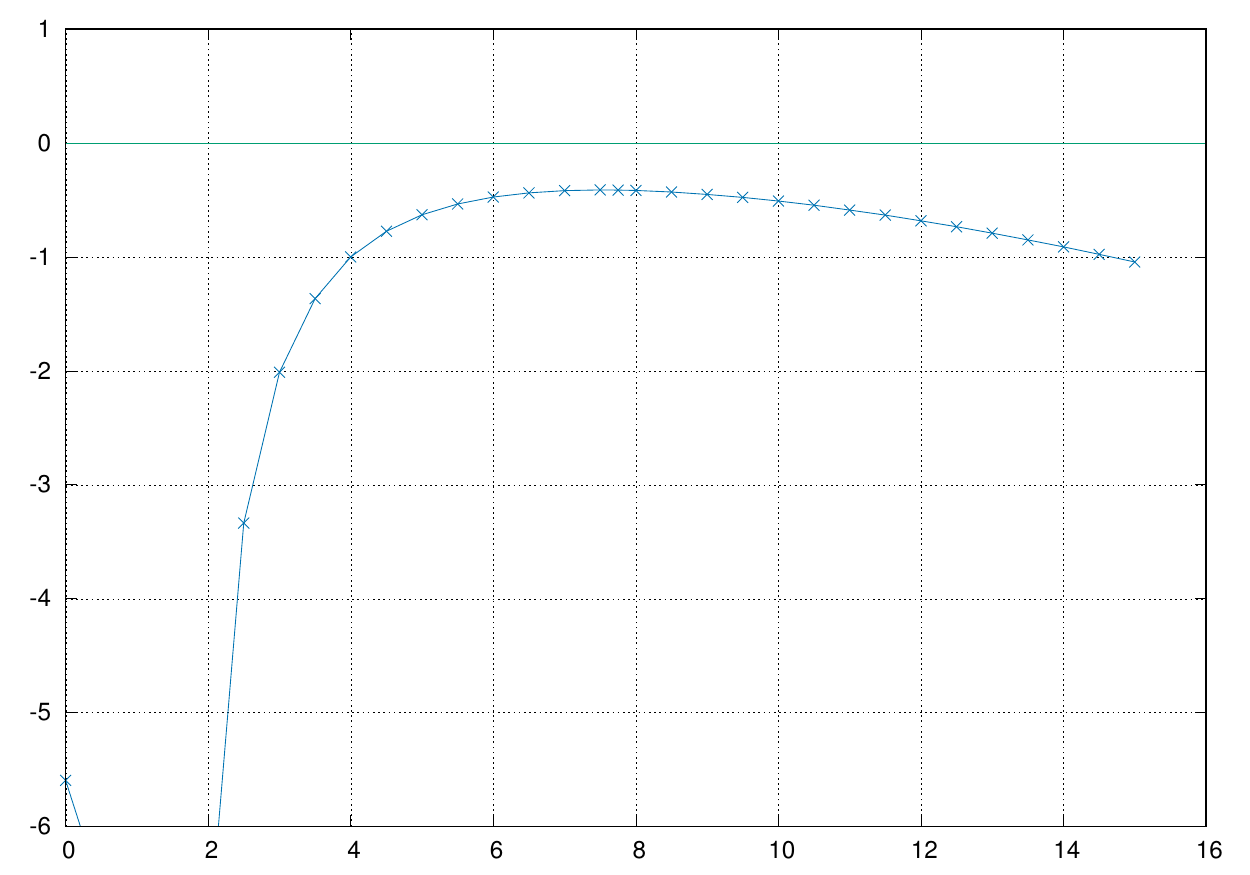} 
\caption{
$f(m_0, z_+, y, w)$ vs. $z_+$ ($y=w=1$):
The consistency condition for the 
SO(10) symmetry breaking in the effective spin model of 
$\tilde S_{\rm EP/WY}$
within the saddle point analysis in the spirit of the large N
expansion.
}
\label{fig:GapEquationSO10_EPWY}
\end{center}
\end{figure}

Given the result that the FM phase is absent 
in the model $\tilde S_{\rm Ov}$,
it is instructive to compare the above result
to 
the situation 
in 
the original Eichten-Preskill models
$S_{\rm EP}$ and $S_{\rm EP/WY}$\cite{Eichten:1985ft, Golterman:1992yha},
where
it was argued by Golterman, Petcher and Rivas\cite{Golterman:1992yha}
that 
the crucial ingredient for the failure of the proposal
is the existence of the FM phase which 
separates the PMS and PMW phases.
In fact, as mentioned before, 
there is an important technical difference in our models,
which is relevant for preserving the (global) SO(10) symmetry:
we choose the product function $F(\omega)$ as given by eq.~(\ref{eq:def-of-F}) and therefore
use the unit SO(10)-vector spin fields, 
omitting their kinetic(hopping) terms. 
This corresponds to 
taking the limits eqs.~(\ref{eq:limit-S10-lambda}) 
and (\ref{eq:limit-S10-kappa}) in $S_{\rm EP/WY}$ 
as given in eq.~(\ref{eq:EP-WY-with-spin-field-E}).
This region in the coupling-constant space of $S_{\rm EP/WY}$
has not been explored by Golterman, Petcher and Rivas.
In applying the saddle point analysis in the spirit of the large N expansion,  
we further
reduce the degrees of freedom of the spin fields through
the identification $E^{a}(x) = \bar E^{a}$, $y=\bar y$, 
$w=\bar w$ and consider the following model.
\begin{eqnarray}
\tilde S_{\rm EP/WY}
&=&
 \sum_{x\in\Lambda} 
\big\{
\bar \psi (x) 
  \gamma_\mu P_-  ([\nabla_\mu - \nabla_\mu^\dagger]/2)
\psi(x)
+
z_+ 
\bar \psi (x) 
 \gamma_\mu P_+  ([\nabla_\mu - \nabla_\mu^\dagger]/2)
 \psi(x)
\big\}
\nonumber\\
&-&
\sum_{x \in \Lambda} 
y \,
E^a(x) \, 
  \{ 
\psi^{\rm T}(x) i \gamma_5 C_D {\rm T}^a P_+ \psi(x)
+ 
\bar \psi(x) P_- i \gamma_5 C_D {{\rm T}^a}^\dagger \bar\psi (x)^T
\} 
\nonumber\\
&-&
\sum_{x \in \Lambda}  
w \,
E^a(x) \,
\{ 
\psi^{\rm T}(x) i \gamma_5 C_D {\rm T}^a 
(\nabla_\mu\nabla_\mu^\dagger/2) 
P_+ \psi(x)
\nonumber\\
&&
\qquad\qquad\qquad
+ 
\bar \psi(x) P_- i \gamma_5 C_D {{\rm T}^a}^\dagger
(\nabla_\mu\nabla_\mu^\dagger/2) 
 \bar\psi (x)^T
\} .
\end{eqnarray}
%
%
In this model, the consistency condition for 
$\langle X^a(x) \rangle \not = 0$
is given by 
\begin{equation}
\label{eq:concistency-condition-X0X0-tilde-S-EPWY}
 f(m_0, z_+, y, w) \equiv 1 - \frac{9}{32}\frac{1}{V} 
\sum_{k \not = 0} 
\frac{4}{-\tilde D'(k \, ; X_0) + \tilde D'(k^0 \, ; X_0)}\, \bigg\vert_{X_0^c X_0^c = 1}  \, \, > \, \, 0 ,
\end{equation}
where $k^0_\mu = 0$ or $\pi_\mu^{(15)}$ depending
on the value of the couplings $z_+, y, w$.
$\tilde D(k \, ; X_0)$ is the fourier transform of the kinetic operator, 
\begin{eqnarray}
D'(x-y \, ; X_0) &=& X_0^c X_0^c B'(x-y) + (z_+/2)^2 A'(x-y) ,
\end{eqnarray}
where $B'(x-y)$ and $A'(x-y)$ are defined 
by
\begin{eqnarray}
B'(x-y)
&=&
\frac{1}{V} \sum_k {\rm e}^{i k x} \,
\frac{1}{V} \sum_q
\Big\{
\tilde W(q+k)^2 + \tilde W(q)^2 + 2 \tilde W(q+k) \tilde W(q)
\Big\} \times 
\nonumber\\
&&\qquad
\frac{\tilde W(q+k)}
{X^a_0 X^a_0 \tilde W(q+k)^2 +(z_+/2)^2 \sin^2(q+k)}
\frac{\tilde W(q)}
{X^a_0 X^a_0 \tilde W(q)^2 +(z_+/2)^2 \sin^2(q)} \, \,, 
\nonumber\\
&&\\
A'(x-y)
&=&
\frac{1}{V} \sum_k {\rm e}^{i k x} \,
\frac{1}{V} \sum_q
\Big\{
\tilde W(q+k)^2 + \tilde W(q)^2 + 2 \tilde W(q+k) \tilde W(q)
\Big\} \times 
\nonumber\\
&&\qquad
\frac{\sin(q+k)_\mu}{X^a_0 X^a_0 \tilde W(q+k)^2 +(z_+/2)^2 \sin^2(q+k)}
\frac{\sin q_\mu}{X^a_0 X^a_0 \tilde W(q)^2 +(z_+/2)^2 \sin^2(q)},
\nonumber\\
\end{eqnarray}
and $W = y + (w/2) \nabla_\mu^\dagger \nabla_\mu$.
In fig.~\ref{fig:GapEquationSO10_EPWY}, 
$f(m_0, z_+, y, w)$ is plotted as the function of $z_+$ for $y=w=1$ and $m_0=1$.
%
The singular behavior of the plots around 
$z_+ \simeq 1.4$ indicates the fact that
at a certain critical value $z_+ = z_+^c (\simeq 1.4)$
the kinetic operator
degenerates: 
$\tilde D'(0) =\tilde D'(k) = \tilde D'(\pi^{(15)})$,
where $\tilde D'(k) \equiv 
\tilde D'(k \, ; X_0)\big\vert_{X_0^2=1}$.
For $z_+ < z_+^c$, 
$\tilde D'(k) \le \tilde D'(\pi^{(15)})$ and the saddle point
is assumed to be Anti-Ferromagnetic, $\langle X^a(x) \rangle = X_0^a (-1)^{x_\mu}$.
For $z_+ > z_+^c$, 
$\tilde D'(k) \le \tilde D'(0)$ and the saddle point
is assumed to be Ferromagnetic, $\langle X^a(x) \rangle = X_0^a$. In both cases,
$f(m_0, z_+, y, w) < 0 $ 
and 
the fluctuation of the spin field $E^a(x)$
is too large to maintain the non-zero expectation value
of the spin field  $\langle E^a(x) \rangle$.
Thus 
the model is in the PMS phase
in the entire region of the coupling $z_+$ ($y=1$)
up to $z_+ \simeq 15$.  

In the case of the model $\tilde S_{\rm Ov}$, 
we found that 
$\tilde D(k) \le \tilde D(0)$ for the entire region $z_+ \ge 0$, 
where $\tilde D(k) \equiv  \tilde D(k \, ; X_0)\big\vert_{X_0^2=1}$. 
And 
the saddle point is assumed to be Ferromagnetic, $\langle X^a(x) \rangle = X_0^a$.
Therefore the coupling-constant space of the model $\tilde S_{\rm Ov}$ 
should correspond to the region of the weaker  Majorana-Yukawa coupling,  
$z_+ > z_+^c$, within the coupling-constant space of the model $\tilde S_{\rm EP/WY}$. 
This fact is also supporting the picture that the PMS phase extends
all the way to the limit of the weak Majorana-Yukawa coupling
$y/z_+ =0$ in our SO(10) model  $\tilde S_{\rm Ov}$.

\begin{figure}[htb]
\begin{center}
\includegraphics[width =75mm]{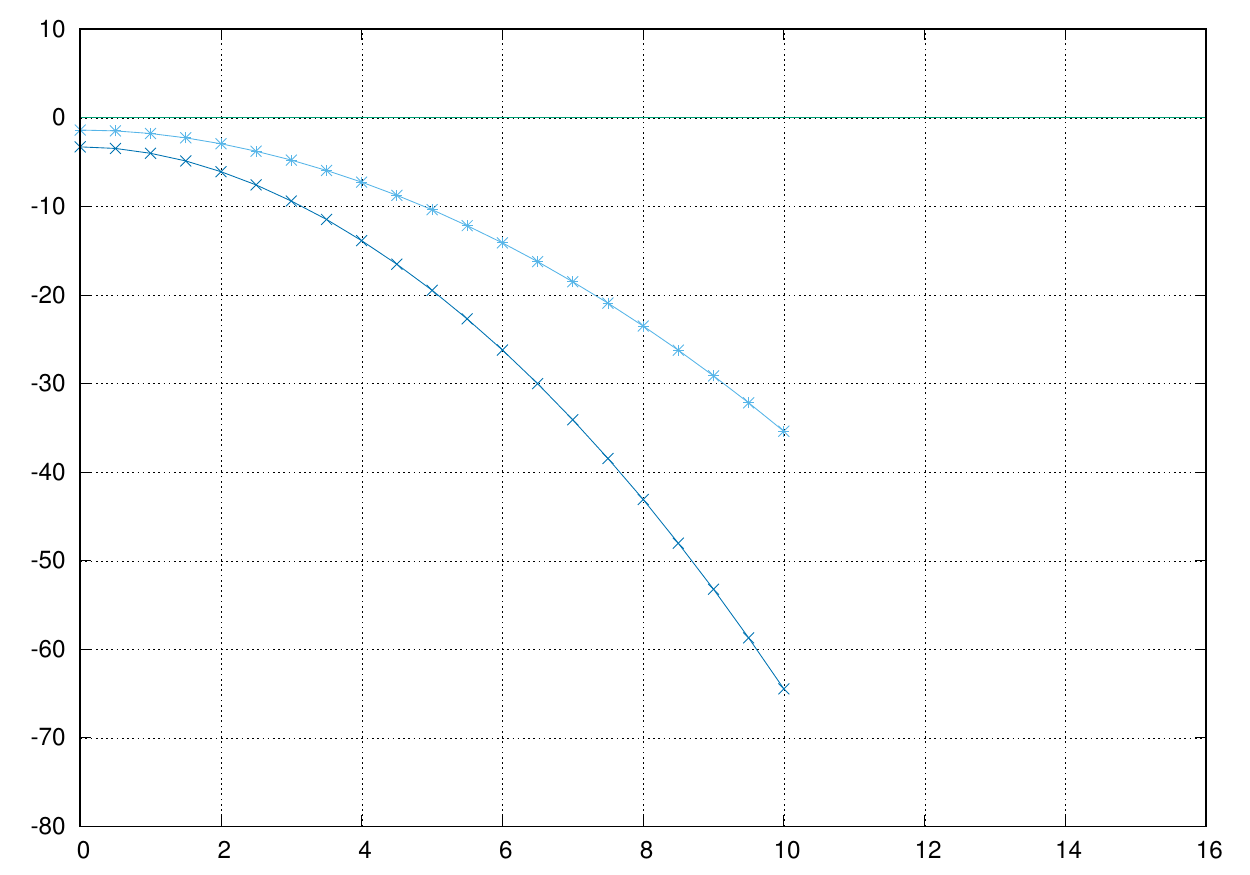} 
\includegraphics[width =75mm]{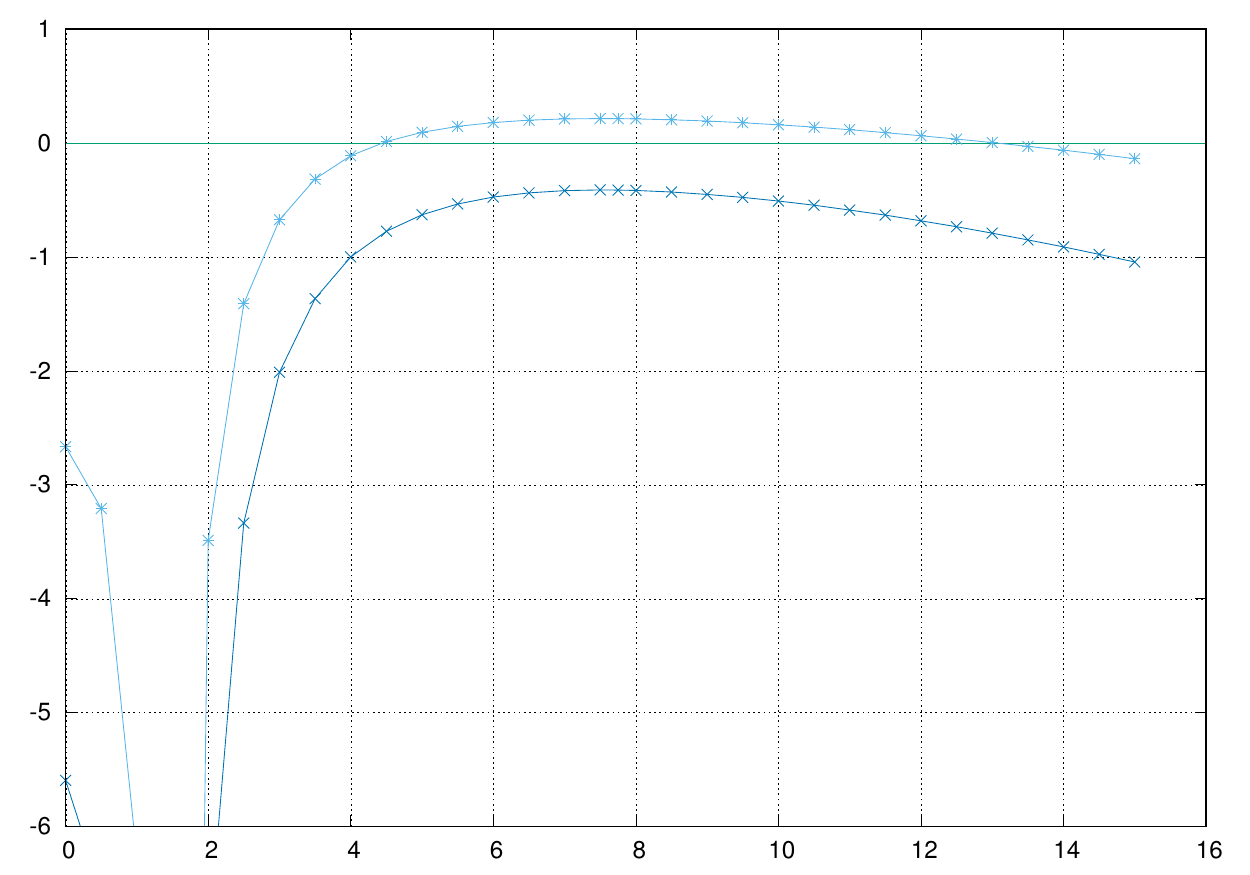} 
\caption{$f(m_0, z_+, y)$ and $f(m_0, z_+, y, w)$ vs. $z_+$ ($y=1$ and $y=w=1$, respectively):
The consistency condition for the 
SO(6) symmetry breaking in the effective spin models of 
$\tilde S_{\rm Ov}$ and 
$\tilde S_{\rm EP/WY}$
in the cases with the reduced symmetry SO(6).
}
\label{fig:GapEquationSO10_SO6_EPWY}
\end{center}
\end{figure}

One may study these models,  $\tilde S_{\rm Ov}$ and
$\tilde S_{\rm EP/WY}$,  as the counter part of the 
reduced staggered fermion model with 
the SU(4)/SO(4) and Z$_4$ symmetries by reducing
SO(10) symmetry to SO(6)(Spin(6)=SU(4)) and
SO(4)(=SU(2)$_+$ $\times$ SU(2)$_-$) 
(or SO(3)(=SU(2)$_+$)).\footnote{For the models with the
symmetries reduced from SO(10), 
one may reformulate
the model $S_{\rm Ov}$ more simply in terms of overlap Majorana fields as
\begin{eqnarray}
\tilde S_{\rm Ov/Mj}[\psi,  E^{a'}]
&=&
 \sum_{x\in\Lambda} 
\big\{
z \, \psi (x)^T \check {\rm C} C_D 
D \psi(x) 
 -
y \,
E^{a'}(x) 
\psi(x)^T i \gamma_5 C_D \check{\rm T}^{a'} \psi(x)
\} .
\end{eqnarray}
In both cases, one can show rigorously 
that
the path-integral measures of the fermion fields 
$\psi_+(x), \bar \psi_+(x)$ and $\psi(x)$ are saturated completely in the limits $z/y \rightarrow 0$.
%
}
In fig.~\ref{fig:GapEquationSO10_SO6_EPWY},
the results of the similar saddle point analysis
for the models with the SO(6) symmetry are shown
in comparison with those with the SO(10) symmetry.
In the model $\tilde S_{\rm Ov}$,  
the FM phase does not appear for the SO(6) symmetry
and
the PMS phase extends all the way to the limit of the 
weak Majorana-Yukawa coupling up to $y/z_+ =0.1$. 
In the model $\tilde S_{\rm EP/WY}$, the FM phase
appears between the PMS and PMW phases. It is because
the effect of the fluctuation modes 
$\tilde X^{a'}(x) (a'=1, \cdots, 6)$
is reduced by the factor $(6-1)/(10-1)$ in the consistency
condition eq.~(\ref{eq:concistency-condition-X0X0-tilde-S-EPWY}).
But the effect still remains rather large. 
Then the full quantum fluctuations
can reduce the region of the coupling-constant $z_+$ 
where the FM phase appears, 
restoring the broken SO(6) symmetry.
Thus our results here seems quite consistent with the observations
and arguments made by these authors
about 
the reduced staggered fermion model with 
the quartic interaction term which respects
the SU(4)/SO(4) and Z$_4$ symmetries,
and about ''Mass without Symmetry Breaking''.




\subsection{cf. Domain wall fermions with the boundary Eichten-Preskill term 
}
\label{subsec:cf-DW-EP}
 
In the proposal by Creutz, Tytgat, Rebbi, Xue\cite{Creutz:1996xc}
to formulate the standard model plus the right-handed neutrinos
by the domain wall fermion, 
the authors have considered 
the quartic term with the symmetry SU(4) $\times$ SU(2)$_L$ $\times$ SU(2)$_R$ as boundary interaction terms.
In fact, this type of the boundary interaction term can be obtained
from the SO(10) interaction term by 
reducing the symmetry to SO(6) $\times$ SO(4) (= SU(4) $\times$ SU(2)$_L$ $\times$ SU(2)$_R$).
Then, it is straightforward to lift their proposal to the SO(10) chiral gauge theory. 

In fact, we can show that the following action 
defines such a domain wall fermion model
for the SO(10) chiral gauge theory:
\begin{eqnarray}
\label{eq:action-DW-EP-SO10}
S_{\rm DW/Mi} &=& \sum_{t=1}^{L_5} \sum_{x \in \Lambda}
\bar \psi(x,t) \big\{ [1+ a'_5 (D_{4 \rm w} - m_0)] \delta_{t t'}  - P_- \delta_{t+1, t'} -P_+ \delta_{t,t'+1} \big\} \psi(x, t') 
\nonumber\\
&+&
\sum_{x \in \Lambda}
(z_+ -1) \, \bar \psi(x, L_5)P_- [1+a'_5(D_{4 \rm w} - m_0) ] \psi(x,L_5)
\nonumber\\
&-&
\sum_{x \in \Lambda}  \{ 
y \,
X^a(x) 
\psi^{\rm T}(x, L_5) 
 i \gamma_5 C_D {\rm T}^a 
\psi(x, L_5)
\nonumber\\
&& \qquad\qquad\qquad
+ 
 \bar y \,
\bar X^a(x)
\bar \psi(x, L_5) P_- i \gamma_5 C_D {{\rm T}^a}^\dagger 
\bar \psi(x, L_5)^T
\} 
\nonumber\\
&+&
S_{X}[X^a] ,
\end{eqnarray}
where the Dirichlet b.c. is assumed, 
\begin{eqnarray}
&&
P_+ \psi(x,0) = 0, \quad \bar \psi(x,0) P_- = 0 
\quad ; \quad
P_- \psi(x, L_5 + 1) =0, \quad \bar \psi(x, L_5 +1) P_+ = 0 ,
\nonumber\\
\end{eqnarray}
and
$a'_5 (= a_5 /a)$ is the lattice spacing of extra dimension in the lattice unit.
In this action, the second term in the r.h.s. is introduced so that all the terms which involve the field 
$ \bar \psi(x, L_5) P_- (= \bar q_+(x))$ in the original action of the domain wall fermion (the five-dimensional Wilson fermion) are rescaled by the factor $z_+$ and made vanished in the limit 
$z_+ \rightarrow 0$. Then the forth term with the Yukawa coupling 
$\bar y$ is required so that it saturates the path-integral measure of that field 
$ \bar \psi(x, L_5) P_-$.
On the other hand,  the field $\psi(x,L_5)$ is 
related to the (truncated) overlap fermion field $\psi(x)$ by the relation 
$\psi(x,L_5) = (- \gamma_5)\big(1+e^{a_5 L_5 \widetilde H}\big)^{-1} \psi(x)$ in the subtraction scheme using the five-dimensional Wilson fermion
subject to the anti-periodic b.c., 
and is projected
to the right-handed Weyl field $\psi_+(x) = \hat P_+ \psi(x)$ in the limit $L_5 \rightarrow \infty$ (plus $a_5 \rightarrow 0$)\cite{Kikukawa:1999sy}. Thus it ends up
with the overlap fermion model with $S_{\rm Ov/Mi}$
in the limit 
(\ref{eq:limit-S10-lambda}) and
(\ref{eq:limit-S10-kappa}), 
which is the similar model with $\tilde S_{\rm Ov}$, but
before 
reducing the degrees of freedom of the spin fields through
the identification $E^{a}(x) = \bar E^{a}$, $y=\bar y$.
The global U(1) symmetry of the five-dimensional fermion 
fields is broken to $Z_4$ by the boundary Yukawa couplings.
The CR$_5$, P, and CPR$_5$ symmetries 
are all broken by the boundary Yukawa couplings and the term
with the coupling $(1-z_+)$.
%

Taking the limit
eqs.~(\ref{eq:limit-S10-y}), 
(\ref{eq:limit-S10-lambda}) and 
(\ref{eq:limit-S10-kappa}) first, the model provides
the five-dimensional implementation of the path-integral measure for the left-handed Weyl field 
eq.~(\ref{eq:def-effective-action-again}). That is, 
\begin{eqnarray}
\label{eq:DW-EP-10}
S_{\rm DW/Ov} &=& \sum_{t=1}^{L_5} \sum_{x \in \Lambda}
\bar \psi(x,t) \big\{ [1+ a'_5 (D_{4 \rm w} - m_0)] \delta_{t t'}  - P_- \delta_{t+1, t'} -P_+ \delta_{t,t'+1} \big\} \psi(x, t') 
\nonumber\\
&-&
\sum_{x \in \Lambda}
\, \bar \psi(x, L_5)P_-[1+ a'_5(D_{4 \rm w} - m_0)] \psi(x,L_5)
\nonumber\\
&-&
\sum_{x \in \Lambda}  \frac{1}{2}\,\{ 
E^a(x) 
\psi^{\rm T}(x, L_5) 
 i \gamma_5 C_D {\rm T}^a 
\psi(x, L_5)
\nonumber\\
&& \qquad\qquad\qquad\qquad
+ 
\bar E^a(x)
\bar \psi(x, L_5) P_- i \gamma_5 C_D {{\rm T}^a}^\dagger 
\bar \psi(x, L_5)^T
\} 
\end{eqnarray}
and
\begin{eqnarray}
\big\langle 1 \big\rangle_F
&=&
\int 
{\cal D}[\psi] {\cal D}[\bar \psi] \,
{\cal D}[E] {\cal D}[\bar E] \,
\, {\rm e}^{- S_{\rm Ov}[\psi,\bar \psi, E^a, \bar E^a] }
 \\
&=& 
\frac{
\int 
\prod_{x,t} d \bar \psi(x,t) d \psi(x,t) \ 
{\cal D}[E] {\cal D}[\bar E] \,
\, {\rm e}^{- S_{\rm DW/Ov}[\psi, \bar \psi, E^a,\bar E^a] }
\big\vert_{\rm Dir}
}
{
\int 
\prod_{x,t} d \bar \psi(x,t) d \psi(x,t) \ 
\, {\rm e}^{- S_{\rm DW}[\psi, \bar \psi] }
\big\vert_{\rm AP}
}
\nonumber\\
&&
\nonumber\\
&=&
\frac{
\left\langle
 {\rm pf}
\left(
\begin{array}{cc}
- i \gamma_5 C_D {\rm T}^a E^a 
\delta_{t L_5} \delta_{t' L_5}
& - a'_5 (D_{\rm 5w} - m_0)^{' T} / 2
 \\
a'_5 (D_{\rm 5w} - m_0)' / 2
&  
- i \gamma_5 C_D P_- {\rm T}^{a \dagger} \bar E^a 
\delta_{t L_5} \delta_{t' L_5} 
\end{array}
\right)
\biggl\vert_{\rm Dir}
\right\rangle'_E
}{
{\rm det } \, a'_5(D_{\rm 5w} - m_0) \big\vert_{\rm AP} 
} ,
\end{eqnarray}
where 
$(D_{\rm 5w} - m_0)'_{t t'} = (D_{\rm 5w} - m_0)_{t t'} 
- \delta_{t L_5} P_- (D_{\rm 5w} - m_0)_{t t'} \delta_{t' L_5}$
and the limit $L_5 \rightarrow \infty$ ($a'_5 \rightarrow 0$) is understood.
Then, what we have argued in the previous sections 
about the four-dimensional model $S_{\rm Ov}$ 
implies that the domain wall fermion path-integral measure 
is properly saturated at around the right-handed boundary
with the fields, $\psi(x,L_5)$, $\bar \psi(x,L_5) P_-$, even
when the spin fields $E^a(x)$, $\bar E^a(x)$ have the disordered nature.
Moreover, the CP symmetry is restored in the limit
$L_5 \rightarrow \infty$. 

Thus the five-dimensional domain wall fermion model
defined by the action eq.~(\ref{eq:DW-EP-10})
provides a very explicit and well-defined implementation of 
the proposal by Creutz, Tytgat, Rebbi, Xue for the (more general) case of
the SO(10) chiral gauge theory.
And our four-dimensional lattice model 
defined with
the path-integration measure for the left-handed Weyl field
eq.~(\ref{eq:def-effective-action-again}) 
is nothing but the low energy effective theory of the five-dimensional domain wall model in the limit 
$L_5 \rightarrow \infty$ ($a'_5 \rightarrow 0$). 

In this repect, we note that one may define
the action of such a SO(10) domain wall fermion model simply by 
\begin{eqnarray}
S'_{\rm DW/Mi} &=& \sum_{t=1}^{L_5} \sum_{x \in \Lambda}
\bar \psi(x,t) \big\{ [1+ a'_5 ( D_{\rm 4w} - m_0)] \delta_{t t'}  - P_- \delta_{t+1, t'} -P_+ \delta_{t,t'+1} \big\} \psi(x, t') 
\nonumber\\
&-&
\sum_{x \in \Lambda}  \{ 
y \,
X^a(x) 
q_+^{\rm T}(x) i \gamma_5 C_D {\rm T}^a P_+ q_+(x)
+ 
\bar y \,
\bar X^a(x)
\bar q_+(x)  P_-  i \gamma_5 C_D {{\rm T}^a}^\dagger \bar q_+(x)^T
\} 
\nonumber\\
&+&
S_{X}[X^a] .
\end{eqnarray}
Note here that the bounary interaction terms 
are formulated solely with the boundary field variables, 
$q(x) = \psi_-(x, 1) + \psi_+(x, L_5)$, 
$\bar q(x) = \bar \psi_-(x, 1) + \bar \psi_+(x, L_5)$,
which are first introduced by Shamir and Furman\cite{Shamir:1993zy, Furman:ky}.
In this action,
the global U(1) symmetry of the five-dimensional Wilson fermion 
fields is broken to $Z_4$ by the boundary Yukawa couplings.
The CR$_5$ and P symmetries 
are also broken to the CPR$_5$ symmetry in the same manner.
We note, however, that this model ends up with the overlap fermion model $S'_{\rm Mi/Ov}$ with the Yukawa couplings eq.~(\ref{eq:singular-yukawa-coupling}) in the limit $L_5 \rightarrow \infty$ 
in the same subtraction scheme.
Therefore, this type of the Majorana-Yukawa couplings 
at the boundary are singular in the large limit.
%

\subsection{cf. Topological Insulators/Superconductors
with gapped boundary phases 
}
\label{subsec:cf-4DTI}

It has been proposed
by Wen, 
by You, BenTov and Xu, and 
by  You and Xu\cite{Wen:2013ppa,Wang:2013yta,You:2014oaa,You:2014vea}
to use the 4D Topological Insulators(TIs)/Superconductors(TSCs)
with the gapped boundary phases 
in order to formulate the 3+1D chiral gauge theories
in the Hamiltonian formalism.
%
These authors have considered 
the same 
4D TI with the time-reversal symmetry
defined by the following quantum Hamiltonian,
\begin{eqnarray}
\label{eq:4D-TI-nu=16}
\hat H_{\rm 4D TI} &=&
\sum_{i=1}^{\nu} \sum_{p} 
\hat a_i(p)^\dagger
\Big\{
\sum_{k=1}^{4} \alpha_k \sin(p_k) + \beta  \Big(
\big[\sum_{k=1}^{4} \cos(p_k) -4 \big] + m
\Big)
\Big\}
\hat a_i(p) ,
\end{eqnarray}
where $\hat a_i(p)$ and  $\hat a_i(p)^\dagger$ are fermionic annihilation-creation operators
in momentum space, satisfying the canonical anti-commutation relations,
$\hat a_i(p) \hat a_j(p')^\dagger + \hat a_j(p')^\dagger \hat a_i(p) = \delta_{p, p'} \delta_{i,j} $.
The alpha and beta matrices are chosen here as
$\alpha_k = \sigma_3 \otimes \sigma_k$ $(k=1,2,3)$, 
$\alpha_4 = \sigma_2 \otimes I$, 
and
$\beta = - \sigma_1 \otimes I$.
The generator of the time-reversal symmetry transformation is given as 
${\cal T} =K \, ( i  I \otimes \sigma_2 ) $, where $K$ stands for complex conjugation.
%
This 
4D quantum lattice fermion model is
nothing but the Hamiltonian formulation of 
Kaplan's 5-dim. domain wall fermion defined with the Wilson 
term.\cite{Kaplan:1992bt, Golterman:1992ub}
It was first examined by Creutz and Horvath\cite{Creutz:1994ny} to study the chiral property of the massless lattice fermions realized as Shockley surface states, and later by
X.-L. Qi, Hughes and S.H. Zhang\cite{Qi:2008ew} as a 4D extension of the 2D Integer Quantum Hall Effect (IQHE).

The insulator is in topological phase for $m>0$ and in trivial phase for $m<0$.  
On the 3D boundary of the domain wall due to the change of
the mass parameter from  $m >0$ to $m< 0$, there appear 
$\nu (\in \mathbb{Z})$ copies of 
two-component (right-handed) Weyl fermions at low energy $\vert p_l \vert \ll 0$ ($l=1,2,3$) assuming 
the thermodynamic limit of the 4D space.
These Weyl fermions are protected from acquiring mass
by the topological index defined by 
the second Chern character of the 
U(1) bundle associated with the connection 
$\sum_k \psi_k^\dagger \delta \psi_k$
and the time reversal symmetry.
This gapless boundary phase
can be described by the low energy effective Hamiltonian,
\begin{eqnarray}
\hat H^{\rm (bd)}_{\rm 3D}=
\sum_{i=1}^{\nu}
\int d^3 x \, 
\hat \psi_i(x)^\dagger 
\Big\{
\sum_{l=1}^{3} (-i) \sigma_l \partial_l 
\Big\}
\hat \psi_i(x).
\end{eqnarray}
The generator of the time-reversal symmetry transformation 
acting the effective Hamiltonian is given as 
${\cal T} =K \, ( i \sigma_2 )$.
%

For the case $\nu=16$, the authors have proposed 
the boundary interaction terms to fully gap the boundary phase
with the sixteen massless Weyl fermions, or the bulk
interaction terms
to be able to interpolate between the topological and trivial phases without closing the mass gap nor breaking the symmetries.
In fact, 
the boundary/bulk interaction terms 
introduced in these works 
are the SO(10)-invariant quartic (or Yukawa) term
\begin{eqnarray}
\hat {\cal O}(x) &=&
\frac{1}{2}
\big[ 
\hat \psi(x)^{\rm T} C_{\rm D} 
\check{{\rm T}}^a \hat \psi(x) 
-
\hat \psi(x)^\dagger C_{\rm D} 
\check{{\rm T}}^{a \dagger} \hat\psi(x)^{\dagger {\rm T}}
\big]^2 
\end{eqnarray}
assuming that the sixteen massless Weyl fermions are 
in the $\underbar{16}$ of SO(10)
and its descendants with reduced symmetries,
SO(7)$\times$SO(3) and SO(6)$\times$SO(4)(=SU(4)$\times$SU(2)$\times$SU(2)).
%
It is quite interesting to see that these are essentially identical to 
the SO(10)-invariant quartic terms of the 't Hooft vertices,
$T_+(x)$, $\bar T_+(x)$, 
\begin{eqnarray}
{\cal O}_{\rm T}(x) &=&
T_+(x) + \bar T_+(x)
\\
&=&
\frac{1}{2^3}
\big[ 
\psi^{\rm T}(x)  C_D {\rm T}^a \psi(x) 
\big]^2
+ 
\frac{1}{2^3}
\big[
\bar \psi(x)  C_D {{\rm T}^a}^\dagger \bar\psi(x)^T
\big]^2 ,
\end{eqnarray}
and their descendants.

%

Wen, in particular, have considered the SO(10) chiral gauge theory
as a target theory\cite{Wen:2013ppa}. The author
have proposed to use
the following SO(10)-invariant boundary interaction terms, 
\begin{eqnarray}
\hat H_{\rm 3D, 10} 
&=& 
\int d^3 x \, 
\Big\{
\hat \psi(x)^T 
i \sigma_2 
{\check {\rm T}}^a \phi^a(x) 
\hat \psi (x)
\nonumber\\
&&\qquad\qquad
-
\hat \psi(x)^\dagger
i \sigma_2 
{\check {\rm T}}^{a \dagger} \phi^a(x) 
\hat \psi (x)^\dagger
+
{\cal H}[\phi^a(x)] 
\Big\} ,
\end{eqnarray}
where the Weyl field $\hat \psi(x) = \{ \hat \psi_s(x) \}$($s=1,\cdots,16$) is assumed to form the irreducible spinor representation $\underbar{16}$ of SO(10), 
and ${\cal H}[\phi^a(x)]$ stands for  the kinetic and potential terms of the SO(10) vector real scalar field $\phi^a(x)$.
It is assumed that ${\cal H}[\phi^a(x)]$ is chosen
to make $\phi^a(x) \phi^a(x) = M^2 \not = 0 $ without
breaking the SO(10) symmetry, $\langle \phi^a(x) \rangle = 0$.
It is also assumed that the correlation length of 
the field $\phi^a(x)$, $\xi_\phi$, is much larger than the lattice spacing so that one can then expect
the Yukawa coupling to generate a (Majorana-type) mass for all
the sixteen Weyl fermions.
The author then discusses that topological defects 
with $\phi^a(x) =0$ at some points or in some regions 
of the 3+1D space-time, which can give rise to 
massless (gapless) fermionic excitations, do not exist
for the SO(10) vector real scalar field $\phi^a(x)$
satisfying  $\phi^a(x) \phi^a(x) = M^2 = {\rm const.}$,
because $\pi_d(S^9) = 0$ $( 0 \le d < 9 )$: 
$\pi_3(S^9) = 0$ against point-defects like the instantons,
$\pi_2(S^9) = 0$ against line-defect like hedgehog solitons,
$\pi_1(S^9) = 0$ against membrane-defect like vortex lines,
and 
$\pi_0(S^9) = 0$ against 3-brane-defect like domain walls. 
The author also points out that the WZW term does not exist,
because $\pi_5(S_9)=0$.
Based on these assumptions and considerations, the author have argued that 
the boundary interaction $\hat H_{\rm 3D, 10} $
can make the boundary phase with the sixteen massless Weyl fermions fully gapped without breaking the SO(10) 
and time-reversal symmetries.

%


As mentioned above, 
the 4D TI, eq.~(\ref{eq:4D-TI-nu=16}), 
with the boundary phases
of the $\nu$ massless Weyl fermions
is nothing but the Hamiltonian formulation of 
Kaplan's 5-dim. domain wall fermion defined with 
the Wilson term.
Then 
the 4D TI(TSC) 
can be formulated 
in the framework of 4+1D Euclidean path-integral quantization
using the five-dimensional lattice domain wall fermion
including suitable boundary/bulk interaction terms.
Using
this 5~dim. lattice formulation of the 4D TI(TSC),
one can 
study the effect of the boundary/bulk multi-fermion (Yukawa) interactions on the properties of the 4D TI(TSC),
in particular,
the behaviors of the {\em proposed} 3D gapped boundary phases of the 4D TI(TSC) of $\nu =16$, 
by using the various perturbative/non-perturbative methods 
in the framework of lattice field theory.

The domain wall fermion models $S_{\rm DW/Mi}$ 
and $S_{\rm DW/Ov}$ discussed in the previous sec.~\ref{subsec:cf-DW-EP}
provide such a formulation.
%
In fact, the partition function of the 4D TI(TSC) of $\nu=16$
with the SO(10)-invariant boundary interaction terms 
can be defined precisely by
the domain wall fermion model $S_{\rm DW/Mi}$:
\begin{eqnarray}
Z_{\rm 4D TI/\nu=16} 
&\equiv& 
\int 
\prod_{t=-L_5+1}^{L_5} {\cal D}[\psi(t)]{\cal D}[\bar \psi(t)] \, 
{\cal D}[E] {\cal D}[\bar E] \,
\, {\rm e}^{- S_{\rm DW/Mi}[\psi, \bar \psi, E^a,\bar E^a] }
\big\vert_{\rm Dir}
\nonumber\\
&=& \det \, a'_5 ( D_{\rm 5w} - m_0 ) \big\vert_{\rm AP} \, 
Z_{\rm Ov/Mi}
 \qquad
[ L_5 \rightarrow \infty \, (a'_5 \rightarrow 0) ] .
\end{eqnarray}
In this 4+1D lattice model, 
one can fix the radii of the SO(10) spin fields to unity as
\begin{eqnarray}
E^a(x) E^a(x) =1 , \quad \bar E^a(x)\bar E^a(x)=1
\end{eqnarray}
from the beginning
by taking the limit
eqs.~(\ref{eq:limit-S10-y}), 
(\ref{eq:limit-S10-lambda}) and 
(\ref{eq:limit-S10-kappa}).
Moreover, one can take the limit of the large
Majorana-Yukawa couplings,
\begin{eqnarray}
\label{eq:limit-S10-y-second}
&& y = \bar y, \quad \frac{z_+}{\sqrt{y \bar y}} \rightarrow 0 .
\end{eqnarray}
Then one ends up with the domain wall fermion model 
$S_{\rm DW/Ov}$
and the four-dimensional lattice model 
$S_{\rm Ov}$ as a low energy effective lattice theory for
the 
edge modes at the boundaries.
The partition function of the 4D TI(TSC) of $\nu=16$ then reads
\begin{eqnarray}
Z_{\rm 4D TI/\nu=16} 
&=& \det \, a'_5 ( D_{\rm 5w} - m_0 ) \big\vert_{\rm AP} \, 
Z_{\rm Ov}
 \qquad\qquad
[ L_5 \rightarrow \infty \, (a'_5 \rightarrow 0) ]
\nonumber\\
&=& \det \, a'_5 ( D_{\rm 5w} - m_0 ) \big\vert_{\rm AP} \, \,
\det ( \bar v D v  ) \,\,
\big\langle 
{\rm pf} (u^T i \gamma_5 C_{\rm D}{\rm T}^a E^a u) 
\big\rangle'_{E} \, .
\end{eqnarray}
Thus our four-dimensional lattice model 
of the SO(10) chiral gauge theory, 
defined with
the path-integration measure for the left-handed Weyl field
eq.~(\ref{eq:def-effective-action-again}),
gives a direct and well-defined description of
the $\nu=16$ 3D gapped/gapless boundary phases
proposed by Wen\cite{Wen:2013ppa}
in the framework of 3+1D {\em local} lattice theory.
%
%
In sections~\ref{sec:path-integration-measure} 
and \ref{sec:saturation-right-handed-measure},
we have argued that 
the pfaffian 
${\rm pf} (u^T i \gamma_5 C_{\rm D}{\rm T}^a E^a u) $
is real positive semi-definite
and 
its path-integration over $E^a(x)$ ($a=1, \cdots, 10$) 
is non-vanishing
for the trivial link field $U(x,\mu)=1$ 
in the weak-gauge coupling limit.
This implies that 
the partition function of 
the boundary phase is real-positive,
\begin{eqnarray}
\big\langle 
{\rm pf} (u^T i \gamma_5 C_{\rm D}{\rm T}^a E^a u) 
\big\rangle_{E}   \,  >  0 
\qquad ( U(x,\mu) =1 ) ,
\end{eqnarray}
and
the massless singularity associated with the
sixteen right-handed Weyl fermions is absent.
We have also argued that the auxiliary spin fields show the disordered nature for 
$m_0 < 2$ 
and the correlation functions of the 
right-handed fermions 
and the auxiliary spin fields are short-ranged
(except the unknown one, 
$\big\langle 
   \psi_+ 
   \big[  
  \psi_+^{\rm T}   i \gamma_5 C_D {\rm T}^a E^a \hat P_- 
   \big] \,    \big\rangle_F$ ).
These results 
provide analytical evidences that
the $\nu=16$ 3D boundary phase is fully gapped indeed
and respects the SO(10) symmetry.

We will discuss the details of the description
of the gapped boundary phases of 1-4D TI/TSC 
in terms of overlap fermions elsewhere\cite{Kikukawa:2017c}.

%

%

\section{Conclusion}
\label{sec:conclusion}

In this paper, we formulated  
the SO(10) chiral gauge theory with Weyl fermions in the sixteen dimensional spinor representation $\underbar{16}$ within the framework of the Overlap fermion/the Ginsparg-Wilson relation.
We defined the path-integral measure of the left-handed Weyl fermions
with all the components of the Dirac field, but the right-handed part of which is saturated completely by inserting a suitable product of the SO(10)-invariant  't Hooft vertices in terms of the right-handed field.  The definition of the measure applies to all possible topological sectors.
In fact, we examined in detail 
the two cases of the trivial link field
and of the SU(2) link fields with non-vanishing topological charge $Q (\not = 0)$. 
We argued that 
the path-integration of the pfaffian 
${\rm pf}( u^{\rm T}\, i \gamma_5 C_D {\rm T}^a E^a u )$ over the auxiliary spin field $E^a(x)$ gives a non-zero result,
\begin{equation*}
\int {\cal D}[E] \, 
{\rm pf}  \big(u^{\rm T} \,  i \gamma_5 C_D {\rm T}^a E^a u \big)
= c \,[ U(x,\mu) ] \not = 0 
\end{equation*}
and that
the measure of the right-handed field, 
${\cal D}_\star[\psi_+]$, is indeed saturated 
completely by inserting the product of the 't Hooft vertex 
$T_+(x) [\psi_+]$,
while
the SO(10) symmetry does not break spontaneously in the thermodynamic limit.
The measure possesses all required transformation properties under lattice symmetries
and 
we gave a proof of the CP-invariance of the  induced effective action.
The global U(1) symmetry of the left-handed 
field is anomalous due to the non-trivial trasformation of the measure, while that of the right-handed field is explicitly  broken by the 't Hooft vertices.
%

There remains the issue of smoothness/locality in the gauge-field dependence of the Weyl fermion measure term,
\begin{equation}
 -i \mathfrak{T}_\eta 
\equiv
- \,{\rm Tr} \big\{ 
\delta_\eta \hat P_+  
 \big\langle 
\psi_+ \big[ \psi_+^{\rm T} i \gamma_5 C_D {\rm T}^a E^a 
\big] 
\big\rangle_F  \big\}\, 
\big\slash 
\big\langle 1 \big\rangle_F .
\end{equation}
This question is of not quite dynamical but non-perturbative nature in our model, 
involving
the path-integration of the spin field $E^a(x)$ with the 
weight of the pfaffian 
${\rm pf}( u^T  i \gamma_5 C_D  T^a E^a u) $, 
which is complex in general. 
But 
the question is well-defined and 
it is highly desirable to 
establish these properties rigorously, 
if possible. 
It can be addressed 
in the weak gauge-coupling expansion at least 
because
the pfaffian 
is positive semi-definite 
and 
Monte Carlo methods are applicable 
to evaluate the correlation functions and the vertex functions. 
We leave this important and interesting  question
for our future study.\footnote{
Two-dimensional abelian chiral gauge theories
can be formulated in the similar manner.
Those include 
the $1^4(-1)^4$ axial gauge model
and 
the $21(-1)^3$ chiral gauge model.
In these models, the two-point vertex function
of the U(1) gauge field in the mirror sector was computed
through
Monte Carlo simulations.
The simulation results indeed showed 
a numerical evidence  
that the two-point vertex function is regular 
and 
that the induced measure term
$-i \mathfrak{T}_\eta$ is a local functional of the link field.
See \cite{Kikukawa:2017gvk} for detail.
}

We discussed the relations of our formulation to
other approaches/proposals
to decouple the species-doublers and mirror fermions
such as
the Eichten-Preskill model\cite{Eichten:1985ft, Golterman:1992yha}, 
the Mirror-fermion model
using Ginsparg-Wilson fermions\cite{Bhattacharya:2006dc,Giedt:2007qg,Poppitz:2007tu,Poppitz:2008au,Poppitz:2009gt,Poppitz:2010at,Chen:2012di,Giedt:2014pha}, 
Domain wall fermions with the boundary Eichten-Preskill term\cite{Creutz:1996xc}, 
the recent studies on the PMS phase/``Mass without symmetry breaking''\cite{Ayyar:2014eua, Ayyar:2015lrd, Ayyar:2016lxq, Catterall:2015zua, Catterall:2016dzf, Catterall:2017ogi,Schaich:2017czc}
and 
4D TI/TSCs with gapped boundary 
phases\cite{Wen:2013ppa,Wang:2013yta,You:2014oaa,You:2014vea,DeMarco:2017gcb}. 
These discussions, we hope, 
clarified the similarity and
the difference in technical detail
and showed
that our proposal is a well-defined testing ground for that basic question.










\appendix

\section{Dirac gamma matrices}
\label{app:rep-the-gamma-matrix}

Dirac gamma matrices (The Clifford algebra of $2^{[5/2]}$ dimensions):
\begin{eqnarray}
&&
\gamma_0=\left( \begin{array}{cc} 0 & I \\ I & 0 \end{array}\right), 
\gamma_1=\left( \begin{array}{cc} 0 & i \sigma_1 \\ -i \sigma_1 & 0 \end{array}\right), 
\gamma_2=\left( \begin{array}{cc} 0 & i \sigma_2 \\ -i \sigma_2 & 0 \end{array}\right), 
\gamma_3=\left( \begin{array}{cc} 0 & i \sigma_3 \\ -i \sigma_3 & 0 \end{array}\right), \\
&&
\gamma_5=\left( \begin{array}{cc} I & 0 \\ 0 & -I  \end{array}\right), \\
&& C_D = \left( \begin{array}{cc}  i \sigma_2 & 0 \\ 0 & -i \sigma_2  \end{array}\right) . 
\end{eqnarray}
\begin{eqnarray}
&& 
\{ \gamma_\mu, \gamma_\nu \} = 2 \delta_{\mu \nu} \, ; \quad 
\gamma_\mu^\dagger =\gamma_\mu \quad (\mu=0,1,2,3) , \\
&&
\gamma_5 = \gamma_0 \gamma_1 \gamma_2 \gamma_3, \quad  \{ \gamma_5 , \gamma_\nu \} = 0 \quad (\mu=0,1,2,3) ,\\
&&
C_D=\gamma_2\gamma_0 ,  \\
&& 
C_D \gamma_\mu C_D^{-1} = - \gamma_\mu^{\rm T},  \,
C_D \gamma_5 C_D^{-1} = \gamma_5 \, , \, 
C_D^{\rm T} = C_D^{-1}= C_D^\dagger = -C_D \, .
\end{eqnarray}

%
%

\section{SO(10) gamma matrices}
\label{sec:appendix-so10-representation}

\noindent
SO(10) gamma matrices (The Clifford algebra of $2^{[11/2]}$ dimensions):

\begin{eqnarray}
\Gamma^1 &=& \tau_1 \times \tau_1 \times \tau_1 \times \tau_1 \times \tau_1 , \\
\Gamma^2 &=& \tau_2 \times \tau_1 \times \tau_1 \times \tau_1 \times \tau_1  , \\
\Gamma^3 &=& \tau_3 \times \tau_1 \times \tau_1 \times \tau_1 \times \tau_1 , \\
\Gamma^4 &=& I               \times \tau_2 \times \tau_1 \times \tau_1 \times \tau_1  , \\
\Gamma^5 &=& I                \times \tau_3 \times \tau_1 \times \tau_1 \times \tau_1 , \\
\Gamma^6 &=& I               \times I  \times \tau_2 \times \tau_1 \times \tau_1 , \\
\Gamma^7 &=& I                \times I  \times \tau_3 \times \tau_1 \times \tau_1 , \\
\Gamma^8 &=& I               \times I  \times I \times \tau_2 \times \tau_1 , \\
\Gamma^9 &=& I                \times I  \times I \times \tau_3 \times \tau_1 , \\
\Gamma^{10} &=& I               \times I  \times I \times I \times \tau_2 , \\
\Gamma^{11} &=& I                \times I  \times I \times I \times \tau_3, \\
{\rm C} &=& i \tau_2 \times \tau_3 \times \tau_2 \times \tau_3 \times \tau_2 . 
\end{eqnarray}
\begin{eqnarray}
&&
\Gamma^a \Gamma^b + \Gamma^b \Gamma^a = 2 \delta^{ab} \, ; \quad 
{\Gamma^a}^\dagger = \Gamma^a  \quad (a=1, \cdots, 10) , \\
&&
\Gamma^{11}= -i \Gamma^1 \Gamma^2 \cdots \Gamma^{10}, \quad
[ \Gamma^{11}, \Gamma^a ] =0 \quad (a=1, \cdots, 10) , \\
&&
{\rm C} \Gamma^a {\rm C}^{-1} = - \{ \Gamma_a \}^T, \,  \,   
{\rm C} \Gamma^{11} {\rm C}^{-1} =  - \Gamma_{11} ,  \, \, ; \, 
{\rm C}^T = {\rm C}^{-1} = {\rm C}^\dagger = - {\rm C} .
\end{eqnarray}

\noindent
The ${\rm T}$ matrices

\begin{equation}
{\rm T}^a = {\rm C} \Gamma^a \, ; \qquad 
{{\rm T}^a}^T = {\rm T}^a
\end{equation}

\begin{eqnarray*}
{\rm T}^1 &=&
 i (-i)(+i) (-i)(+i)(-i)  \, 
 \tau_3 \times \tau_2 \times \tau_3 \times \tau_2 \times \tau_3 , \\
{\rm T}^2 &=& 
i (+1)(+i) (-i)(+i)(-i)  \,
 I \times \tau_2 \times \tau_3 \times \tau_2 \times \tau_3 , \\
{\rm T}^3 &=&
i (+i)(+i) (-i)(+i)(-i)  \,
 \tau_1 \times \tau_2 \times \tau_3 \times \tau_2 \times \tau_3 , \\
{\rm T}^4 &=& 
 i (+1)(-i) (-i)(+i)(-i)  \, 
 \tau_2 \times \tau_1 \times \tau_3 \times \tau_2 \times \tau_3 , \\
{\rm T}^5 &=&
 i (+1)(+1) (-i)(+i)(-i)  \, 
 \tau_2 \times I \times \tau_3 \times \tau_2 \times \tau_3 , \\
{\rm T}^6 &=&
i (+1)(+1) (+1)(+i)(-i)  \, 
 \tau_2 \times \tau_3 \times I \times \tau_2 \times \tau_3 , \\
{\rm T}^7 &=&
i (+1)(+1) (+i)(+i)(-i)  \, 
 \tau_2 \times \tau_3 \times \tau_1 \times \tau_2 \times \tau_3 , \\
{\rm T}^8 &=&
i (+1)(+1) (+1)(-i)(-i)  \, 
 \tau_2 \times \tau_3 \times \tau_2 \times \tau_1 \times \tau_3 , \\
{\rm T}^9 &=&
i (+1)(+1) (+1)(+1)(-i)  \, 
 \tau_2 \times \tau_3 \times \tau_2 \times I \times \tau_3 , \\
{\rm T}^{10} &=& 
i (+1)(+1) (+1)(+1)(+1)  \, 
 \tau_2 \times \tau_3 \times \tau_2 \times \tau_3 \times I 
\end{eqnarray*}

\noindent
The reduced Clliford algebra of $2^{[9/2]}$ 

\begin{eqnarray}
\Gamma^{a'} &=& \check \Gamma^{a'} \times \tau_1 \quad(a'=1, \cdots, 9),\\
{\rm C} &=& \check {\rm C} \times \tau_2 .
\end{eqnarray}

\noindent
The reduced $T$ matrices

\begin{eqnarray}
{\rm T}^{a'} &=& \check {\rm T}^{a'} \times \tau_3, \\
{\rm T}^{10} &=& \check {\rm T}^{10} \times I = \check {\rm C} \times I .
\end{eqnarray}

\begin{eqnarray}
{{\rm T}^{10}}^\dagger {\rm T}^{a'} &=& \Gamma^{10} \Gamma^{a'} = - i \, \check \Gamma^{a'} \times \tau_3 .
\end{eqnarray}

\section{Chiral basis in the weak coupling limit}
\label{app:chiral-basis-free}

\begin{equation}
H=\gamma_5 (D_w - m_0) = \frac{1}{L^4} \sum_p  \,{\rm e}^{ip(x-y)}
\left( \begin{array}{cc}   b(p) I  & c(p) \\ c^\dagger(p) & -b(p) I  \end{array} \right), 
\end{equation}
where
\begin{eqnarray}
b(p) &=& \big\{ \sum_\mu (1- \cos p_\mu) - m_0 \big\}, \\
c(p) &=& I  \{ i \sin p_0 \} - \sum_k \sigma_k \sin p_k . 
\end{eqnarray}

\noindent
Orthonormal chiral basis:  $ j = (p,s) $
\begin{eqnarray}
u_{j}(x)  &=& {\rm e}^{i p x} \, u(p, s) \, ; \quad
u(0, s) = \left( \begin{array}{c} \chi_s \\ 0 \end{array} \right) , \,\,
u(p, s) =
\left( \begin{array}{c} c \chi_s \\ 
                                  -( \omega+ b )\chi_s      \end{array}\right) 
                                  / \sqrt{2 \omega ( \omega+ b )}   \quad (p \not = 0 ) , 
                                  \,
                                                                    \nonumber \\
                                                                    && \\
v_{j}(x)  &=& {\rm e}^{i p x}  \, v(p,s) \, ;  \quad
v(0, s) = \left( \begin{array}{c} 0 \\ \chi_s  \end{array} \right) , \,\,
v(p,s) = \left( \begin{array}{c} +(\omega+b)\chi_s \\ 
                                        c^\dagger \chi_s      \end{array}\right) 
                                        / \sqrt{2 \omega( \omega+b )}   \quad (p \not = 0 ) ,
                                                                    \nonumber \\
\end{eqnarray}
where 
\begin{eqnarray}
\omega(p) &=& \sqrt{\sum_\mu \sin^2 ( p_\mu )   
+ \big\{ \sum_\mu (1- \cos (p_\mu) ) - m_0 \big\}^2} .
\end{eqnarray}

\noindent
Majorana-mass-type inner products:

\begin{eqnarray}
( u^T_j , \gamma_5 C_D u_{j'}) 
&=& \delta_{p+p', 0} \,  u^T(-p',s) \gamma_5 C_D u(p',s')  \\
&=& \delta_{p+p', 0} \, \chi_s^T i\sigma_2 \chi_{s'} \\
&=& \delta_{p+p', 0} \,\epsilon_{s s'} , 
\end{eqnarray}

\acknowledgments
The author would like to thank M.~Sato, H.~Fujii, M.~Kato, Y.~Okawa, T.~Okuda for enlightening discussions.
This work is supported in part by JSPS KAKENHI Grant Numbers 24540253, 16K05313 and 25287049.

\end{document}